\documentclass[11pt]{article}
\usepackage{amsmath,amssymb,epsfig,color,bbold}
\usepackage{feynmf}
\makeatletter
\newcommand{\vast}{\bBigg@{4}}
\newcommand{\Vast}{\bBigg@{5}}
\makeatother

\usepackage{cite}
\usepackage{feynmf}
\usepackage{psfrag}
\usepackage{graphicx}
\usepackage{bm}

\newcommand{\be}{\begin{equation}}  
\newcommand{\ee}{\end{equation}} 
\def\slash#1{#1\!\!\!/}  
\newcommand{\vslash}{v\!\!\!/\,}
\newcommand{\dslash}{\partial\!\!\!/\,\,}
\newcommand{\Dslash}{D\!\!\!\!/\,\,}
\newcommand{\nl}{\nonumber \\ }

\newcommand{\order}{{\cal O}}
\newcommand{\mym}{m}
\newcommand{\mymu}{m_1}
\newcommand{\mymd}{m_2}
\DeclareMathOperator{\arccot}{arccot}

\long\def\symbolfootnote[#1]#2{\begingroup%
\def\thefootnote{\fnsymbol{footnote}}\footnote[#1]{#2}\endgroup}

\allowdisplaybreaks

\begin{document}

\begin{fmffile}{feynmffile} 
\fmfcmd{%
vardef middir(expr p,ang) = dir(angle direction length(p)/2 of p + ang) enddef;
style_def arrow_left expr p = shrink(.7); cfill(arrow p shifted(4thick*middir(p,90))); endshrink enddef;
style_def arrow_left_more expr p = shrink(.7); cfill(arrow p shifted(6thick*middir(p,90))); endshrink enddef;
style_def arrow_right expr p = shrink(.7); cfill(arrow p shifted(4thick*middir(p,-90))); endshrink enddef;}

\begin{titlepage}

\begin{flushright}
EFI Preprint 13-34\\
January 14, 2014 
\end{flushright}

\vspace{0.7cm}
\begin{center}
\Large\bf 
Standard Model anatomy of WIMP dark matter direct detection 
\nl
I: weak-scale matching
\end{center}

\vspace{0.8cm}
\begin{center}
{\sc   Richard J. Hill\symbolfootnote[1]{
richardhill@uchicago.edu, 
} and Mikhail P. Solon\symbolfootnote[2]{mpsolon@uchicago.edu}}\\
\vspace{0.4cm}
{\it Enrico Fermi Institute and Department of Physics \\
The University of Chicago, Chicago, Illinois, 60637, USA
}
\end{center}
\vspace{1.0cm}
\begin{abstract}
\vspace{0.2cm}
\noindent We present formalism necessary to determine weak-scale matching coefficients in the computation of scattering cross sections for putative dark matter candidates interacting with the Standard Model. Particular attention is paid to the heavy-particle limit. A consistent renormalization scheme in the presence of nontrivial residual masses is implemented. Two-loop diagrams appearing in the matching to gluon operators are evaluated. Details are given for the computation of matching coefficients in the universal limit of WIMP-nucleon scattering for pure states of arbitrary quantum numbers, and for singlet-doublet and doublet-triplet mixed states.
\end{abstract}
\vfil

\end{titlepage}

\section{Introduction} 

The compelling evidence for dark matter (DM) inconsistent with
Standard Model (SM) particles has motivated many theoretical studies
and experimental searches to elucidate its particle nature. In
particular, the paradigm of Weakly Interacting Massive Particles
(WIMPs) continues to play a prominent role, and experiments in the
present decade should explore a significant region of remaining WIMP
parameter space \cite{Cushman:2013zza}. Given the multitude of WIMP
candidates and search strategies, it is imperative to develop
theoretical formalism to delineate the possible interactions of DM
with known particles, making clear which uncertainties are inherently
model dependent and which can, at least in principle, be improved by
further SM analysis. 

Even in many seemingly simple cases, determination of WIMP-nucleon
cross sections demands an intricate analysis of competing amplitudes
mediated by SM particles (see
e.g.,~\cite{Cirelli:2005uq,Hisano:2010ct, Freytsis:2010ne,
Hisano:2011cs, Hill:2011be, Klasen:2013btp}). In this paper we set out
the formalism for electroweak-scale matching computations for
application both to theories with specified ultraviolet (UV)
completion (e.g., supersymmetric
models~\cite{Jungman:1995df,Feng:2013pwa}), and to the heavy WIMP
limit where theoretical control is maintained in the absence of a
specified UV completion~\cite{Hill:2011be}.  We review relevant
aspects of techniques such as the background field method for matching
to gluon operators \cite{Novikov:1983gd, Hisano:2010ct}, the extension
of the onshell renormalization scheme for WIMP couplings to the
electroweak SM, and the treatment of effective theory
subtractions. Direct detection experimental
constraints~\cite{Aprile:2012nq,Akerib:2013tjd}, together with other
phenomenological bounds such as LHC searches, may plausibly indicate
that new particles must have mass somewhat above the mass of
electroweak-scale particles ($M \gg m_W$).  In this regime, the
prospects for direct detection become more challenging, but in a
precise sense more constrained due to heavy particle
universality. Extending the particle content of the SM by one or a few
electroweak multiplets, the heavy particle limit implies highly
predictive cross sections with minimal parametric input beyond the SM.
This limit is thus both physically interesting, as well as a
useful pedagogical illustration.  Within the heavy WIMP framework, we
present a complete reduction of the required one- and two-loop amplitudes into a
basis of heavy-particle loop integrals with nonzero residual mass. 

Although we aim for generality, for definiteness throughout the paper,
we illustrate these methods for the case where the lightest,
electrically neutral particle of the new sector corresponds to a
self-conjugate field (e.g., a Majorana fermion or real scalar)
stabilized by a $Z_2$ symmetry, deriving from a theory consisting of
one or two $SU(2)_W\times U(1)_Y$ multiplets beyond the SM particle
content~\cite{Cirelli:2005uq, Hisano:2011cs, Hill:2011be, Klasen:2013btp,
Bagnasco:1993st, Hambye:2009pw, Kilic:2009mi, Frandsen:2009mi, 
Bai:2010qg, Hur:2007uz, Cohen:2011ec, Campbell:2011iw, 
JosseMichaux:2012wj, Cheung:2012qy, Chang:2013oia, Bai:2013iqa,
DiFranzo:2013vra, Earl:2013fpa}. An important
simplification occurs when a scale separation exists between SM masses
($\sim\!m_W$) and the lightest new particle mass ($\sim\!M$),
allowing an expansion in $m_W/M$. We consider in detail the limit $M
\gg m_W$ where universal behavior appears, and present the necessary
heavy particle effective theory tools for such an analysis. For these
SM extensions, we present details of the first complete computation of
the matching at leading order in perturbation theory onto the full
basis of operators at the electroweak scale
\cite{Hill:2011be}.  

The field of DM direct detection is by now a mature subject.%
\footnote{A subset of recent work in the field may be found in the Snowmass review~\cite{Cushman:2013zza}.}  
Early treatments of QCD effects in neutralino-nucleon scattering 
include the works of Drees and Nojiri~\cite{Drees:1992rr}.
Basic aspects of formalism may be found in the review of Jungman, 
Griest and Kamionkowski~\cite{Jungman:1995df}. However, the last 
few years have witnessed the discovery and mass measurement for a 
SM-like Higgs boson~\cite{Aad:2012tfa,Chatrchyan:2012ufa}, new constraints on 
the mass scale of particles beyond the SM~\cite{Gershtein:2013iqa}, 
and important computational advances in lattice QCD~\cite{Ellis:2008hf,Giedt:2009mr}. 
A complete description of DM-SM interactions is now possible in many 
SM extensions but demands the systematic treatment of QCD effects 
and uncertainties, including the consideration of loop amplitudes 
that are typically neglected in the $m_W\sim M$ regime, but which 
contribute at leading order in the general case.

In this work we extend some aspects of heavy particle formalism
familiar from heavy quark effective theory~\cite{Neubert:1993mb} for 
DM applications, and we hope that a detailed treatment will serve as pedagogy for the DM
practitioner unfamiliar with heavy particle tools. Both within and
beyond the heavy particle limit, distinguishing between different DM
candidates in direct detection experiments demands careful
treatment of QCD corrections when passing from a theory renormalized
at the electroweak scale to a low-energy theory of quarks and gluons
where hadronic matrix elements are evaluated. A companion paper treats
this separate problem for applications involving a range of dark
matter candidates~\cite{part2}.

The remainder of the paper is structured as follows. In
Sec.~\ref{sec:heavymethods}, we briefly review aspects of heavy
particle effective theory relevant for DM
applications. Section~\ref{sec:weakscaleEFT} specifies the
operator basis for DM-SM interactions at the weak scale relevant for
spin-independent, low-velocity scattering with nucleons. In
Sec.~\ref{sec:lagrangians}, we construct the effective theory for
one or two heavy electroweak multiplets interacting with SM Higgs and
electroweak gauge fields, accounting for masses induced by electroweak
symmetry breaking (EWSB), 
and presenting the lagrangian in terms of mass eigenstate fields from 
which the complete set of Feynman rules may be easily derived. 
In Sec.~\ref{sec:renscheme}, we define an extension of the
onshell renormalization scheme for the electroweak SM for a consistent
loop-level evaluation of amplitudes. Section~\ref{sec:matchingcalculation} presents the details of the matching
calculation, including the systematic reduction of heavy-particle
integrals, and the implementation of background field techniques for
gluon operators. We present the bare matching coefficients in
Sec.~\ref{sec:pheno}, and conclude with a summary in
Sec.~\ref{sec:summary}. 

\section{Heavy particle effective field theory for dark matter
applications\label{sec:heavymethods} } 

Heavy particle methods may be used to efficiently
describe the interactions of DM, of mass $M$, with much lighter
degrees of freedom such as those of $n_f=5$ flavor QCD (in the case
$m_b \ll M$, where $m_b$ is the bottom quark mass) or those of the SM
electroweak sector (in the case $m_W \ll M$, where $m_W$ is the
$W^\pm$ boson mass). Let us briefly review a few aspects of heavy
particle effective theory relevant for the DM applications in
Secs.~\ref{sec:weakscaleEFT} and \ref{sec:lagrangians}.

A heavy-particle field, $h_v$, is identified with a representation of
the little group for massive particles, and carries a label $v$
associated with the time-like unit vector $v^\mu$ that defines the
little group~\cite{Heinonen:2012km}. The little group for massive
particles is isomorphic to $SO(3)$, and therefore has field
representations carrying spin $s=0,1/2,1, \dots \ $. We may write such
fields in covariant notation using a Dirac spinor-vector with
appropriate constraints. For example, a spin-$1/2$ heavy-particle
field has $2(1/2)+1$ degrees of freedom and can be written as a Dirac
spinor, $h_v$, obeying $\slash{v} h_v = h_v$ as a projection
constraint.%
\footnote{The case of arbitrary spin is discussed in
Appendix A.1 of Ref.~\cite{Heinonen:2012km}.} 
For integer spin we define $\bar{h}_v \equiv h_v^\dagger$, while for half-integer spin $h_v$ carries spinor indices and we define $\bar{h}_v
\equiv h_v^\dagger \gamma^0$.
Additionally, for
self-conjugate fields the simultaneous operations 
\be\label{eq:vparity}  v^\mu \to -v^\mu\,, \quad h_v \to h_v^c \equiv
{\cal C} h_v^* \,,  \ee
where ${\cal C}$ is the charge conjugation matrix,  implement a
symmetry of the heavy-particle lagrangian.%
\footnote{A discussion of
this invariance is given in Appendix A.2 of
Ref.~\cite{Heinonen:2012km}; for a heavy Majorana fermion, see also
\cite{Kopp:2011gg}.}

Having specified the building blocks, interactions with heavy-particle
fields can be constructed in the usual way. We write down the most
general set of gauge-invariant and Lorentz-covariant operators in
terms of the heavy field $h_v$, the time-like unit vector $v^\mu$, and
other relativistic degrees of freedom up to a given order in the $1/M$
power counting. In the case of a
self-conjugate heavy particle, such as that derived from a Majorana
fermion or a real scalar of the underlying UV completion, the
invariance (\ref{eq:vparity}) is additionally imposed. 

Lorentz invariance should also be implemented using the heavy-particle
boost transformation rules that follow from the little group
representation. The implementation of Lorentz invariance in heavy
particle effective theories is formally
interesting, and has important consequences for
applications involving higher-order $1/M$
expansions~\cite{Heinonen:2012km, Luke:1992cs, Hill:2012rh}.
In the present paper, we focus on the leading order in $1/M$.

\section{Effective theory at the weak scale \label{sec:weakscaleEFT}}

Let us construct the effective theory of DM with mass $M \gtrsim m_W$
interacting with $n_f=5$ flavor QCD. The hierarchy of scales between
the DM mass and the relevant low-energy degrees of freedom,
$\Lambda_{\rm QCD}, m_c, m_b \ll m_W$, allows us to use heavy particle
effective theory to describe the DM field. The most general lagrangian
relevant for spin-independent, low-velocity scattering with nucleons,
is then given at energies $E \ll m_W$ by,%
\footnote{General bases including spin-dependent interactions, and non-self-conjugate WIMPs are presented in~\cite{part2}.}
\begin{align}\label{eq:LWIMPSM}
{\cal L}_{\chi_v, \, {\rm SM}} &= 
 \bar{\chi}_v\chi_v  
\bigg\{ 
\sum_{q=u,d,s,c,b} \bigg[ 
c_{q}^{(0)} O_{q}^{(0)}   + c_{q}^{(2)} v_\mu v_\nu O_{q}^{(2)\mu\nu} 
\bigg]
+ c_{g}^{(0)} O_{g}^{(0)} + c_{g}^{(2)} v_\mu v_\nu O_{g}^{(2)\mu\nu} 
\bigg\}  + \dots \,,
\end{align} 
where $\chi_v$ is the lightest, electrically neutral,
self-conjugate WIMP of arbitrary spin.
The ellipsis in the above equation includes higher-dimension operators suppressed by powers of $1/m_W$. The assumed self-conjugacy of $\chi_v$ implies that (\ref{eq:LWIMPSM}) is invariant under (\ref{eq:vparity}). The SM component of (\ref{eq:LWIMPSM}) is expressed in terms of quark and gluon fields as
\begin{align}\label{eq:ops}
O^{(0)}_{q} &= m_q \bar{q} q \,,
\nl
O^{(0)}_g &= (G^A_{\mu\nu})^2 \,,
\nl
O^{(2)\mu\nu}_{q} &= \frac12 \bar{q}\left( \gamma^{\{\mu} iD_-^{\nu\}} 
- \frac{1}{d} g^{\mu\nu} i\Dslash_- \right) q \,,
\nl
O^{(2)\mu\nu}_g &= -G^{A \mu \lambda} G^{A \nu}_{\phantom{A \nu} \lambda} + \frac{1}{d} g^{\mu\nu} (G^A_{\alpha\beta})^2 \,.
\end{align} 
Here $D_- \equiv \overrightarrow{D} - \overleftarrow{D}$,
and $A^{\{\mu}B^{\nu\}} \equiv (A^\mu B^\nu + A^\nu B^\mu)/2$ denotes
symmetrization. The operators in (\ref{eq:ops}) are expressed in terms
of bare lagrangian fields, where we employ dimensional regularization
with $d=4-2\epsilon$ spacetime dimensions. We use the background field
method for gluons in the effective theory thus ignoring gauge-variant
operators, and assume that appropriate field redefinitions are
employed to eliminate operators that vanish by leading order equations
of motion. We ignore flavor non-diagonal operators, whose nucleon
matrix elements have an additional weak-scale suppression relative to
those considered. We will not be concerned here with leptonic
interactions. 

For a self-conjugate WIMP, $\chi_v$, with mass $M \gtrsim m_W$ and
arbitrary spin, Eq.~(\ref{eq:LWIMPSM}) represents the most general
effective lagrangian at leading order in $1/m_W$, relevant for
spin-independent, low-velocity scattering with nucleons. Details of
the UV completion are encoded in the twelve matching coefficients
$c_{q}^{(0)}$, $c_{q}^{(2)}$, $c_g^{(0)}$ and $c_g^{(2)}$. Matching
onto the effective theory (\ref{eq:LWIMPSM}) is in general dependent
on the specific SM extension. Although much of the
formalism applies more generally, for definiteness we focus on the
heavy WIMP limit, $M \gg m_W$, where universal features
appear~\cite{Hill:2011be}.

\section{Effective theory for one or two heavy electroweak multiplets \label{sec:lagrangians} }

In place of a specified UV theory for DM, let us use heavy
particle effective theory to describe extensions of the SM consisting
of one or two electroweak multiplets with masses large compared to the
mass of electroweak-scale particles, $M, M^\prime \gg m_W$.
The extension to more than two multiplets is straightforward. 
We will construct the effective theory describing interactions of such heavy WIMPs with the SM in the regime $|M^\prime - M|, \, m_W \ll M,\, M^\prime$. In the case $|M^\prime - M| \gg m_W$ the effects of the heavier multiplet
appear as power corrections in the effective theory for the lighter multiplet.
For notational clarity, below we omit the subscript $v$ labeling a heavy-particle field.

Consider one or two multiplets of heavy-particle fields with arbitrary spin,
transforming under irreducible representations 
of electroweak $SU(2)_W \times U(1)_Y$. Let us collect the heavy
fields in a column vector $h$, and their masses in a diagonal matrix
$M_{}$. The precise specification of $M$ beyond tree level is
described in Sec.~\ref{sec:renscheme}. 
At leading order in the $1/M$ expansion, the most general gauge- and
Lorentz-invariant lagrangian, bilinear in $h$, and written in terms of
the building-blocks $h$, $v^\mu$, and SM fields, takes the form
\be\label{eq:heavy}
{\cal L} = \bar{h} \left[ iv\cdot D -\delta \mym - f(H)  \right] h + \order(1/M) \,,
\ee
where $iD_\mu = i\partial_\mu + g_1 Y B_\mu + g_2 W_\mu^a T^a $, and
$f(H)$ is a linear matrix function of $H$ (and $H^*$).
For pure states gauge invariance
implies $f(H)=\mathbb{0}$, while for mixed states $f(H)$ describes the
mixing of the pure-state constituents through the Higgs field.
In terms of a reference mass $M_{\rm ref}$, the residual mass matrix is 
\be 
\delta \mym =
M_{} - M_{\rm ref} \mathbb{1}.  
\ee 
Note that if the masses composing $M_{}$ are degenerate, as for a single ``pure" electroweak multiplet, we may choose $M_{\rm ref}$ appropriately to set $\delta \mym=\mathbb{0}$. In the case of two ``mixed" electroweak multiplets $M_{}$ will have non-degenerate entries in general. 

Upon accounting for EWSB we may write (\ref{eq:heavy}) as 
\begin{align}\label{eq:heavy2}
{\cal L} &= \bar{h} \bigg[
 iv\cdot \partial + e Q v\cdot A + {g_2\over c_W} v\cdot Z (T^3 - s_W^2 Q) 
\nl
&\quad 
+ {g_2\over\sqrt{2}} ( v\cdot W^+ T^+ + v\cdot W^- T^- )  -\delta M(v_{\rm wk}) -f(\phi) \bigg] h + \order(1/M)  \,,
\end{align} 
where $T^\pm = T^1 \pm i T^2$, the charge matrix is $Q=T^3
+ Y$ in units of the proton charge, and $\phi$ denotes the fluctuation
of the Higgs field about $\langle H \rangle$, 
\be 
H = {v_{\rm wk}\over\sqrt{2}}
\left(\begin{array}{c}
0 \\ 1 \end{array}
\right) +
\left( 
\begin{array}{c} 
\phi_W^+ \\ {1\over \sqrt{2}}(h + i \phi_Z) 
\end{array}\right) \,.  
\ee 
The residual mass matrix now includes EWSB contributions, 
\be 
\delta M (v_{\rm wk}) = \delta \mym + f( \langle H \rangle), 
\ee
and in the mass eigenstate basis for $\delta M (v_{\rm wk})$, we will set the residual mass of the lightest, (assumed) electrically neutral WIMP, $\chi$, to zero by appropriate choice of $M_{\rm ref}$. Other states may have non-vanishing residual masses. In the following, we will suppress the subscript in $v_{\rm wk}$; the resulting $v$ is not to be confused with the velocity $v^\mu$. 

The heavy-particle lagrangian (\ref{eq:heavy}) can also be obtained at
tree level from a manifestly relativistic lagrangian by performing
field redefinitions. We illustrate this for the singlet-doublet
mixture in Appendix~\ref{sec:SDfull}.%
\footnote{We remark that the
consistency of an effective description for the one-heavy particle
sector for a self-conjugate field follows from the identification of
lowest-lying states odd under a $Z_2$ symmetry. In contrast, the
one-heavy particle sector for a heavy field carrying $U(1)$ global
symmetry (e.g., heavy-quark number in a heavy quark effective theory)
is identified by this quantum number.} 
Let us now have a detailed look
at extensions with one (pure states) or two (mixed states) electroweak
multiplets.

\subsection{Pure states}
The pure-state heavy-particle lagrangian is completely specified by
electroweak quantum numbers since $\delta \mym= \mathbb{0}$ and $f(H) =\mathbb{0}$. We may
proceed in generality, assuming a multiplet of fields in the isospin
$J$ representation of $SU(2)_W$ with hypercharge $Y$. The amplitudes
for weak-scale matching in Sec.~\ref{sec:matchingcalculation} will be given in terms of $Y^2$ and the Casimir
$J(J+1)$. In particular, amplitudes with two $W^\pm$
bosons or two $Z^0$ bosons carry the respective factors 
\be\label{eq:WZcoeff}
{\cal C}_W = J(J+1) - Y^2 \,,\quad {\cal C}_Z = Y^2.
\ee
For extensions consisting of electroweak multiplets with non-zero
hypercharge, we assume that higher-dimension operators cause the mass
eigenstates after EWSB to be self-conjugate combinations. This forbids
a phenomenologically disfavored tree-level vector coupling between the
lightest, electrically neutral state, $\chi$, and $Z^0$.

As specific illustrations we consider the cases of an $SU(2)_W$
triplet ($J=1$) with $Y=0$, and a pair of $SU(2)_W$ doublets ($J=1/2$)
with opposite hypercharge $Y=\pm 1/2$. In supersymmetric extensions,
these represent pure wino and pure higgsino states, respectively. Let
us look at these cases in some detail.

\subsubsection{Pure triplet}
Let the column vector $h_T = (h^1, h^2, h^3)$, with subscript $T$ for triplet, be a heavy, self-conjugate, $SU(2)_W$ triplet with $Y=0$. The heavy-particle lagrangian for $h_T$ is given by (\ref{eq:heavy}) with $(T^a)^{bc} = i\epsilon^{bac}$, $f(H)=\mathbb{0}$, and $\delta \mym=\mathbb{0}$. The electric charge eigenbasis is given by
\be
\left(\begin{array}{c} {h}^1 \\ {h}^2 \\ {h}^3 \end{array} \right) 
\equiv \left( \begin{array}{ccc} 0 & {1\over \sqrt{2}} & {1\over \sqrt{2}} \\ 0 & {i\over \sqrt{2}} & {-i\over \sqrt{2}} \\
1 & 0 & 0 \end{array} \right) \, \left(\begin{array}{c} h_0 \\ h_+ \\ h_- \end{array} \right).
\ee
In terms of the column vector $h = (h_0,\, h_+,\, h_-)$, where $h_0 \equiv \chi$, the lagrangian is given by (\ref{eq:heavy2}) with
\be 
Q = T^3 = {\rm diag}(0,1,-1) \,, \quad 
T^+ = \left(\begin{array}{ccc} 0 & 0 & \sqrt{2} \\ -\sqrt{2} & 0 & 0 \\ 0 & 0 & 0 \end{array} \right) \,, \quad
T^- = \left(\begin{array}{ccc} 0 & -\sqrt{2} & 0 \\ 0 & 0 & 0 \\ \sqrt{2} & 0 & 0 \end{array} \right)  \,. 
\ee

\subsubsection{Pure doublet}
Let $h_\psi$ and $h_{\psi^c}$ be heavy-particle doublets in the $(\bm{2},1/2)$ 
and $(\bm{\bar{2}},-1/2)$ representations of $SU(2)_W \times U(1)_Y$.%
\footnote{This construction is analogous to that appearing in 
applications of heavy quark effective theory to
 processes where both a heavy quark and a heavy anti-quark are active degrees of freedom.}
Anticipating perturbations that cause the mass eigenstates to be self-conjugate 
fields, let us introduce the linear combinations
\be\label{eq:selfconj}
h_{D_1} = {h_\psi+h_{\psi^c} \over \sqrt{2} } = \left(\begin{array}{c} h_1 \\ h_0 \end{array} \right)  \, , \quad  h_{D_2} =  { i(h_\psi - h_{\psi^c}) \over \sqrt{2} } = \left(\begin{array}{c}  h_2 \\ h_0^\prime \end{array} \right) \, ,
\ee
with subscript $D$ for doublet. The heavy-particle lagrangian for the column vector $h=\left( h_{D_1},h_{D_2} \right)$ is given by (1), with $f(H)=\mathbb{0}$, and gauge couplings
\be\label{eq:higgsinogauge}
T^a = \left(\begin{array}{cc} {\tau^a - \tau^{aT} \over 4} &  {-i(\tau^a + \tau^{aT}) \over 4 } \\  {i(\tau^a + \tau^{aT}) \over 4 }  & {\tau^a - \tau^{aT} \over 4} \end{array} \right)\, , \quad  
Y = \frac{i}{2} \left(\begin{array}{cc} 0 & -\mathbb{1} \\ \mathbb{1} & 0 \end{array} \right)\, ,
\ee
where $\tau^a$ are the Pauli isospin matrices. Neglecting the small mass perturbation mentioned above, the tree-level mass eigenstates are degenerate, and we may choose $\delta \mym = \mathbb{0}$. The charge eigenstates are given by
\be\label{eq:higgsinobasis}
 \left(\begin{array}{c} h_1 \\ h_0 \\ h_2 \\ h_0^\prime \end{array} \right) 
\equiv \left(\begin{array}{cccc} 0 & 0 & {1\over \sqrt{2}} & {1\over\sqrt{2}} \\ 
1 & 0 & 0 & 0 \\
0 & 0 & {i\over\sqrt{2}} & -{i\over\sqrt{2}} \\
0 & 1 & 0 & 0 
\end{array}
\right)  \left(\begin{array}{c} h_0 \\ h_0^\prime \\ h_+ \\ h_- \end{array} \right) \,.
\ee
In terms of the column vector $h= (h_{0}, h_0^\prime, h_{+}, h_{-})$, where $h_0 \equiv \chi$, the lagrangian is given by (\ref{eq:heavy2}) with $Q = {\rm diag}(\mathbb{0}_2,1,-1)$ and
\begin{align}
T^3 = \left(\begin{array}{cccc}
0 & {i\over 2} & 0 & 0 \\ 
-{i\over 2} & 0 & 0 & 0 \\
0 & 0 & \frac12 & 0 \\
0 & 0 & 0 & -\frac12 
\end{array}
\right) \,, \quad
T^+ = \left(\begin{array}{cccc} 
0 & 0 & 0 & -{1\over \sqrt{2}} \\
0 & 0 & 0 & {i\over \sqrt{2}} \\
{1\over\sqrt{2}} & -{i\over\sqrt{2}} & 0 & 0 \\
0 & 0 & 0 & 0 
\end{array}
\right) \,, \quad
T^- = \left(\begin{array}{cccc} 
0 & 0 & {1\over\sqrt{2}} & 0 \\
0 & 0 & {i\over\sqrt{2}} & 0 \\
0 & 0 & 0 & 0 \\
-{1\over\sqrt{2}} & -{i\over\sqrt{2}} & 0 & 0 
\end{array}
\right) \,.
\end{align}

\subsection{Admixtures}

As an example of mixed states, let us consider in detail the
singlet-doublet admixture. Results for the triplet-doublet admixture
will also be given below.

\subsubsection{Singlet-doublet admixture \label{sec:SDmatrices}}
Let $h_S$, with subscript $S$ for singlet, be a heavy, self-conjugate,
$SU(2)_W$ singlet with $Y=0$ and mass $M_S$. Consider an admixture of
$h_S$ and the previously defined self-conjugate doublets $h_{D_1}$ and
$h_{D_2}$, with mass $M_D$. At leading order in the $1/M$ expansion,
the gauge-invariant interactions of $h_S$, $h_{D_1}$ and $h_{D_2}$
involving the Higgs field are
\begin{align}\label{eq:SDhiggs}
{\cal L}_{H \bar{h}h} &= -\bar{h}_S \left[ y H^\dagger {(h_{D_1} -i h_{D_2} ) \over \sqrt{2} }+ y^* H^T {(h_{D_1} + i h_{D_2}) \over \sqrt{2} } \right] + {\rm h.c.}
= -\bar{h} f(H)h \,,
\end{align}
where we have imposed the invariance (\ref{eq:vparity}), and collected
the heavy-particle fields in a column vector $h = \left( h_S , h_{D_1} , h_{D_2}
\right) =  \left( h_S , h_1 , h_0 , h_2 , h_0^\prime \right)$. The
Higgs coupling matrix is given by
\begin{align}\label{eq:fSD}
f(H) &= 
{a_1 \over \sqrt{2}}\left(\begin{array}{ccc} 
0 & H^\dagger + H^T & i(H^T-H^\dagger) \\
H+H^* & \mathbb{0}_2 & \mathbb{0}_2 \\
i(H-H^*) & \mathbb{0}_2 & \mathbb{0}_2 
\end{array}
\right)
+ {a_2 \over \sqrt{2}}\left(\begin{array}{ccc} 
0 & -i(H^T - H^\dagger) & H^T+H^\dagger \\
-i(H-H^*) & \mathbb{0}_2 & \mathbb{0}_2 \\
H+H^* & \mathbb{0}_2 & \mathbb{0}_2 
\end{array}
\right) \,,
\end{align}
with real parameters $a_1 = {\rm Re}(y)$ and $a_2 = {\rm Im}(y)$. For comparison, the derivation in
Appendix~\ref{sec:SDfull} obtains (\ref{eq:SDhiggs}) at tree level
starting from a manifestly relativistic lagrangian.
The residual mass matrix is $\delta \mym = {\rm diag}\left( M_S ,\, M_D \mathbb{1}_4 \right) - M_{\rm ref} \mathbb{1}_5$, and we define $M_S$ and $M_D$ to be real and positive.%
\footnote{An additional phase redefinition of $h_\psi$, $h_{\psi^c}$ 
could be used to enforce the vanishing of $a_1$ or $a_2$.}  
The gauge couplings are obtained by trivially extending
(\ref{eq:higgsinogauge}) to include the singlet. This completely
specifies the heavy-particle lagrangian given in (\ref{eq:heavy}).

The mass induced by EWSB is accounted for at tree level by including 
contributions from (\ref{eq:fSD}),
\be
\delta M(v) = \delta M +  v \left( \begin{array}{ccccc} 
0 & 0 & a_1  & 0& a_2 \\ 
0 & 0 & 0 & 0 & 0\\
a_1 & 0 & 0 & 0 & 0\\
0 & 0 & 0 & 0 & 0\\
a_2 & 0 & 0 & 0 & 0\\
\end{array} \right) \,.  \ee
In the following, we use subscripts to denote the electric charge and
bracketed superscripts to label the mass eigenstate. For neutral
states we find the residual mass eigenvalues
\be
\delta^{(0)}_0 = M_D - M_{\rm ref} \,, \quad  
\delta^{(\pm)}_0 = {M_D + M_S \over 2} \pm \sqrt{ \Delta^2 + (a v)^2} -M_{\rm ref} \,,
\ee
where we define 
\be\label{eq:SDparameters} \Delta \equiv { M_S -  M_D \over 2} \,,
\quad  a \equiv \sqrt{a_1^2 + a_2^2} \,.  \ee
By definition $a>0$, and regardless of the sign of $\Delta$, the
smallest eigenvalue is $\delta_0^{(-)}$. Let us set this eigenvalue to
zero by appropriately choosing the reference mass $M_{\rm ref}$. The
corresponding normalized eigenvectors in the $(h_S,h_0,h_0^\prime)$
basis of electrically neutral states are then 
\be
\vec{\rm v}^{(0)}_0 = {1\over a} \left(\begin{array}{c} 0 \\ a_2 \\ -a_1 \end{array} \right) \,, \quad
\vec{\rm v}^{(\pm)}_0 = 
{1\over \left[ \left( \Delta \pm \sqrt{\Delta^2+(a v)^2} \right)^2 + (a v)^2 \right]^{\frac12} } 
\left(\begin{array}{c} \Delta \pm \sqrt{\Delta^2+(a v)^2} \\ a_1 v \\ a_2 v \end{array} \right) \,,
\ee
and we may construct the unitary matrix $U_0$ (on the three-dimensional
neutral subspace) to translate to the mass eigenbasis,
\be
U_0 = \left( \vec{\rm v}^{(0)}_0 \quad \vec{\rm v}^{(+)}_0 \quad \vec{\rm v}^{(-)}_0 \right) \,,
\quad 
\left(\begin{array}{c} h_S \\ h_0 \\ h_0^\prime \end{array}\right) 
= U_0 \left(\begin{array}{c} h_0^{(0)} \\ h_0^{(+)} \\ h_0^{(-)} \end{array}\right) \, ,
\quad U_0^\dagger 
\delta M(v) U_0 
= {\rm diag }\left( \delta^{(0)}_0, \delta^{(+)}_0, \delta^{(-)}_0 \right)  \,.
\ee
The tree-level masses for the electrically charged sector are
unchanged by EWSB, given by $\delta^{(0)}_\pm = \delta^{(0)}_0$, and the
corresponding charge eigenstates are given by
\be\label{eq:lambdacharged}
\left(\begin{array}{c} h_1 \\ h_2 \end{array} \right)
= {1\over \sqrt{2} } \left(\begin{array}{cc} 1 & 1 \\ i & -i \end{array} \right) 
\left(\begin{array}{c} h_+^{(0)} \\ h_-^{(0)} \end{array} \right) \,.
\ee
The basis of mass eigenstates is thus given by the column vector
$h = \left(h_0^{(0)} ,
h_0^{(+)} , h_0^{(-)} , h_+^{(0)} , h_-^{(0)} \right)$, where
$h_0^{(-)} \equiv \chi$, and the lagrangian is given by (\ref{eq:heavy2}) with
\begin{align} \label{eq:SDfinal}
\delta M(v) & = {\rm diag} \left( \delta_0^{(0)}, \delta_0^{(+)}, \delta_0^{(-)}, \delta_+^{(0)}, \delta_-^{(0)} \right) = a v \, {\rm diag}\left( t_\frac{\rho}{2} , 2 s_\rho^{-1}, 0 , t_\frac{\rho}{2} , t_\frac{\rho}{2} \right) \,, 
\nl
\quad Q &= {\rm diag}(\mathbb{0}_3,1,-1) \,, 
\nl
T^3 &- s_W^2 Q = \left(\begin{array}{ccccc}  
0 & {i\over 2}|s_{\rho\over 2}| & {i\over 2} |c_{\rho\over 2}| & 0 & 0 \\
-{i\over 2}|s_{\rho\over 2}| & 0 & 0 & 0 & 0 \\
-{i\over 2} |c_{\rho\over 2}| & 0 & 0 & 0 & 0 \\
0 & 0 & 0 & \frac12 - s_W^2 & 0 \\
0 & 0 & 0 & 0 & -\frac12 + s_W^2 
\end{array} 
\right)
\,,
 \nl
 T^+ &= {e^{- i \xi} \over \sqrt{2} } \left( \begin{array}{ccccc} 
 0 & 0 & 0 & 0 & -i \\
 0 & 0 & 0 & 0 & -|s_{\rho\over 2}| \\
 0 & 0 & 0 & 0 & -|c_{\rho\over 2}| \\
 i & |s_{\rho\over 2}| & |c_{\rho\over 2}| & 0 & 0 \\
 0 & 0 & 0 & 0 & 0 
 \end{array}
 \right) \,, 
\quad 
 T^- = {e^{+ i \xi} \over \sqrt{2}  } \left( \begin{array}{ccccc} 
 0 & 0 & 0 & -i & 0 \\
 0 & 0 & 0 & |s_{\rho\over 2}| & 0 \\
 0 & 0 & 0 & |c_{\rho\over 2}| & 0 \\
 0 & 0 & 0 & 0 & 0 \\
 i & -|s_{\rho\over 2}| & -|c_{\rho\over 2}| & 0 & 0 \\
 \end{array}
 \right) \,, 
\nl
f(\phi) &= a \left( \begin{array}{ccccc} 
0 & |c_{\rho\over 2}| \phi_Z & - |s_{\rho\over 2}|\phi_Z & 0 & 0 \\
|c_{\rho\over 2}| \phi_Z & s_\rho \,h & c_\rho\, h & |c_{\rho\over 2}| e^{+ i \xi} \phi^-_W & 
|c_{\rho\over 2}|e^{- i \xi} \phi^+_W \\
 - |s_{\rho\over 2}|\phi_Z & c_\rho\, h & -s_\rho \, h &  -|s_{\rho\over 2}| e^{+i \xi} \phi^-_W & 
-|s_{\rho\over 2}| e^{- i \xi} \phi^+_W \\
0 & 
|c_{\rho\over 2}| e^{-i \xi} \phi^+_W & 
-|s_{\rho\over 2}| e^{- i \xi} \phi^+_W & 0 & 0 \\
0 & |c_{\rho\over 2}| e^{+ i \xi} \phi^-_W & 
-|s_{\rho\over 2}| e^{+i \xi} \phi^-_W & 0 & 0
\end{array}
\right) \,,
\end{align}
where we have introduced
\be\label{eq:defangles}
\sin\rho \equiv {a v \over \sqrt{(av)^2 + \Delta^2} }  \,, \quad
\cos\rho \equiv {\Delta \over \sqrt{(av)^2 +\Delta^2} }  \,, \quad e^{\pm i \xi}   \equiv {(a_1 \pm ia_2)\over a}.
\ee
The shorthand notation $c_x \equiv \cos x$, $s_x \equiv \sin x$, and
$t_x \equiv \tan x$ is used throughout this paper. Note that $s_\rho$
is positive, and that $c_\rho$ can have either sign depending on the
hierarchy between $M_S$ and $M_D$. It is straightforward to extract
Feynman rules from the lagrangian (\ref{eq:heavy2}) and the matrices
(\ref{eq:SDfinal}). For example, the propagator for $\chi$, and its
coupling to the physical Higgs boson, $h$, are
\vspace{4mm}
\be
\parbox{25mm}{
\begin{fmfgraph*}(60,40)
  \fmfleftn{l}{3}
  \fmfrightn{r}{3}
  \fmf{dbl_plain_arrow,label=$k$,label.side=left}{l2,r2}
  \fmffreeze
  \fmflabel{$\chi$}{l2}
  \fmflabel{$\chi$}{r2}
\end{fmfgraph*}
} \ \
= {i \over v \cdot k - \delta_0^{(-)} +i0 }\, ,
 \quad \quad \quad 
\parbox{25mm}{
\begin{fmfgraph*}(80,40)
  \fmfleftn{l}{2}
  \fmfrightn{r}{2}
  \fmftopn{t}{3}
  \fmfbottomn{b}{3}
  \fmf{double,label=$\chi$,label.side=left}{l2,t2}
  \fmf{double,label=$\chi$,label.side=left}{t2,r2}
  \fmffreeze
  \fmf{dashes,label=$h$,label.side=left}{t2,b2}
\end{fmfgraph*}
} 
= \quad i a s_\rho \,.
\ee

\subsubsection{Triplet-doublet admixture \label{sec:TDmatrices}}
The construction for the triplet-doublet case follows closely that for
the singlet-doublet case, with a heavy triplet $h_T$ in place of the
singlet $h_S$. Using $\bm{\tau} = (\tau^1, \tau^2, \tau^3)$ and
$\bar{\bm{\tau}} = -(\tau^{1T}, \tau^{2T}, \tau^{3T})$, the
gauge-invariant interactions of $h_T$, $h_{D_1}$ and $h_{D_2}$
involving the Higgs field can be written in the form ${\cal L}_{H
\bar{h} h} =  -\bar{h} f(H)h$, where we collect fields in a
seven-component column vector $h = (h_T, h_{D1}, h_{D2} )$, and the
matrix $f(H)$ is given by
\begin{align}
f(H) &= 
{a_1 \over \sqrt{2}}\left(\begin{array}{ccc} 
\mathbb{0}_3 & H^\dagger\bm{\tau} - H^T\bar{\bm{\tau}} & \ \ i(-H^T\bar{\bm{\tau}}  -H^\dagger\bm{\tau}) \\
-\bar{\bm{\tau}}H^* + \bm{\tau} H & \mathbb{0}_2 & \mathbb{0}_2 \\
i(\bm{\tau}H + \bar{\bm{\tau}} H^*) & \mathbb{0}_2 & \mathbb{0}_2 
\end{array}
\right)
\nl
&\quad
+ {a_2 \over \sqrt{2}}\left(\begin{array}{ccc} 
\mathbb{0}_3 & i(H^T\bar{\bm{\tau}}  +H^\dagger\bm{\tau}) & \ \ H^\dagger\bm{\tau} - H^T\bar{\bm{\tau}}  \\
i(-\bm{\tau}H - \bar{\bm{\tau}} H^*) & \mathbb{0}_2 & \mathbb{0}_2 \\
-\bar{\bm{\tau}}H^* + \bm{\tau} H & \mathbb{0}_2 & \mathbb{0}_2 
\end{array}
\right) \, ,
\end{align}
with real parameters $a_1$ and $a_2$. Upon accounting for mass
contributions from EWSB, the basis of mass eigenstates is given by the column
vector $h= \left( h_0^{(0)} , h_0^{(+)}, h_0^{(-)}, h_+^{(+)},
h_+^{(-)}, h_-^{(+)}, h_-^{(-)} \right)$, where $h_0^{(-)} \equiv \chi$, and the lagrangian
is given by (\ref{eq:heavy2}) with
\begin{align} \label{eq:TDfinal}
\delta M(v) &= {\rm diag} \left( \delta_0^{(0)}, \delta_0^{(+)}, \delta_0^{(-)}, \delta_+^{(+)}, \delta_+^{(-)}, \delta_-^{(+)}, \delta_-^{(-)} \right)= av \, {\rm diag}\left( t_\frac{\rho}{2} ,  2 s_\rho^{-1}, 0 ,  2 s_\rho^{-1},0,2 s_\rho^{-1},0 \right) \,, 
\nl
Q &= {\rm diag}(0,0,0,1,1,-1,-1) \,,
\nl
T^3 &= \left(\begin{array}{ccccccc} 
 0 & \frac{i}{2}|s_{\rho\over 2}| & \frac{i}{2} |c_{\rho\over 2}| & 0 & 0 & 0 & 0 \\
-\frac{i}{2}|s_{\rho\over 2}| & 0 & 0 & 0 & 0 & 0 & 0 \\
-\frac{i}{2}|c_{\rho\over 2}| & 0 & 0 & 0 & 0 & 0 & 0 \\
0 & 0 & 0 & 1-\frac12s_{\rho\over 2}^2 & -\frac14 s_\rho & 0 & 0 \\
0 & 0 & 0 & -\frac14 s_\rho & 1 -\frac12 c_{\rho\over 2}^2 & 0 & 0 \\
0 & 0 & 0 & 0 & 0 & -1+\frac12 s_{\rho\over 2}^2 & \frac14 s_\rho \\
0 & 0 & 0 & 0 & 0 & \frac14 s_\rho & -1 + \frac12 c_{\rho\over 2}^2 
\end{array}
\right) \,,
\nl
T^+ &= {1\over \sqrt{2}} \left(\begin{array}{ccccccc}
0 & 0 & 0 & 0 & 0 & {i}|s_{\rho\over 2}| & {i}|c_{\rho\over 2}| \\
0 & 0 & 0 & 0 & 0 & 1+ c_{\rho\over 2}^2  & -{1\over 2}s_\rho \\
0 & 0 & 0 & 0 & 0 & -{1\over 2} s_\rho & 1+s_{\rho\over 2}^2 \\
-{i}|s_{\rho\over 2}| & \ -1- c_{\rho\over 2}^2 & {1\over 2}s_\rho & 0 & 0 & 0 & 0 \\
-{i}|c_{\rho\over 2}| & {1\over 2}s_\rho &\  -1-s_{\rho\over 2}^2 & 0 & 0 & 0 & 0 \\
0 & 0 & 0 & 0 & 0 & 0 & 0 \\
0 & 0 & 0 & 0 & 0 & 0 & 0 
\end{array} 
\right) \,,
\nl
T^- &= {1\over \sqrt{2}} \left(\begin{array}{ccccccc} 
0 & 0 & 0 & {i}|s_{\rho\over 2}| & {i}|c_{\rho\over 2}| & 0 & 0 \\
0 & 0 & 0 & -1-c_{\rho\over 2}^2 & {1\over 2}s_\rho & 0 & 0 \\
0 & 0 & 0 & {1\over 2}s_\rho& -1-s_{\rho\over 2}^2 & 0 & 0 \\
0 & 0 & 0 & 0 & 0 & 0 & 0 \\
0 & 0 & 0 & 0 & 0 & 0 & 0 \\
-{i}|s_{\rho\over 2}| & \ 1+c_{\rho\over 2}^2 & -{1\over 2}s_\rho & 0 & 0 & 0 & 0 \\
-{i}|c_{\rho\over 2}| & -{1\over 2}s_\rho & \ 1+s^2_{\rho\over 2} & 0 & 0 & 0 & 0 
\end{array}
\right) \,,
\nl
f(\phi) &= a\left(\begin{array}{ccccccc} 
0 & |c_{\rho\over 2}|\phi_Z & -|s_{\rho\over 2}|\phi_Z & -i|c_{\rho\over 2}|\phi^-_W & i|s_{\rho\over 2}|\phi^-_W 
& i|c_{\rho\over 2}|\phi^+_W & -i|s_{\rho\over 2}|\phi^+_W \\
 |c_{\rho\over 2}| \phi_Z & s_\rho h & c_\rho h & 0 & \phi^-_W & 0 & \phi^+_W \\
-|s_{\rho\over 2}| \phi_Z & c_\rho h &  -s_\rho h & -\phi^-_W & 0 & -\phi^+_W & 0 \\
i|c_{\rho\over 2}| \phi^+_W & 0 & -\phi^+_W & s_\rho h & c_\rho h-i\phi_Z & 0 & 0 \\
-i|s_{\rho\over 2}|\phi^+_W & \phi^+_W & 0 & c_\rho h + i\phi_Z & -s_\rho h & 0 & 0 \\
-i|c_{\rho\over 2}|\phi^-_W & 0 & -\phi^-_W & 0 & 0 & s_\rho h & c_\rho h + i\phi_Z \\
i|s_{\rho\over 2}| \phi^-_W & \phi^-_W & 0 & 0 & 0 & c_\rho h -i\phi_Z & -s_\rho h 
\end{array}
\right) \,,
\end{align}
where $s_\rho$ and $c_\rho$ are as defined in (\ref{eq:defangles}), with $a =
\sqrt{a_1^2 + a_2^2}$ and $\Delta = (M_T-M_D)/2$. Again, $s_\rho$ is
positive and $c_\rho$ can have either sign depending on the hierarchy
between $M_T$ and $M_D$. 

\subsection{Pure-case limits \label{sec:limits}}
Appropriate parametric limits can be taken to decouple the pure state constituents of an admixture. This can be used to check the consistency of matching computations in Sec.~\ref{sec:matchingcalculation}. 
From the singlet-doublet admixture, we may recover the pure doublet
(singlet) case by taking $a \to 0$ or $|\Delta| \to \infty$, with
$\Delta >0$ ($\Delta < 0$), or by taking $\rho \to 0$ ($\rho \to \pi$). Similarly, to recover the pure doublet (triplet) case
from the triplet-doublet admixture, we decouple the triplet (doublet)
component by taking $a \to 0$ or $|\Delta| \to \infty$, with $\Delta
> 0$ ($\Delta < 0$), or by taking $\rho \to 0$ ($\rho \to \pi$).

\section{Onshell renormalization scheme \label{sec:renscheme}}
A consistent evaluation of amplitudes beyond tree level demands
renormalization of the Higgs-WIMP vertex, $h \bar{\chi} \chi$, that
appears for admixtures. We define an extension of the onshell
renormalization scheme for the electroweak SM (e.g., see~\cite{Hollik:1988ii}) by expressing the vertex amplitude in terms of physical masses in the SM
and  DM sectors. We begin by studying the singlet-doublet mixture, and
will later quote the analogous results for the triplet-doublet
mixture.

To avoid confusion with standard notation for counterterms, in this
section (only) we denote a residual mass by $\mu$, and
a residual mass counterterm by $\delta \mu$. We keep the notation
introduced in Sec.~\ref{sec:lagrangians} for the residual mass
eigenvalues, $\delta_0^{(0)}$, $\delta_0^{(\pm)}$, etc.

\subsection{Singlet-doublet counterterm lagrangian} 
Let us write the bare lagrangian as the sum of renormalized and
counterterm contributions
\begin{align}\label{eq:barelag}
{\cal L} &= \bar{h}^{\rm bare} \left[
 iv\cdot D - \mu^{\rm bare} 
- f^{\rm bare}(H^{\rm bare}) 
 \right] h^{\rm bare} 
\nl
&= \bar{h} \left[ 
 iv\cdot D + \delta Z_h iv\cdot D - \mu  - \delta \mu 
- f(H^{\rm bare}) 
- \delta f(H^{\rm bare}) \right] h 
\,,
\end{align}
where the bare quantities are given by 
\begin{align}
\mu^{\rm bare} 
&= {\rm diag}(\mu_S^{\rm bare}, \mu_D^{\rm bare}, \mu_D^{\rm bare}, \mu_D^{\rm bare}, \mu_D^{\rm bare} )
\,, \nl
f^{\rm bare} (H) 
&= 
{a^{\rm bare}_1 \over \sqrt{2}}\left(\begin{array}{ccc} 
0 & H^\dagger + H^T & i(H^T-H^\dagger) \\
H+H^* & \mathbb{0}_2 & \mathbb{0}_2 \\
i(H-H^*) & \mathbb{0}_2 & \mathbb{0}_2 
\end{array}
\right)
\nl
&\quad
+ {a^{\rm bare}_2 \over \sqrt{2}}\left(\begin{array}{ccc} 
0 & -i(H^T - H^\dagger) & H^T+H^\dagger \\
-i(H-H^*) & \mathbb{0}_2 & \mathbb{0}_2 \\
H+H^* & \mathbb{0}_2 & \mathbb{0}_2 
\end{array}
\right)
\nl
&\equiv a_1^{\rm bare} f_1(H) + a_2^{\rm bare} f_2(H)\, ,
\end{align}
and the expression for $f^{\rm bare}(H)$ above is valid for
arbitrary $H$ (in particular, for $H^{\rm bare}$). The gauge symmetry
preserving counterterms are given by 
\begin{align}\label{eq:ct}
Z_h &= 1 + \delta Z_h  = 1 + {\rm diag}(\delta Z_S, \delta Z_D \mathbb{1}_4) \,, 
\nl
\mu + \delta \mu &= Z_h^{\frac12} \mu^{\rm bare} Z_h^{\frac12} = {\rm diag}(\mu_S + \delta \mu_S , 
(\mu_D+\delta \mu_D) \mathbb{1}_4 )
  \,,
\nl
f(H^{\rm bare}) + \delta f(H^{\rm bare}) &= Z_h^\frac12 f^{\rm bare}(H^{\rm bare}) Z_h^\frac12 
= (a_1+\delta a_1) f_1(H^\prime) + (a_2 + \delta a_2) f_2(H^\prime) \,.
\end{align}
We have introduced $H^\prime$ to absorb the renormalization of $v$:
\be
H^{\rm bare} 
= Z_H^{\frac12} \left( 
\begin{array}{c} 
\phi_W^+ \\ {1\over \sqrt{2}}(v - \delta v + h + i \phi_Z) 
\end{array}\right)
=  
Z_H^\frac12 \left( 1 - {\delta v\over v} \right) H^\prime \,. 
\ee
Note that the renormalization of $v$ introduces 
a coupling $\sim {\delta v \over v} \, h \bar{\chi} \chi$ through the
$a_1 f_1(H^\prime) + a_2 f_2(H^\prime)$ term in (\ref{eq:ct}). 
We will fix the counterterms by enforcing
renormalization conditions on the residual mass matrix (two point
functions). Three point functions involving the Higgs interaction will
then be determined.

\subsection{Propagator corrections} 
Anticipating renormalization conditions that preserve the basis $h = \left(h_0^{(0)},h_0^{(+)} , h_0^{(-)} , h_+^{(0)} , h_-^{(0)} \right)$
of mass eigenstates introduced in Sec.~\ref{sec:SDmatrices}, let us express the counterterms in this basis,
\begin{align} 
\delta \mu 
&= \delta \mu_D \mathbb{1}_5 +  \left( 
\begin{array}{ccccc} 
0 & |c_{\rho\over 2}| {v\over a}(a_2\delta a_1 - a_1\delta a_2) & |s_{\rho\over 2}| {v\over a}( a_1\delta a_2 - a_2\delta a_1) & 0 & 0 \\
\cdot &  2  c^2_{\rho\over 2} (\delta \Delta) + s_\rho {v\over a}(a_1\delta a_1 + a_2 \delta a_2)
& \ \ - s_\rho (\delta \Delta) + c_\rho {v\over a}(a_1\delta a_1 + a_2 \delta a_2)  
& 0 & 0 \\
\cdot & \cdot &  2  s_{\rho\over 2}^2 (\delta \Delta)  - s_\rho {v\over a}(a_1\delta a_1 + a_2 \delta a_2) & 0 & 0 \\
\cdot & \cdot & \cdot &0 & 0 \\
\cdot & \cdot & \cdot & \cdot & 0 
\end{array} 
\right) \,,
\nl
\delta Z_h  
&= \delta Z_D \mathbb{1}_5+ \left( 
\begin{array}{ccccc} 
0 & 0 & 0 & 0 & 0 \\
\cdot & c^2_{\rho \over 2} (\delta Z_S - \delta Z_D) 
& - \frac12  s_\rho (\delta Z_S - \delta Z_D)    
& 0 & 0 \\
\cdot & \cdot &s^2_{\rho \over2} (\delta Z_S - \delta Z_D)   
& 0 & 0 \\
\cdot & \cdot & \cdot & 0 & 0 \\
\cdot & \cdot & \cdot & \cdot & 0 
\end{array} 
\right) \,,
\end{align} 
where the above matrices are symmetric, and $ ( \delta \Delta ) =
(\delta \mu_S - \delta \mu_D)/2$. Due to the masslessness of the
photon, the onshell renormalization factor for the electrically
charged state, $\delta Z_D$, is infrared (IR) divergent. To avoid the
associated complications, we may turn off $\delta Z_D$, corresponding
to an additional overall renormalization of the fields with $\delta
Z_S = \delta Z_D$. This overall renormalization will not impact the
determination of physical masses or mass eigenstates. However, we will
of course need to include additional wavefunction renormalization
factors when computing physical amplitudes. In the following, we allow
for arbitrary $\delta Z_D$.
 
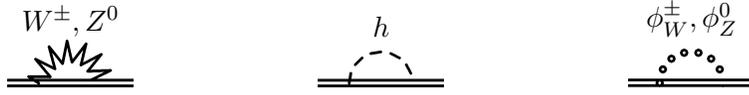
\begin{figure}[htb]
\begin{center}
\parbox{40mm}{
\begin{fmfgraph*}(60,40)
  \fmfleftn{l}{3}
  \fmfrightn{r}{3}
  \fmf{double}{l1,v,x,w,r1}
  \fmffreeze
  \fmf{zigzag,tension=0,left,label=$W^\pm,,Z^0$}{v,w}
\end{fmfgraph*}
}
\parbox{40mm}{
\begin{fmfgraph*}(60,40)
  \fmfleftn{l}{3}
  \fmfrightn{r}{3}
  \fmf{double}{l1,v,x,w,r1}
  \fmffreeze
  \fmf{dashes, label=$h$, left,tension=0}{v,w}
\end{fmfgraph*}
}
\parbox{40mm}{
\begin{fmfgraph*}(60,40)
  \fmfleftn{l}{3}
  \fmfrightn{r}{3}
  \fmf{double}{l1,v,x,w,r1}
  \fmffreeze
  \fmf{dbl_dots, label=$\phi_W^\pm,,\phi_Z^0$, left,tension=0}{v,w}
\end{fmfgraph*}
}
\caption{\label{fig:2point} 
One-loop corrections to two-point functions. Double lines denote heavy WIMPs, zigzag lines denote gauge bosons, $W^\pm$ or $Z^0$, dashed lines denote the physical Higgs boson, $h$, and dotted lines denote Goldstone bosons, $\phi_W^\pm$ or $\phi_Z^0$. 
}
\end{center}
\end{figure}

We compute the one-loop corrections to the amputated two-point
function, $\Sigma_2$, from virtual $Z^0$, $W^\pm$, $h$, $\phi_Z^0$ and
$\phi_W^\pm$ exchange, as illustrated in Fig.~\ref{fig:2point}. In the
following results, we set the external momentum to zero (i.e., we compute $\Sigma_2(0)$), and the first
(second) subscript denotes the final (initial) state, with values
$(1,2,3,4,5)$ corresponding to the mass eigenstates $(h_0^{(0)}, h_0^{(+)}, h_0^{(-)},
h_+^{(0)}, h_-^{(0)})$. 
Using Feynman-t'Hooft gauge, and expressing results in terms of the basic 
integral $I_3(\delta,m)$ of Appendix~\ref{sec:self}, we find
\begin{align}\label{eq:Sigma2} 
-i[\Sigma_2(0)]_{11} &= 
-{g_2^2\over 4 c_W^2} c^2_{\rho\over 2} I_3(\delta_0^{(-)}, m_Z) 
-{g_2^2\over 4 c_W^2} s^2_{\rho\over 2} I_3(\delta_0^{(+)}, m_Z) 
- {g_2^2\over 2} I_3(\delta_\pm^{(0)}, m_W) 
\nl
&\quad 
+ a^2c^2_{\rho\over 2} I_3(\delta_0^{(+)}, m_Z) 
+ a^2s^2_{\rho\over 2} I_3(\delta_0^{(-)}, m_Z) \, ,
\nl
-i[\Sigma_2(0)]_{22} &= 
-{g_2^2\over 4 c_W^2} s^2_{\rho\over 2} I_3(\delta_0^{(0)}, m_Z) 
- {g_2^2\over 2}s^2_{\rho\over 2} I_3(\delta_\pm^{(0)}, m_W) 
+ a^2s^2_\rho I_3(\delta_0^{(+)}, m_h) 
\nl
&\quad + a^2c^2_\rho I_3(\delta_0^{(-)}, m_h) 
+ a^2 c^2_{\rho\over 2} I_3(\delta_0^{(0)}, m_Z) 
+ 2 a^2c^2_{\rho\over 2} I_3(\delta_\pm^{(0)}, m_W) \, ,
\nl
-i[\Sigma_2(0)]_{23} &= -i[\Sigma_2(0)]_{32} 
\nl
&= -{g_2^2\over 8c_W^2} s_\rho  I_3(\delta_0^{(0)},m_Z)
-{g_2^2\over 4} s_ \rho I_3(\delta_\pm^{(0)}, m_W) 
+ a^2s_\rho c_\rho I_3(\delta_0^{(+)},m_h) 
\nl
&\quad 
- a^2s_\rho c_\rho I_3(\delta_0^{(-)},m_h) 
- { a^2 \over 2} s_\rho I_3(\delta_0^{(0)},m_Z) 
- a^2 s_ \rho I_3(\delta_\pm^{(0)}, m_W) \,,
\nl
-i[\Sigma_2(0)]_{33} &= 
-{g_2^2\over 4 c_W^2} c^2_{\rho\over 2} I_3(\delta_0^{(0)}, m_Z) 
- {g_2^2\over 2}c^2_{\rho\over 2} I_3(\delta_\pm^{(0)}, m_W) 
+ a^2s^2_\rho I_3(\delta_0^{(-)}, m_h) 
\nl
&\quad + a^2c^2_\rho I_3(\delta_0^{(+)}, m_h) 
+ a^2 s^2_{\rho\over 2} I_3(\delta_0^{(0)}, m_Z) 
+ 2 a^2s^2_{\rho\over 2} I_3(\delta_\pm^{(0)}, m_W) \, ,
\nl
-i[\Sigma_2(0)]_{44} &= -i [\Sigma_2(0)]_{55}
\nl
&= 
-e^2 I_3(\delta_\pm^{(0)}, \lambda) 
-{g_2^2\over 4 c_W^2} (1-2s_W^2)^2 I_3( \delta_\pm^{(0)}, m_Z) 
- {g_2^2\over 4} I_3(\delta_0^{(0)}, m_W) 
\nl
&\quad
- {g_2^2\over 4}s^2_{\rho\over 2} I_3(\delta_0^{(+)},m_W)
- {g_2^2\over 4}c^2_{\rho\over 2} I_3(\delta_0^{(-)},m_W) 
+ a^2c^2_{\rho\over 2} I_3(\delta_0^{(+)},m_W) 
\nl
&\quad
+ a^2s^2_{\rho\over 2} I_3(\delta_0^{(-)},m_W) \, ,
\end{align} 
where $\lambda$ is a fictitious photon mass, and the self-energy
components not displayed above vanish. We may evaluate $\Sigma (v
\cdot k)$ by the substitution $I_3(\delta, m) \to I_3(\delta - v \cdot
k, m)$.

\subsection{Renormalization conditions} 

Let us fix the counterterms $\delta a_1$, $\delta a_2$, $\delta
\mu_S$, $\delta \mu_D$ and $\delta Z_S$ by enforcing that the physical
residual masses of the neutral states are given by the renormalized parameters of the
lagrangian,
\begin{align}\label{eq:condition1}
[\delta \mu]_{11} + {\rm Re}[\Sigma_2(\delta_\pm ^{(0)})]_{11} -\delta_0^{(0)} [\delta Z_h]_{11} &= 0 \,, 
\nl
[\delta \mu]_{22} + {\rm Re}[\Sigma_2(\delta_0^{(+)})]_{22} - \delta_0^{(+)} [\delta Z_h]_{22} &= 0 \,,
\nl
[\delta \mu]_{33} + {\rm Re}[\Sigma_2(0)]_{33} &= 0 \,, 
\end{align}
and that the lightest mass eigenstate is proportional to the vector $(0,0,1,0,0)$,
\begin{align}\label{eq:condition2}
[\delta \mu]_{13} + {\rm Re}[\Sigma_2(0)]_{13} &= 0 \,,
\nl
[\delta \mu]_{23} + {\rm Re}[\Sigma_2(0)]_{23} &= 0 \,. 
\end{align} 
This scheme defines renormalized values for $a$ and $t_{\rho\over 2}$
through the physical mass differences between neutral states,
\begin{align}
M_{h_0^{(+)}} - M_{h_0^{(-)}} 
&= 2av s_\rho^{-1}
\,,
\nl
M_{h_0^{(0)}} - M_{h_0^{(-)}} 
&= av t_{\rho\over 2} \,,
\end{align}
where the mass of the neutral mass eigenstate $h_{0}^{( \cdot )}$ is
denoted $M_{h_{0}^{( \cdot )}}$.  Note also that the presence of
$\delta Z_S \ne \delta Z_D$ is required to maintain the orientation of
the lightest mass eigenstate under renormalization. Solving for the
counterterms, we find from $[\delta \mu]_{13}$,
\be
{\delta a_1 \over a_1} = {\delta a_2 \over a_2} 
\implies 
a_1 \delta a_1 + a_2 \delta a_2  = a^2 {\delta a_1 \over a_1} \,. 
\ee
The remaining system of equations involving 
$[\delta \mu]_{23}$, $[\delta \mu]_{11}$, $[\delta \mu]_{22}$ and $[\delta \mu]_{33}$  
then yields 
\begin{align} \label{eq:deltaaSD}
a v{\delta a_1 \over a_1} 
&= - [\delta \mu]_{23} + t_{\rho\over 2}^{-1} \left( [\delta \mu]_{11} - [\delta \mu]_{33} \right) 
\nl
&= [\Sigma_2(0)]_{23} + t_{\rho\over 2}^{-1} 
\left( [\Sigma_2(0)]_{33} - [\Sigma_2(\delta_0^{(0)})]_{11} + \delta_0^{(0)} [\delta Z_h]_{11} \right) \,,
\nl
\delta Z_S &= 
\delta Z_D + {1\over av} \bigg\{ 
t_{\rho\over 2} [ \Sigma_2(\delta_0^{(+)}) ]_{22} + 2 [\Sigma_2(0)]_{23} + t_{\rho\over 2}^{-1} [\Sigma_2(0)]_{33}
-2 s_\rho^{-1} [\Sigma_2(\delta_0^{(0)})]_{11} 
\bigg\} \,. 
\end{align} 
We focus here on the counterterms $\delta a_1$, $\delta a_2$, and
$\delta Z_S$ which enter in the calculation of amplitudes relevant for
WIMP-nucleon scattering. Explicit expressions for 
the remaining counterterms $\delta \mu_S$ and $\delta \mu_D$ 
may be similarly obtained. We note that the degeneracy between the mass of the $h_0^{(0)}$ state
and the $h_\pm^{(0)}$ states is lifted by a finite amount, predicted
in terms of renormalized parameters as
\begin{align}
M_{h_\pm^{(0)}} - M_{h_0^{(0)}}
&= [\Sigma_2(\delta_\pm^{(0)})]_{44} - [\Sigma_2(\delta_0^{(0)})]_{11}\, , 
\end{align} 
where we have used that $[\delta \mu]_{11} = [\delta \mu]_{44}$,
$[\delta Z_h]_{11}=[\delta Z_h]_{44}$ and
$\delta_0^{(0)}=\delta_\pm^{(0)}$.

\subsection{Extension to triplet-doublet}

The extension to the triplet-doublet case is straightforward. The
counterterms $\delta a_1$, $\delta a_2$, $\delta \mu_T$, $\delta
\mu_D$, $\delta Z_T$ and $\delta Z_D$ are introduced in an analogous
manner. In terms of the mass eigenbasis $h= \left( h_0^{(0)} ,
h_0^{(+)}, h_0^{(-)}, h_+^{(+)}, h_+^{(-)}, h_-^{(+)}, h_-^{(-)}
\right)$ introduced in
Sec.~\ref{sec:TDmatrices}, the counterterms are given by the $7 \times
7$ matrices,
\begin{align}
& \delta \mu = 
\delta \mu_D \mathbb{1}_7 + \left(\begin{array}{ccc} 
\delta \mu_{0} & 0 & 0 \\
0 & \delta \mu_{+} & 0 \\
0 & 0 & \delta \mu_{-}  
\end{array}
\right) \,,
\quad 
\delta Z_h = \delta Z_D\mathbb{1}_7 + 
\left(\begin{array}{ccc} 
\delta Z_{0} & 0 & 0 \\
0 & \delta Z_{+ } & 0 \\
0 & 0 & \delta Z_{ -}  
\end{array}
\right) \,,
\end{align}
where the submatrices for the neutral and charged sectors are
specified by the following symmetric matrices,
\begin{align}
&\delta \mu_{0} = 
 \left(\begin{array}{ccc}
0 & |c_{\rho\over 2}| {v\over a}(a_2 \delta a_1-a_1\delta a_2) 
& |s_{\rho\over 2}|{v\over a} (a_1 \delta a_2 - a_2 \delta a_1) 
\\
\cdot &  2c^2_{\rho\over 2} ( \delta \Delta ) + s_\rho{v\over a}(a_1\delta a_1 + a_2\delta a_2) 
& -s_\rho ( \delta \Delta ) + c_\rho {v\over a} (a_1 \delta a_1 + a_2 \delta a_2) 
\\
\cdot & \cdot & 2s^2_{\rho\over 2} ( \delta \Delta ) - s_\rho{v\over a} (a_1 \delta a_1 + a_2 \delta a_2)
\end{array}
\right)\,, 
\nl
&\delta \mu_{\pm} =  
\left(\begin{array}{cc} 
2c_{\rho\over 2}^2 ( \delta \Delta ) + s_\rho {v\over a} (a_1 \delta a_1 + a_2 \delta a_2) 
& \quad -s_\rho ( \delta \Delta ) + c_\rho{v\over a} (a_1\delta a_1 + a_2 \delta a_2) \pm i{v\over a}
( a_1\delta a_2 - a_2\delta a_1) 
\\
\cdot & 2 s_{\rho\over 2}^2 ( \delta \Delta ) -s_\rho {v\over a} (a_1\delta a_1 + a_2\delta a_2) 
\end{array}
\right) \,,
\nl
&\delta Z_{0} = 
\left(\begin{array}{ccc}
0  & 0 & 0 
\\
\cdot &  c^2_{\rho\over 2} (\delta Z_T - \delta Z_D ) & -\frac12 s_\rho ( \delta Z_T - \delta Z_D)  
\\
\cdot & \cdot & s^2_{\rho\over 2} ( \delta Z_T - \delta Z_D )
\end{array}
\right)\,, 
\nl
&\delta Z_{\pm} =  
\left(\begin{array}{cc} 
c^2_{\rho\over 2}  ( \delta Z_T - \delta Z_D ) & -\frac12 s_\rho ( \delta Z_T - \delta Z_D)  
\\
\cdot & s^2_{\rho\over 2} ( \delta Z_T - \delta Z_D )
\end{array}
\right) \,,
\end{align}
with $( \delta \Delta ) = (\delta \mu_T - \delta \mu_D)/2$. To fix
counterterms, we impose the same renormalization conditions given in
(\ref{eq:condition1}) and (\ref{eq:condition2}). We
again require the one-loop corrections to the two-point function,
$\Sigma_2$. In the following results, 
the first (second) subscript denotes the final (initial) state, with values $(1,2,3,4,5,6,7)$ corresponding to the mass eigenstates $\left( h_0^{(0)} ,
h_0^{(+)}, h_0^{(-)}, h_+^{(+)}, h_+^{(-)}, h_-^{(+)}, h_-^{(-)}
\right)$. Using Feynman-t'Hooft gauge and expressing results in
terms of the basic integral $I_3(\delta,m)$ of Appendix~\ref{sec:self}, we
find
\begin{align} \label{eq:selfTD}
-i[\Sigma_2(0)]_{11} &= 
-{g_2^2\over 4 c_W^2} s^2_{\rho\over 2} I_3(\delta_0^{(+)},m_Z) 
-{g_2^2\over 4 c_W^2} c^2_{\rho\over 2} I_3(\delta_0^{(-)},m_Z) 
-{g_2^2\over 2} s^2_{\rho\over 2} I_3(\delta_\pm^{(+)},m_W) 
\nl
&\quad
-{g_2^2\over 2} c^2_{\rho\over 2} I_3(\delta_\pm^{(-)},m_W) 
+ a^2c^2_{\rho\over 2} I_3(\delta_0^{(+)},m_Z) 
+ a^2s^2_{\rho\over 2} I_3(\delta_0^{(-)},m_Z) 
\nl
&\quad
+ 2a^2 c^2_{\rho\over 2} I_3(\delta_\pm^{(+)},m_W) 
+ 2a^2 s^2_{\rho\over 2} I_3(\delta_\pm^{(-)},m_W) \,,
\nl
-i[\Sigma_2(0)]_{22} &= 
-{g_2^2\over 4 c_W^2}s^2_{\rho\over 2} I_3(\delta_0^{(0)},m_Z) 
-{g_2^2\over 2} \left(1+c^2_{\rho\over 2} \right)^2 I_3(\delta_\pm^{(+)},m_W) 
-{g_2^2\over 8} s^2_\rho I_3(\delta_\pm^{(-)},m_W) 
\nl
&\quad
+ a^2 c^2_{\rho\over 2} I_3(\delta_0^{(0)},m_Z) 
+ 2 a^2 I_3(\delta_\pm^{(-)},m_W) 
+ a^2c^2_\rho I_3(\delta_0^{(-)},m_h)
+ a^2s^2_\rho I_3(\delta_0^{(+)},m_h) 
\,,
\nl
-i[\Sigma_2(0)]_{23} &= -i[\Sigma_2(0)]_{32} 
\nl &= 
-{g_2^2 \over 8 c_W^2} s_\rho I_3(\delta_0^{(0)},m_Z) 
+ {g_2^2\over 4} s_\rho \left(1+c^2_{\rho\over 2} \right) I_3(\delta_\pm^{(+)},m_W) 
+ {g_2^2\over 4} s_\rho \left(1+s^2_{\rho\over 2} \right) I_3(\delta_\pm^{(-)},m_W)
\nl
&\quad
-{a^2\over 2}s_\rho I_3(\delta_0^{(0)},m_Z) 
+ a^2c_\rho s_\rho I_3(\delta_0^{(+)},m_h) 
- a^2 c_\rho s_\rho I_3(\delta_0^{(-)},m_h) 
\,,
\nl
-i[\Sigma_2(0)]_{33} &= 
-{g_2^2\over 4 c_W^2} c^2_{\rho\over 2} I_3(\delta_0^{(0)},m_Z) 
-{g_2^2\over 2} \left(1+s^2_{\rho\over 2} \right)^2 I_3(\delta_\pm^{(-)},m_W) 
-{g_2^2\over 8} s^2_\rho I_3(\delta_\pm^{(+)},m_W) 
\nl
&\quad 
+a^2s^2_{\rho\over 2} I_3(\delta_0^{(0)},m_Z) 
+ 2 a^2 I_3(\delta_\pm^{(+)},m_W) 
+ a^2c^2_\rho I_3(\delta_0^{(+)},m_h) 
+ a^2s^2_\rho I_3(\delta_0^{(-)},m_h) 
\,,
\nl
-i[\Sigma_2(0)]_{44} &= -i[\Sigma_2(0)]_{66} \nl
&= 
-{g_2^2\over c_W^2} \left(c_W^2 -\frac12 s^2_{\rho\over 2} \right)^2 I_3(\delta_\pm^{(+)},m_Z) 
-{g_2^2 \over 16 c_W^2} s^2_\rho I_3(\delta_\pm^{(-)},m_Z) 
- {g_2^2\over 4} s^2_{\rho\over 2} I_3(\delta_0^{(0)},m_W) 
\nl
&\quad 
-e^2 I_3(\delta_\pm^{(+)},\lambda) -{g_2^2\over 4} \left( 1+c^2_{\rho\over 2} \right)^2 I_3(\delta_0^{(+)},m_W) 
-{g_2^2\over 16} s^2_\rho I_3(\delta_0^{(-)},m_W) 
+ a^2 I_3(\delta_\pm^{(-)},m_Z) 
\nl
&\quad 
+ a^2 I_3(\delta_0^{(-)},m_W) 
+ a^2c^2_{\rho\over 2} I_3(\delta_0^{(0)},m_W) 
+ a^2s^2_\rho I_3(\delta_\pm^{(+)},m_h) 
+ a^2c^2_\rho I_3(\delta_\pm^{(-)},m_h) 
\,,
\nl
-i[\Sigma_2(0)]_{45} &= -i[\Sigma_2(0)]_{54} =  -i[\Sigma_2(0)]_{67} =  -i[\Sigma_2(0)]_{76} 
\nl
&= {g_2^2\over 4 c_W^2} s_\rho \left( c_W^2 -\frac12 s^2_{\rho\over 2} \right) I_3(\delta_\pm^{(+)},m_Z) 
+ {g_2^2\over 4 c_W^2} s_\rho \left( c_W^2 -\frac12 c^2_{\rho\over 2} \right) I_3(\delta_\pm^{(-)},m_Z) 
\nl
&\quad
- {g_2^2\over 8} s_\rho I_3(\delta_0^{(0)},m_W) 
+ {g_2^2\over 8}s_\rho \left( 1+c^2_{\rho\over 2} \right) I_3(\delta_0^{(+)},m_W) 
+ {g_2^2\over 8}s_\rho \left( 1+s^2_{\rho\over 2} \right) I_3(\delta_0^{(-)},m_W) 
\nl
&\quad 
-{a^2\over 2} s_\rho I_3(\delta_0^{(0)},m_W) 
+ a^2 c_\rho s_\rho I_3(\delta_\pm^{(+)},m_h) 
-a^2c_\rho s_\rho I_3(\delta_\pm^{(-)},m_h) 
\,,
\nl
-i[\Sigma_2(0)]_{55} &= -i[\Sigma_2(0)]_{77} 
\nl
&= 
-{g_2^2\over 16 c_W^2} s^2_\rho I_3(\delta_\pm^{(+)},m_Z) 
-{g_2^2\over c_W^2} \left( c_W^2 -\frac12 c^2_{\rho\over 2} \right)^2 I_3(\delta_\pm^{(-)},m_Z) 
-{g_2^2\over 4} c^2_{\rho\over 2} I_3(\delta_0^{(0)},m_W) 
\nl
&\quad
-e^2 I_3(\delta_\pm^{(-)},\lambda) -{g_2^2\over 16} s^2_\rho I_3(\delta_0^{(+)},m_W) 
-{g_2^2\over 4} \left( 1+s^2_{\rho\over 2} \right)^2 I_3(\delta_0^{(-)},m_W) 
+ a^2 I_3(\delta_\pm^{(+)},m_Z) 
\nl
&\quad 
+ a^2 I_3(\delta_0^{(+)},m_W) 
+ a^2 s^2_{\rho\over 2} I_3(\delta_0^{(0)},m_W) 
+ a^2c^2_{\rho} I_3(\delta_\pm^{(+)},m_h) 
+ a^2 s^2_\rho I_3(\delta_\pm^{(-)},m_h) \,,
\end{align}  
where $\lambda$ is a fictitious photon mass, and the self-energy components 
not displayed above vanish. The remainder of the renormalization program 
proceeds as for the singlet-doublet system. In particular, the similarity 
of the neutral sectors implies relations similar to (\ref{eq:deltaaSD}),
\begin{align} \label{eq:deltaaTD}
a v{\delta a_1 \over a_1} &= av {\delta a_2 \over a_2} 
= [\Sigma_2(0)]_{23} + t_{\rho\over 2}^{-1}
\left( [\Sigma_2(0)]_{33} - [\Sigma_2(\delta_0^{(0)})]_{11} + \delta_0^{(0)} [\delta Z_h]_{11} \right) \,,
\nl
\delta Z_T &= 
\delta Z_D + {1\over av} \bigg\{ 
t_{\rho\over 2} [ \Sigma_2(\delta_0^{(+)}) ]_{22} + 2 [\Sigma_2(0)]_{23} + t_{\rho\over 2}^{-1} [\Sigma_2(0)]_{33}
-2 s_\rho^{-1} [\Sigma_2(\delta_0^{(0)})]_{11} 
\bigg\} \,,  
\end{align} 
where the self-energy components are those of the 
triplet-doublet system given in (\ref{eq:selfTD}).   

\section{Matching at the weak scale\label{sec:matchingcalculation}}
This section describes the matching of the effective theory described
by (\ref{eq:heavy2}) onto the effective theory described by
(\ref{eq:LWIMPSM}), through integrating out weak-scale particles,
$W^\pm$, $Z^0$, $h$, $\phi_Z^0$, $\phi_W^\pm$, and $t$. The complete
basis of twelve bare matching coefficients, $c_{q}^{(0)}$,
$c_{q}^{(2)}$, $c_{g}^{(0)}$, and $c_{g}^{(2)}$, are determined at
leading order in perturbation theory.

We may write the quark and gluon matching coefficients in terms of
contributions from one-boson exchange (1BE) and two-boson exchange
(2BE) diagrams,
\begin{align}\label{eq:cBE}
&c_{q}^{(0)} = c_{q}^{(0)} {}_{\rm 1BE} + c_{q}^{(0)} {}_{\rm 2BE} + \dots \, ,
\nl
&c_g^{(0)} = c_g^{(0)} {}_{\rm 1BE} + c_g^{(0)} {}_{\rm 2BE} + \dots  \, ,
\nl
&c_{q}^{(2)} = c_{q}^{(2)} {}_{\rm 2BE} + \dots  \, ,
\nl
&c_g^{(2)} = c_g^{(2)} {}_{\rm 2BE}+ \dots  \, ,
\end{align}
where the ellipses denote subleading contributions with more than two
bosons exchanged. Note that spin-2 coefficients do not receive
contributions from one-boson exchange amplitudes. 

In the following analysis, we denote generic up- and down-type quarks by $U$ and $D$,
respectively, and an arbitrary quark flavor by $q$. We specify the contributions to the matching coefficients in terms of 
the constants
\begin{align}\label{eq:quarkscvca}
c_V^{(U)} = 1-\frac83 s_W^2 \,, \quad
c_V^{(D)} = -1 + \frac43 s_W^2 \,, \quad
c_A^{(U)} = -1 \,, \quad
c_A^{(D)} = 1 \,.
\end{align}
We systematically neglect subleading corrections involving light
quark masses, and use CKM unitarity to simplify sums over quark
flavors. Together with $|V_{tb}|\approx 1$ (and hence $|V_{td}|\approx |V_{ts}|\approx 0$), these assumptions lead to $c_{u}^{(S)} = c_{c}^{(S)}$ and $c_{d}^{(S)} = c_{s}^{(S)}$ for both $S=0,2$, reducing the number of
independent matching coefficients to eight. When the interactions are
isospin-conserving, e.g., as in the pure triplet case, we furthermore
have $c_{u}^{(S)} = c_{d}^{(S)}$ 
and $c_{c}^{(S)} = c_{s}^{(S)}$ 
for both $S=0,2$, leaving only six
independent coefficients. We use Feynman-t'Hooft gauge for the
electroweak sector, and neglect higher-order corrections to the tree-level 
relations between residual masses, $\delta_0^{(0)} = \delta_\pm^{(0)}$ 
for the singlet-doublet system, and $\delta_0^{(\pm)} = \delta_\pm^{(\pm)}$ 
for the triplet-doublet system. In Secs.~\ref{sec:1BEquark} and~\ref{sec:2BEquark}, 
we match to quark operators using onshell external quarks, 
and thus use the equivalence of $m_q u_q(p)$ and $\slash{p} u_q(p)$.

\subsection{Quark matching: one-boson exchange\label{sec:1BEquark}}

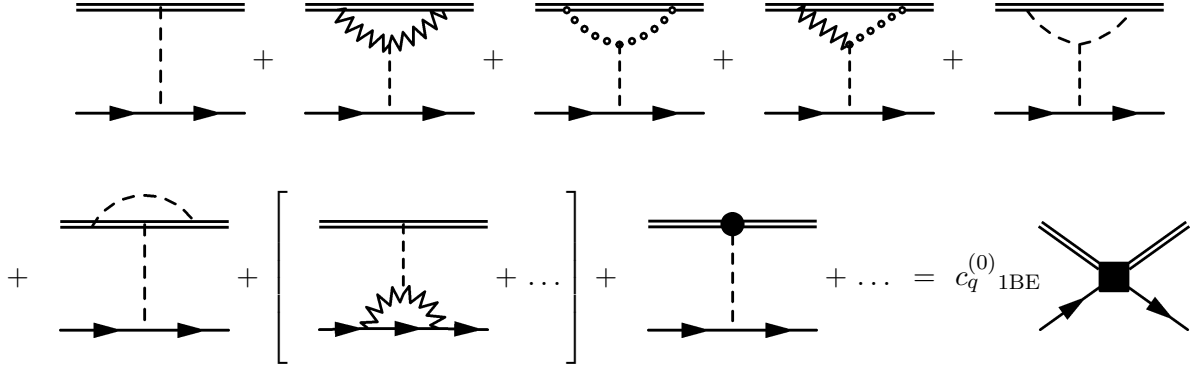
\begin{figure}[htb]
\begin{center}
\parbox{25mm}{
\begin{fmfgraph*}(80,40)
  \fmfleftn{l}{2}
  \fmfrightn{r}{2}
  \fmftopn{t}{3}
  \fmfbottomn{b}{3}
  \fmf{double}{l2,t2,r2}
  \fmf{fermion}{l1,b2,r1}
  \fmffreeze
  \fmf{dashes}{t2,b2}
\end{fmfgraph*}
}
+
\parbox{25mm}{
\begin{fmfgraph*}(80,40)
  \fmfleftn{l}{2}
  \fmfrightn{r}{2}
  \fmftopn{t}{5}
  \fmfbottomn{b}{5}
  \fmf{double}{l2,r2}
  \fmf{fermion}{l1,b3,r1}
  \fmffreeze
  \fmfset{zigzag_width}{1.5thick}
  \fmf{zigzag,right=0.2}{t2,v,t4}
  \fmf{dashes}{v,b3}
\end{fmfgraph*}
} 
+
\parbox{25mm}{
\begin{fmfgraph*}(80,40)
  \fmfleftn{l}{2}
  \fmfrightn{r}{2}
  \fmftopn{t}{5}
  \fmfbottomn{b}{5}
  \fmf{double}{l2,r2}
  \fmf{fermion}{l1,b3,r1}
  \fmffreeze
  \fmf{dbl_dots,right=0.2}{t2,v,t4}
  \fmf{dashes}{v,b3}
\end{fmfgraph*}
} 
+
\parbox{25mm}{
\begin{fmfgraph*}(80,40)
  \fmfleftn{l}{2}
  \fmfrightn{r}{2}
  \fmftopn{t}{5}
  \fmfbottomn{b}{5}
  \fmf{double}{l2,r2}
  \fmf{fermion}{l1,b3,r1}
  \fmffreeze
  \fmfset{zigzag_width}{1.5thick}
  \fmf{zigzag}{t2,v}
  \fmf{dbl_dots}{v,t4}
  \fmf{dashes}{v,b3}
\end{fmfgraph*}
}
+
\parbox{25mm}{
\begin{fmfgraph*}(80,40)
  \fmfleftn{l}{2}
  \fmfrightn{r}{2}
  \fmftopn{t}{5}
  \fmfbottomn{b}{5}
  \fmf{double}{l2,r2}
  \fmf{fermion}{l1,b3,r1}
  \fmffreeze
  \fmf{dashes,right=0.2}{t2,v,t4}
  \fmf{dashes}{v,b3}
\end{fmfgraph*}
} 
\nl
\vspace{10mm}

+
\parbox{25mm}{
\begin{fmfgraph*}(80,40)
  \fmfleftn{l}{2}
  \fmfrightn{r}{2}
  \fmftopn{t}{5}
  \fmfbottomn{b}{5}
  \fmf{double}{l2,r2}
  \fmf{fermion}{l1,b3,r1}
  \fmffreeze
  \fmf{dashes,left=0.6}{t2,t4}
  \fmf{dashes}{t3,b3}
\end{fmfgraph*}
}
+ \Vast[
\parbox{25mm}{
\begin{fmfgraph*}(80,40)
  \fmfleftn{l}{2}
  \fmfrightn{r}{2}
  \fmftopn{t}{5}
  \fmfbottomn{b}{7}
  \fmf{double}{l2,r2}
  \fmf{fermion}{l1,b3,b5,r1}
  \fmffreeze
  \fmf{dashes}{t3,v}
  \fmfset{zigzag_width}{1.5thick}
  \fmf{zigzag,left=0.4}{b3,v,b5}
\end{fmfgraph*}
}
+
 \dots \Vast]
 +
\parbox{25mm}{
\begin{fmfgraph*}(80,40)
  \fmfleftn{l}{2}
  \fmfrightn{r}{2}
  \fmftopn{t}{3}
  \fmfbottomn{b}{3}
  \fmf{double}{l2,t2,r2}
  \fmf{fermion}{l1,b2,r1}
  \fmfv{decor.shape=circle,decor.size=.10w}{t2}
  \fmffreeze
  \fmf{dashes}{t2,b2}
\end{fmfgraph*}
}
+ \dots \,
= 
$\,\,c_{q}^{(0)} {}_{\rm 1BE}\!\!\!\!\!\!$ 
\parbox{25mm}{
\begin{fmfgraph*}(70,40)
  \fmfleftn{l}{2}
  \fmfrightn{r}{2}
  \fmftopn{t}{4}
  \fmfbottomn{b}{4}
  \fmf{double}{l2,v,r2}
  \fmf{fermion}{l1,v,r1}
  \fmffreeze
  \fmfv{decor.shape=square,decor.size=0.15w}{v}
\end{fmfgraph*} 
}
\caption{\label{fig:quark1match} Matching condition for one-boson
exchange contributions to quark operators. The full theory diagrams on
the left-hand side illustrate the possible types of contributions to
the $h \bar{\chi} \chi$ three-point function. Time-reversed diagrams
are not shown. Double lines denote heavy WIMPs, zigzag lines denote gauge bosons, $W^\pm$ or $Z^0$, dotted lines denote Goldstone bosons,  
$\phi_W^\pm$ or $\phi_Z^0$, dashed
lines denote the physical Higgs boson, $h$, and single lines with arrows denote quarks. The solid circle denotes counterterm contributions. The solid square denotes effective theory vertices.
}
\end{center} 
\end{figure}

The matching condition for one-boson exchange is pictured in
Fig.~\ref{fig:quark1match}. The full-theory amplitude is given by
\begin{align}
i{\cal M}_q &= 
i \left( \hat{\cal M}_{\rm tree} + \hat{\cal M}_{\rm vertex,1}  + \hat{\cal M}_{\rm vertex,2} + \hat{\cal M}_{\delta a_1} + \hat{\cal M}_{\delta Z} + \hat{\cal M}_{\delta v} \right)
{i\over -m_h^2}
{-ig_2 m_q \over 2 m_W} \bar{u}_q (p) u_q (p) \,,
\end{align}
where  the $\hat{\cal M}_i$ are contributions to the $h \bar{\chi}
\chi$ three-point function. These come from tree-level Higgs exchange
($\hat{\cal M}_{\rm tree}$), one-loop diagrams with Higgs coupling to
$W^\pm$ or $Z^0$ ($\hat{\cal M}_{\rm vertex,1}$), one-loop vertex
corrections with Higgs coupling to the heavy particle ($\hat{\cal
M}_{\rm vertex,2}$), the $\delta a_1$ counterterm ($\hat{\cal
M}_{\delta a_1}$), wavefunction renormalization ($\hat{\cal
M}_{\delta Z}$), and the renormalization of the Higgs vacuum
expectation value ($\hat{\cal M}_{\delta v}$). Having included the
counterterms, the sum of these contributions is finite. The one-boson
exchange contribution to the spin-0 quark matching coefficient is thus
\be\label{eq:1BEquark}
c_{q}^{(0)} {}_{\rm 1BE} = - {g_2 \over 2 m_h^2 m_W}
\left( \hat{\cal M}_{\rm tree} + \hat{\cal M}_{\rm vertex,1}  
+ \hat{\cal M}_{\rm vertex,2} + \hat{\cal M}_{\delta a_1} 
+ \hat{\cal M}_{\delta Z} + \hat{\cal M}_{\delta v} \right)
 \,.
\ee
We neglect one-boson exchange contributions containing ${\cal
O}(\alpha_2^1)$ corrections to the SM $h \bar{q} q$ coupling, shown in
Fig.~\ref{fig:quark1match} within square brackets. This
gauge-invariant class of diagrams is loop-suppressed relative to the
tree-level diagram for any value of the $h \bar{\chi} \chi$
coupling. On the other hand, the remaining loop diagrams (including those
in Fig.~\ref{fig:quark2match}) may compete with, or even dominate, the 
tree-level contribution depending on the size of the $h \bar{\chi} \chi$ 
coupling. Let us proceed to specify the contributions, $\hat{\cal M}_i$, for 
each SM extension in terms of the integrals $I_1(\delta, m)$, 
$I_2(\delta, m)$, $I_3(\delta, m)$ and $I_4(\delta_1, \delta_2, m)$ 
of Appendix~\ref{sec:self}.

\subsubsection{Pure states}
For pure states the only diagrams are those with Higgs coupling 
to $W^\pm$ and $Z^0$, and in terms of the constants ${\cal C}_W$ 
and ${\cal C}_Z$ specified in (\ref{eq:WZcoeff}) the amplitude is given by
\begin{align}\label{eq:1BEquarkpureAMP}
i\hat{\cal M}_{\rm vertex,1} &= 
 - {\cal C}_Z {g_2^3\over  c_W^3}  m_Z I_1(0,m_Z) - {\cal C}_W g_2^3  m_W I_1(0, m_W)  \,.
\end{align}
Using (\ref{eq:1BEquark}), we find the contribution to the spin-0 quark matching coefficient, 
\be\label{eq:1BEquarkpureCO}
c_{q}^{(0)} {}_{\rm 1BE} = { \pi \Gamma(1+\epsilon) g_2^4 \over (4\pi)^{2-\epsilon}}  
\bigg\{  - { m_W^{-3-2\epsilon} \over 2 x_h^2} \left( {\cal C}_W + { {\cal C}_Z \over  c_W^3}  \right) + \order(\epsilon) \bigg\} \,,
\ee
where $x_h = m_h/m_W$.
The pure triplet (doublet) result is obtained by setting 
${\cal C}_W=2$ and ${\cal C}_Z=0$ (${\cal C}_W=1/2$ and ${\cal C}_Z=1/4$) above. 

\subsubsection{Singlet-doublet admixture}

For the singlet-doublet case, we have the following contributions to the $h\bar{\chi}\chi$ three-point function,
\begin{align}\label{eq:1BEquarkSDAMP}
i\hat{\cal M}_{\rm tree} &= i a s_\rho \,, 
\quad i\hat{\cal M}_{\delta a_1} = i a s_\rho {\delta a_1\over a_1} \,, 
\quad i\hat{\cal M}_{\delta Z} = i a s_\rho \delta Z_\chi \,,
\quad i\hat{\cal M}_{\delta v} = i a s_\rho {\delta v \over v} \,, 
\nl
i\hat{\cal M}_{\rm vertex,1} &= 
 -{g_2^3\over 4 c_W^3} c^2_{\rho\over 2} m_Z I_1(\delta_0^{(0)},m_Z) 
+ {g_2^2 a \over 4 c_W^2} s_\rho I_2(\delta_0^{(0)}, m_Z) 
+ {g_2 a^2\over 2} s^2_{\rho\over 2} {m_h^2\over m_W} I_1(\delta_0^{(0)},m_Z) 
\nl
&\quad
+ {3 g_2 a^2\over 2} {m_h^2\over m_W} \left[ s^2_\rho I_1(\delta_0^{(-)},m_h) + c^2_\rho I_1(\delta_0^{(+)},m_h) \right]
-{g_2^3\over 2} c^2_{\rho\over 2} m_W I_1(\delta_0^{(0)}, m_W) 
\nl
&\quad
+ {g_2^2 a\over 2} s_\rho I_2(\delta_0^{(0)}, m_W) 
+ g_2 a^2 s^2_{\rho\over 2} {m_h^2\over m_W} I_1(\delta_0^{(0)}, m_W)  \,,
\nl
i\hat{\cal M}_{\rm vertex,2} &= 
- a^3 s^3_\rho I_4(\delta_0^{(-)},\delta_0^{(-)}, m_h )
+ a^3s_\rho c^2_\rho I_4(\delta_0^{(+)},\delta_0^{(+)},m_h ) 
-2 a^3s_\rho c^2_\rho I_4(\delta_0^{(-)},\delta_0^{(+)}, m_h) 
\,,
\end{align}
where $\delta a_1$ is given in (\ref{eq:deltaaSD}), the onshell $Z$ factor is given by 
\begin{align}\label{eq:onshellZ}
Z_\chi^{-1} -1 &= -\delta Z_\chi =  
[\delta Z_h]_{33} - {\partial \over \partial v\cdot k} [\Sigma_2(v\cdot k)]_{33} 
= 
\delta Z_D - [\Sigma_2^\prime(0)]_{33} 
\nl
& \quad + {1 \over av} s^2_{\rho\over 2} 
\bigg\{   - 2 s_\rho^{-1} [\Sigma_2(\delta_0^{(0)})]_{11} 
+t_{\rho\over 2} [\Sigma_2(\delta_0^{(+)})]_{22} + 2 [\Sigma_2(0)]_{23} + t_{\rho\over 2}^{-1} [\Sigma_2(0)]_{33}
\bigg\}  \,,
\end{align}
and $\delta v$ is determined by the SM result~\cite{Denner:1991kt},
\begin{align}\label{eq:deltav}
{\delta v\over v} &= \frac12 \Sigma^{AA\prime}(0) - {s_W \over c_W} {\Sigma^{AZ}(0)\over m_Z^2} 
- {c_W^2\over 2 s_W^2} {{\rm Re} [\Sigma^{ZZ}(m_Z^2)] \over m_Z^2} 
+ {c_W^2-s_W^2 \over 2 s_W^2} { {\rm Re} [\Sigma^{WW}(m_W^2)] \over m_W^2}  
-\frac12 {\rm Re}[\Sigma^{HH \prime}(m_h^2)] \, .
\end{align}
The two-point functions required in (\ref{eq:deltav}) are specified in (\ref{eq:SMself}) of Appendix~\ref{sec:self}.%
\footnote{We are here neglecting contributions from states beyond the SM. Renormalization schemes relevant for WIMPs of mass $M \sim m_W$ are discussed in Refs.~\cite{Eberl:2001eu}.}
The one-boson exchange quark matching coefficient is obtained by collecting the above 
amplitudes into (\ref{eq:1BEquark}).
Upon taking the pure-case limits described in Sec.~\ref{sec:limits}, 
we recover the results (\ref{eq:1BEquarkpureAMP}) and (\ref{eq:1BEquarkpureCO}) 
for a pure doublet. In the pure singlet limit, the one-boson exchange amplitudes vanish.

\subsubsection{Triplet-doublet admixture}

For the triplet-doublet case 
we have the following contributions to the $h \bar{\chi}\chi$ three-point function,
\begin{align}\label{eq:1BEquarkTDAMP}
i\hat{\cal M}_{\rm tree} &= i a s_\rho \,, \quad i\hat{\cal M}_{\delta a_1} = i a s_\rho {\delta a_1\over a_1} \,, \quad i\hat{\cal M}_{ \delta Z} = i a s_\rho \delta Z_\chi \,, \quad i\hat{\cal M}_{ \delta v} = i a s_\rho {\delta v \over v},
\nl
i\hat{\cal M}_{\rm vertex,1} &= 
-{g_2^3\over 4 c_W^3} c^2_{\rho\over 2} m_Z I_1(\delta_0^{(0)},m_Z) 
+ {g_2^2 a \over 4 c_W^2} s_\rho I_2(\delta_0^{(0)}, m_Z) 
+ {g_2 a^2\over 2} s^2_{\rho\over 2} {m_h^2\over m_W} I_1(\delta_0^{(0)},m_Z) 
\nl
&\quad
+ {3 g_2 a^2\over 2} {m_h^2\over m_W} \big[ s^2_\rho I_1(\delta_0^{(-)},m_h) + c^2_\rho I_1(\delta_0^{(+)},m_h) \big]
-{g_2^3\over 8} s^2_\rho m_W I_1(\delta_0^{(+)},m_W) 
\nl
&\quad
-{g_2^3\over 2} (1+s^2_{\rho\over 2})^2 m_W I_1(\delta_0^{(-)},m_W) 
+ {g_2^2 a\over 2} s_\rho I_2(\delta_0^{(+)}, m_W) 
+ g_2 a^2{m_h^2 \over m_W} I_1(\delta_0^{(+)},m_W) 
\,,
\nl
i\hat{\cal M}_{\rm vertex,2} &= 
-{g_2^2a\over 8} s_\rho^3 I_4(\delta_0^{(+)},\delta_0^{(+)},m_W) 
+ {g_2^2 a \over 2} (1+s^2_{\rho\over 2}) s_\rho c_\rho I_4(\delta_0^{(-)}, \delta_0^{(+)}, m_W) 
\nl
&\quad 
+ {g_2^2 a \over 2} (1+s^2_{\rho\over 2})^2 s_\rho I_4(\delta_0^{(-)},\delta_0^{(-)},m_W) 
+ 2a^3 s_\rho I_4(\delta_0^{(+)},\delta_0^{(+)},m_W) 
\nl
&\quad
+ a^3 c^2_\rho s_\rho I_4(\delta_0^{(+)},\delta_0^{(+)},m_h) 
-2 a^3 c^2_\rho s_\rho I_4(\delta_0^{(-)},\delta_0^{(+)},m_h)
- a^3 s_\rho^3 I_4(\delta_0^{(-)},\delta_0^{(-)},m_h)\, ,
\end{align}
where $\delta a_1$ is specified in (\ref{eq:deltaaTD}) and $\delta v$
in (\ref{eq:deltav}). The onshell $Z$ factor takes the same form as in
(\ref{eq:onshellZ}), but uses the self-energy components for the triplet-doublet system given in (\ref{eq:selfTD}).  The one-boson
exchange quark matching coefficient is obtained by collecting the
above amplitudes into (\ref{eq:1BEquark}).  Upon taking the pure case
limits described in Sec.~\ref{sec:limits}, we recover the results
(\ref{eq:1BEquarkpureAMP}) and (\ref{eq:1BEquarkpureCO}) for both pure
triplet and pure doublet.

\subsection{Gluon matching: one-boson exchange}

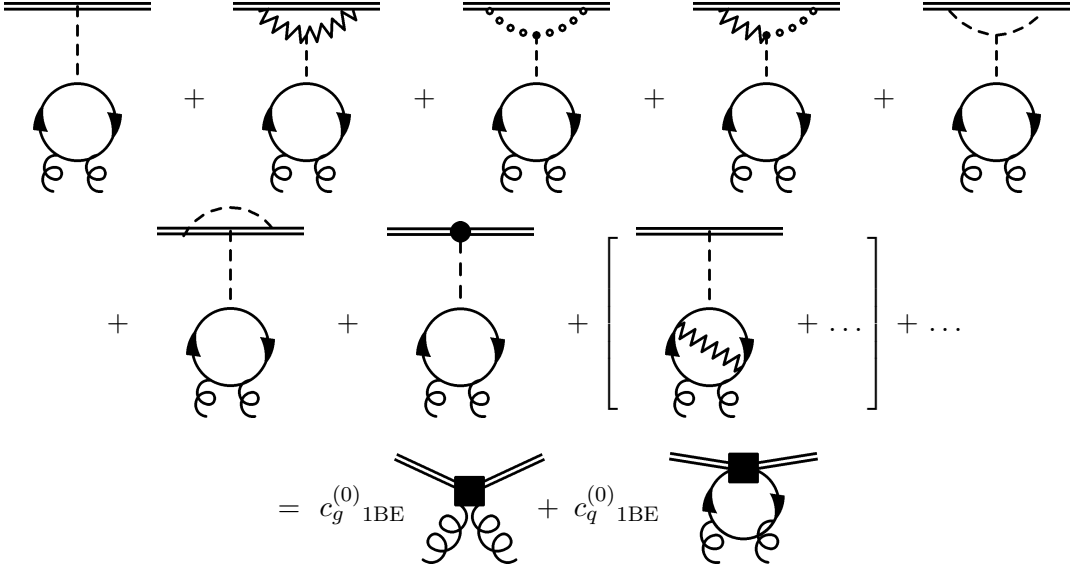
\begin{figure}[htb]
\begin{center}
\parbox{25mm}{
\begin{fmfgraph*}(70,70)
  \fmfleftn{l}{3}
  \fmfrightn{r}{3}
  \fmftopn{t}{5}
  \fmfbottomn{b}{5}
  \fmf{double}{l3,t3}
  \fmf{double}{t3,r3}
  \fmf{phantom,tag=3}{l1,r1}
  \fmf{phantom,tag=4}{l3,r3}
  \fmffreeze
  \fmf{phantom,tension=2}{t3,v1}
  \fmf{phantom,tension=5}{b3,v2}
  \fmf{phantom,left,tag=1}{v1,v2}
  \fmf{phantom,left,tag=2}{v2,v1}
  \fmfposition
  \fmffreeze
  \fmfipath{p[]}
  \fmfiset{p1}{vpath1(__v1,__v2)}
  \fmfiset{p2}{vpath2(__v2,__v1)}
  \fmfiset{p3}{vpath3(__l1,__r1)}
  \fmfiset{p4}{vpath4(__l3,__r3)}
  \fmfi{fermion}{p1}
  \fmfi{fermion}{p2}
  \fmfi{gluon}{point length(p2)/6 of p2 -- point length(p3)/3 of p3 }
  \fmfi{gluon}{point 5length(p1)/6 of p1 -- point 2length(p3)/3 of p3 }
  \fmf{dashes}{t3,v1}
\end{fmfgraph*}
}
+
\parbox{25mm}{
\begin{fmfgraph*}(70,70)
  \fmfleftn{l}{3}
  \fmfrightn{r}{3}
  \fmftopn{t}{5}
  \fmfbottomn{b}{5}
  \fmf{double}{l3,r3}
  \fmf{phantom,tag=3}{l1,r1}
  \fmf{phantom,tag=4}{l3,r3}
  \fmffreeze
  \fmf{phantom,tension=2}{t3,v1}
  \fmf{phantom,tension=5}{b3,v2}
  \fmf{phantom,left,tag=1}{v1,v2}
  \fmf{phantom,left,tag=2}{v2,v1}
  \fmfposition
  \fmffreeze
  \fmfipath{p[]}
  \fmfiset{p1}{vpath1(__v1,__v2)}
  \fmfiset{p2}{vpath2(__v2,__v1)}
  \fmfiset{p3}{vpath3(__l1,__r1)}
  \fmfiset{p4}{vpath4(__l3,__r3)}
  \fmfi{fermion}{p1}
  \fmfi{fermion}{p2}
  \fmfi{gluon}{point length(p2)/6 of p2 -- point length(p3)/3 of p3 }
  \fmfi{gluon}{point 5length(p1)/6 of p1 -- point 2length(p3)/3 of p3 }
  \fmfset{zigzag_width}{1.5thick}
  \fmf{zigzag,right=0.2}{t2,xx,t4}
  \fmf{dashes}{xx,v1}
\end{fmfgraph*}
}
+
\parbox{25mm}{
\begin{fmfgraph*}(70,70)
  \fmfleftn{l}{3}
  \fmfrightn{r}{3}
  \fmftopn{t}{5}
  \fmfbottomn{b}{5}
  \fmf{double}{l3,r3}
  \fmf{phantom,tag=3}{l1,r1}
  \fmf{phantom,tag=4}{l3,r3}
  \fmffreeze
  \fmf{phantom,tension=2}{t3,v1}
  \fmf{phantom,tension=5}{b3,v2}
  \fmf{phantom,left,tag=1}{v1,v2}
  \fmf{phantom,left,tag=2}{v2,v1}
  \fmfposition
  \fmffreeze
  \fmfipath{p[]}
  \fmfiset{p1}{vpath1(__v1,__v2)}
  \fmfiset{p2}{vpath2(__v2,__v1)}
  \fmfiset{p3}{vpath3(__l1,__r1)}
  \fmfiset{p4}{vpath4(__l3,__r3)}
  \fmfi{fermion}{p1}
  \fmfi{fermion}{p2}
  \fmfi{gluon}{point length(p2)/6 of p2 -- point length(p3)/3 of p3 }
  \fmfi{gluon}{point 5length(p1)/6 of p1 -- point 2length(p3)/3 of p3 }
  \fmfset{zigzag_width}{1.5thick}
  \fmf{dbl_dots,right=0.2}{t2,xx,t4}
  \fmf{dashes}{xx,v1}
\end{fmfgraph*}
}
+
\parbox{25mm}{
\begin{fmfgraph*}(70,70)
  \fmfleftn{l}{3}
  \fmfrightn{r}{3}
  \fmftopn{t}{5}
  \fmfbottomn{b}{5}
  \fmf{double}{l3,r3}
  \fmf{phantom,tag=3}{l1,r1}
  \fmf{phantom,tag=4}{l3,r3}
  \fmffreeze
  \fmf{phantom,tension=2}{t3,v1}
  \fmf{phantom,tension=5}{b3,v2}
  \fmf{phantom,left,tag=1}{v1,v2}
  \fmf{phantom,left,tag=2}{v2,v1}
  \fmfposition
  \fmffreeze
  \fmfipath{p[]}
  \fmfiset{p1}{vpath1(__v1,__v2)}
  \fmfiset{p2}{vpath2(__v2,__v1)}
  \fmfiset{p3}{vpath3(__l1,__r1)}
  \fmfiset{p4}{vpath4(__l3,__r3)}
  \fmfi{fermion}{p1}
  \fmfi{fermion}{p2}
  \fmfi{gluon}{point length(p2)/6 of p2 -- point length(p3)/3 of p3 }
  \fmfi{gluon}{point 5length(p1)/6 of p1 -- point 2length(p3)/3 of p3 }
  \fmfset{zigzag_width}{1.5thick}
  \fmf{zigzag,right=0.2}{t2,xx}
    \fmf{dbl_dots,right=0.2}{xx,t4}
  \fmf{dashes}{xx,v1}
\end{fmfgraph*}
}
+
\parbox{25mm}{
\begin{fmfgraph*}(70,70)
  \fmfleftn{l}{3}
  \fmfrightn{r}{3}
  \fmftopn{t}{5}
  \fmfbottomn{b}{5}
  \fmf{double}{l3,r3}
  \fmf{phantom,tag=3}{l1,r1}
  \fmf{phantom,tag=4}{l3,r3}
  \fmffreeze
  \fmf{phantom,tension=2}{t3,v1}
  \fmf{phantom,tension=5}{b3,v2}
  \fmf{phantom,left,tag=1}{v1,v2}
  \fmf{phantom,left,tag=2}{v2,v1}
  \fmfposition
  \fmffreeze
  \fmfipath{p[]}
  \fmfiset{p1}{vpath1(__v1,__v2)}
  \fmfiset{p2}{vpath2(__v2,__v1)}
  \fmfiset{p3}{vpath3(__l1,__r1)}
  \fmfiset{p4}{vpath4(__l3,__r3)}
  \fmfi{fermion}{p1}
  \fmfi{fermion}{p2}
  \fmfi{gluon}{point length(p2)/6 of p2 -- point length(p3)/3 of p3 }
  \fmfi{gluon}{point 5length(p1)/6 of p1 -- point 2length(p3)/3 of p3 }
  \fmfset{zigzag_width}{1.5thick}
  \fmf{dashes,right=0.2}{t2,xx,t4}
  \fmf{dashes}{xx,v1}
\end{fmfgraph*}
}
\nl
\vspace{5mm}
+
\parbox{25mm}{
\begin{fmfgraph*}(70,70)
  \fmfleftn{l}{3}
  \fmfrightn{r}{3}
  \fmftopn{t}{5}
  \fmfbottomn{b}{5}
  \fmf{double}{l3,t3}
  \fmf{double}{t3,r3}
  \fmf{phantom,tag=3}{l1,r1}
  \fmf{phantom,tag=4}{l3,r3}
  \fmffreeze
  \fmf{phantom,tension=2}{t3,v1}
  \fmf{phantom,tension=5}{b3,v2}
  \fmf{phantom,left,tag=1}{v1,v2}
  \fmf{phantom,left,tag=2}{v2,v1}
  \fmfposition
  \fmffreeze
  \fmfipath{p[]}
  \fmfiset{p1}{vpath1(__v1,__v2)}
  \fmfiset{p2}{vpath2(__v2,__v1)}
  \fmfiset{p3}{vpath3(__l1,__r1)}
  \fmfiset{p4}{vpath4(__l3,__r3)}
  \fmfi{fermion}{p1}
  \fmfi{fermion}{p2}
  \fmfi{gluon}{point length(p2)/6 of p2 -- point length(p3)/3 of p3 }
  \fmfi{gluon}{point 5length(p1)/6 of p1 -- point 2length(p3)/3 of p3 }
  \fmfset{zigzag_width}{1.5thick}
  \fmf{dashes,left=0.6}{t2,t4}
  \fmf{dashes}{t3,v1}
\end{fmfgraph*}
}
+
\parbox{25mm}{
\begin{fmfgraph*}(70,70)
  \fmfleftn{l}{3}
  \fmfrightn{r}{3}
  \fmftopn{t}{5}
  \fmfbottomn{b}{5}
  \fmf{double}{l3,t3}
  \fmf{double}{t3,r3}
  \fmf{phantom,tag=3}{l1,r1}
  \fmf{phantom,tag=4}{l3,r3}
  \fmffreeze
  \fmf{phantom,tension=2}{t3,v1}
  \fmf{phantom,tension=5}{b3,v2}
  \fmf{phantom,left,tag=1}{v1,v2}
  \fmf{phantom,left,tag=2}{v2,v1}
  \fmfposition
  \fmffreeze
  \fmfipath{p[]}
  \fmfiset{p1}{vpath1(__v1,__v2)}
  \fmfiset{p2}{vpath2(__v2,__v1)}
  \fmfiset{p3}{vpath3(__l1,__r1)}
  \fmfiset{p4}{vpath4(__l3,__r3)}
  \fmfi{fermion}{p1}
  \fmfi{fermion}{p2}
  \fmfi{gluon}{point length(p2)/6 of p2 -- point length(p3)/3 of p3 }
  \fmfi{gluon}{point 5length(p1)/6 of p1 -- point 2length(p3)/3 of p3 }
  \fmfset{zigzag_width}{1.5thick}
    \fmfv{decor.shape=circle,decor.size=.10w}{t3}
  \fmf{dashes}{t3,v1}
\end{fmfgraph*}
}
+
$\Vast[\!\!$
\parbox{25mm}{
\begin{fmfgraph*}(70,70)
  \fmfleftn{l}{3}
  \fmfrightn{r}{3}
  \fmftopn{t}{5}
  \fmfbottomn{b}{5}
  \fmf{double}{l3,t3}
  \fmf{double}{t3,r3}
  \fmf{phantom,tag=3}{l1,r1}
  \fmf{phantom,tag=4}{l3,r3}
  \fmffreeze
  \fmf{phantom,tension=2}{t3,v1}
  \fmf{phantom,tension=5}{b3,v2}
  \fmf{phantom,left,tag=1}{v1,v2}
  \fmf{phantom,left,tag=2}{v2,v1}
  \fmfposition
  \fmffreeze
  \fmfipath{p[]}
  \fmfiset{p1}{vpath1(__v1,__v2)}
  \fmfiset{p2}{vpath2(__v2,__v1)}
  \fmfiset{p3}{vpath3(__l1,__r1)}
  \fmfiset{p4}{vpath4(__l3,__r3)}
  \fmfi{fermion}{p1}
  \fmfi{fermion}{p2}
  \fmfi{gluon}{point length(p2)/6 of p2 -- point length(p3)/3 of p3 }
  \fmfi{gluon}{point 5length(p1)/6 of p1 -- point 2length(p3)/3 of p3 }
  \fmf{dashes}{t3,v1}
  \fmfset{zigzag_width}{1.5thick}
  \fmfi{zigzag}{point 4length(p1)/6 of p1 -- point 2length(p2)/3 of p2}
\end{fmfgraph*}
}
\!\!\!\!+ \dots
$\Vast]$ + \dots \,
\nl
\vspace{5mm}
 = 
$\,\,c_{g}^{(0)} {}_{\rm 1BE}\!\!\!\!\!\!\!\!$ 
\parbox{20mm}{
\begin{fmfgraph*}(70,40)
  \fmfleftn{l}{2}
  \fmfrightn{r}{2}
  \fmftopn{t}{4}
  \fmfbottomn{b}{5}
  \fmf{double}{l2,v,r2}
  \fmf{gluon,tension=0.5}{b2,v,b4}
  \fmffreeze
  \fmfv{decor.shape=square,decor.size=0.15w}{v}
\end{fmfgraph*} 
} 
+
$\, c_{q}^{(0)} {}_{\rm 1BE}\!\!\!\!$ 
\parbox{20mm}{
\begin{fmfgraph*}(70,40)
  \fmfleftn{l}{3}
  \fmfrightn{r}{3}
  \fmftopn{t}{3}
  \fmfbottomn{b}{3}
  \fmf{phantom}{l3,t2,r3}
  \fmf{phantom}{l1,b2,r1}
  \fmf{phantom,tag=3}{l1,r1}
  \fmf{phantom,tension=5}{t2,b2}
  \fmf{phantom,tension=10}{t2,v1}
  \fmf{phantom,tension=6}{b2,v2}
  \fmf{plain,left,tag=1}{v1,v2}
  \fmf{plain,left,tag=2}{v2,v1}
  \fmf{phantom,tag=4}{l3,v1}  
  \fmf{phantom,tag=5}{v1,r3}  
  \fmfposition
  \fmffreeze
   \fmfipath{p[]}
   \fmfiset{p1}{vpath1(__v1,__v2)}
   \fmfiset{p2}{vpath2(__v2,__v1)}
   \fmfiset{p3}{vpath3(__l1,__r1)}
   \fmfiset{p4}{vpath4(__l3,__v1)}
   \fmfiset{p5}{vpath5(__v1,__r3)}
   \fmfi{fermion}{subpath (0,length(p1)) of p1}
   \fmfi{fermion}{subpath (0,length(p2)) of p2}
   \fmfi{gluon}{point 2length(p1)/3 of p1 -- point 2length(p3)/3 of p3 }
   \fmfi{gluon}{point length(p2)/3 of p2 -- point length(p3)/3 of p3 }
   \fmfi{double}{subpath (0,length(p4)) of p4}
   \fmfi{double}{subpath (0,length(p5)) of p5}
  \fmfv{decor.shape=square,decor.size=0.15w}{v1}
\end{fmfgraph*}
}
\caption{\label{fig:gluon1match} Matching condition for one-boson
exchange contributions to gluon operators. The notation for the
different lines and vertices is as in Fig.~\ref{fig:quark1match}. 
All active quark
flavors, such as the top quark in the full theory, are included in the
loops.  }
\end{center}
\end{figure}

One-boson exchange contributions to gluon matching are pictured in
Fig.~\ref{fig:gluon1match}. The two-loop diagrams factorize into
separate one-loop diagrams: the boson loop given by the
amplitudes $\hat{\cal M}_i$ determined in the previous section, and
the fermion loop familiar from, e.g., the top quark contribution to
the effective $h(G^A_{\mu \nu})^2$ vertex (e.g., see~\cite{Chetyrkin:1997un}). In terms of quark matching
coefficients from one-boson exchange, $c_{q}^{(0)} {}_{\rm 1BE}$, the leading contribution to the
bare gluon matching coefficient is thus
\be\label{eq:1BEglue}
c_g ^{(0)} {}_{\rm 1BE} = - { g^2  \over (4\pi)^{2} } {1 \over 3}  c_{q}^{(0)} {}_{\rm 1BE} +\order(\epsilon) \ .
\ee
For the same reason discussed after Eq.~(\ref{eq:1BEquark}), we
neglect the one-boson exchange contributions containing ${\cal
O}(\alpha_2^1)$ corrections to the effective $h (G_{\mu \nu}^A)^2$
coupling, shown within square brackets in Fig.~\ref{fig:gluon1match}.
In the above result for $c_g ^{(0)} {}_{\rm 1BE}$, the light quark
contributions cancel between the full and effective theory amplitudes, 
leaving only contributions from the top quark. Further discussion of effective theory contributions can be found in Sec.~\ref{sec:EFTsubtract}. 

\subsection{Quark matching: two-boson exchange\label{sec:2BEquark}}

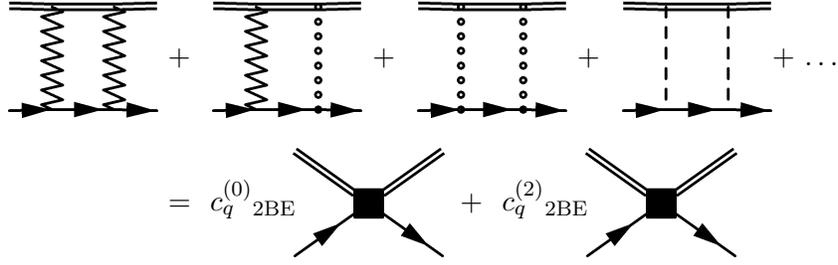
\begin{figure}[htb]
\begin{center}
\parbox{21mm}{
\begin{fmfgraph*}(70,40)
  \fmfleftn{l}{2}
  \fmfrightn{r}{2}
  \fmftopn{t}{4}
  \fmfbottomn{b}{4}
  \fmf{double}{l2,t2}
  \fmf{double}{t2,t3}
  \fmf{double}{t3,r2}
  \fmf{fermion}{l1,b2}
  \fmf{fermion}{b2,b3}
  \fmf{fermion}{b3,r1}
  \fmffreeze
  \fmf{zigzag}{t2,b2}
  \fmf{zigzag}{t3,b3}
\end{fmfgraph*}
}
\ +\!
\parbox{21mm}{
\begin{fmfgraph*}(70,40)
  \fmfleftn{l}{2}
  \fmfrightn{r}{2}
  \fmftopn{t}{4}
  \fmfbottomn{b}{4}
  \fmf{double}{l2,t2}
  \fmf{double}{t2,t3}
  \fmf{double}{t3,r2}
  \fmf{fermion}{l1,b2}
  \fmf{fermion}{b2,b3}
  \fmf{fermion}{b3,r1}
  \fmffreeze
  \fmf{zigzag}{t2,b2}
  \fmf{dbl_dots}{t3,b3}
\end{fmfgraph*}
}
\ +\!
\parbox{21mm}{
\begin{fmfgraph*}(70,40)
  \fmfleftn{l}{2}
  \fmfrightn{r}{2}
  \fmftopn{t}{4}
  \fmfbottomn{b}{4}
  \fmf{double}{l2,t2}
  \fmf{double}{t2,t3}
  \fmf{double}{t3,r2}
  \fmf{fermion}{l1,b2}
  \fmf{fermion}{b2,b3}
  \fmf{fermion}{b3,r1}
  \fmffreeze
  \fmf{dbl_dots}{t2,b2}
  \fmf{dbl_dots}{t3,b3}
\end{fmfgraph*}
}
\ +\!
\parbox{21mm}{
\begin{fmfgraph*}(70,40)
  \fmfleftn{l}{2}
  \fmfrightn{r}{2}
  \fmftopn{t}{4}
  \fmfbottomn{b}{4}
  \fmf{double}{l2,t2}
  \fmf{double}{t2,t3}
  \fmf{double}{t3,r2}
  \fmf{fermion}{l1,b2}
  \fmf{fermion}{b2,b3}
  \fmf{fermion}{b3,r1}
  \fmffreeze
  \fmf{dashes}{t2,b2}
  \fmf{dashes}{t3,b3}
\end{fmfgraph*}
}
+
\dots \,
\nl
\vspace{5mm}
\quad \quad \quad = 
$\,\,c_{q}^{(0)} {}_{\rm 2BE}\!\!\!\!\!\!$ 
\parbox{25mm}{
\begin{fmfgraph*}(70,40)
  \fmfleftn{l}{2}
  \fmfrightn{r}{2}
  \fmftopn{t}{4}
  \fmfbottomn{b}{4}
  \fmf{double}{l2,v,r2}
  \fmf{fermion}{l1,v,r1}
  \fmffreeze
  \fmfv{decor.shape=square,decor.size=0.15w}{v}
\end{fmfgraph*} 
}
\!\!\!+
$\,\,c_{q}^{(2)} {}_{\rm 2BE}\!\!\!\!\!\!$ 
\parbox{25mm}{
\begin{fmfgraph*}(70,40)
  \fmfleftn{l}{2}
  \fmfrightn{r}{2}
  \fmftopn{t}{4}
  \fmfbottomn{b}{4}
  \fmf{double}{l2,v,r2}
  \fmf{fermion}{l1,v,r1}
  \fmffreeze
  \fmfv{decor.shape=square,decor.size=0.15w}{v}
\end{fmfgraph*} 
}
\end{center}
\caption{\label{fig:quark2match} Matching condition for two-boson
exchange contributions to quark operators. The notation for the
different lines and vertices is as in Fig.~\ref{fig:quark1match}. The full theory
diagrams illustrate the possible types of two-boson exchange.
Crossed diagrams and time-reversed diagrams are not shown.}
\end{figure}

\noindent Let us now consider quark matching from two-boson exchange,
as displayed in Fig.~\ref{fig:quark2match}. In covariant gauges, in
particular Feynman-t'Hooft gauge employed here, the full theory
contributions include diagrams with exchange of two gauge bosons
($W^\pm$ or $Z^0$), two Goldstone bosons ($\phi_Z^0$ or $\phi_W^\pm$),
one gauge and one Goldstone boson ($Z^0$ and $\phi_Z^0$, or $W^\pm$
and $\phi_W^\pm$), or two Higgs bosons. In terms of these
contributions the total amplitude is
\be\label{eq:quarkM}
{\cal M}_q = {\cal M}_q^{ZZ} + {\cal M}_q^{WW} + {\cal M}_q^{W\phi_W} + {\cal M}_q^{Z\phi_Z} 
+ {\cal M}_q^{\phi_W\phi_W} 
+ {\cal M}_q^{\phi_Z\phi_Z} + {\cal M}_q^{hh} \,, 
\ee
where the superscripts denote which bosons are exchanged, and the
contributions from crossed diagrams and time-reversed diagrams are
included in each amplitude. Upon expressing the amplitudes in terms of
the integrals $J(m_V, M, \delta)$, $J^\mu(p,m_V, M,\delta)$, $J_-(p,m_V,M,\delta)$ and $J_-^\mu(m_V,M,\delta)$ defined in
Appendix~\ref{sec:box}, we may write each amplitude in the form
\be\label{eq:2BEquark}
{\cal M}_q^{B B^\prime} =\bar{u}_q(p) \left[ m_q \, c_{q}^{(0) {B B^\prime}}   + \left( \slash{v} v \cdot p - {\slash{p} \over d} \right) \, c_{q}^{(2) {B B^\prime}}   \right] u_q(p)\,,
\ee
where the superscript $B B^\prime$ denotes the type of two-boson exchange. The contributions to spin-0 and spin-2
quark matching coefficients can then be read off as $c_{q}^{(0) {B B^\prime}} $ and $c_{q}^{(2) {B B^\prime}} $, respectively.

\subsubsection{Pure states}

For pure states the contributions come from diagrams with exchange of $W^\pm$ or $Z^0$ bosons.
In terms of ${\cal C}_W$ and
${\cal C}_Z$ specified in (\ref{eq:WZcoeff}), the amplitudes are
\begin{align} \label{eq:2BEquarkpureAMP}
i{\cal M}_q^{ZZ} &=  {g_2^4 {\cal C}_Z \over 16 c_W^4}  \,
\bar{u}_q(p)
\bigg[  \big[ c_V^{(q)2} + c_A^{(q)2} \big] \slash{v} \big[ \slash{J}(p,m_Z,0,0) + 
\slash{p}J(m_Z,0,0)  \big] \slash{v} 
\nl
&\quad 
 + m_q \big[ c_V^{(q)2} - c_A^{(q)2} \big] J(m_Z,0,0)  \bigg] 
u_q(p) 
\,,
\nl
i{\cal M}_U^{WW} &= 
{g_2^4  {\cal C}_W \over 8} \,
\bar{u}_U(p)
\slash{v} \big[ \slash{J}(p,m_W,0,0) + \slash{p} J(m_W,0,0) \big] \slash{v}
u_U(p)  \,,
\nl
i{\cal M}_D^{WW} &=  \sum_U {g_2^4 {\cal C}_W \over 8}   |V_{UD}|^2 \,
\bar{u}_D(p) 
\slash{v} \big[ \slash{J}(p,m_W,m_U,0) + \slash{p} J(m_W,m_U,0) \big] \slash{v} 
u_D(p) \,. 
\end{align}
Upon writing these amplitudes in the form of (\ref{eq:2BEquark}) and
evaluating integrals, we find the contributions to the matching
coefficients,
\begin{align}\label{eq:2BEquarkpureCO}
c_{U}^{(0)} {}_{\rm 2BE} &= { \pi \Gamma(1+\epsilon) g_2^4 \over (4\pi)^{2-\epsilon} } \bigg\{ 
 { m_Z^{-3-2\epsilon} {\cal C}_Z \over 8 c_W^4}   \big[ c_V^{(U)2 }-c_A^{(U)2} \big]  
 + \order(\epsilon)
\bigg\} \,,
\nl
c_{D}^{(0)} {}_{\rm 2BE} &= { \pi \Gamma(1+\epsilon)  g_2^4 \over (4\pi)^{2-\epsilon} } \bigg\{ 
 { m_Z^{-3-2\epsilon} {\cal C}_Z \over 8 c_W^4}  \big[ c_V^{(D)2}-c_A^{(D)2}\big]
 + \delta_{Db} {m_W^{-3-2\epsilon} {\cal C}_W \over 2}
  \bigg[ 
- {x_t \over 4(x_t+1)^3}   \bigg]
 + \order(\epsilon)
\bigg\}
\,,
\nl
c_{U}^{(2)} {}_{\rm 2BE} &= {\pi \Gamma(1+\epsilon)  g_2^4 \over (4\pi)^{2-\epsilon} }  \bigg\{
\bigg[ m_W^{-3-2\epsilon} {\cal C}_W + { m_Z^{-3-2\epsilon} {\cal C}_Z \over 2 c_W^4} 
\big[ c_V^{(U)2}+c_A^{(U)2} \big]  \bigg]  \bigg[ 
{1\over 3} + \left( {11\over 9} -{2\over 3}\log 2  \right) \epsilon 
\bigg]+ \order(\epsilon^2) 
\bigg\}\, ,
\nl
c_{D}^{(2)} {}_{\rm 2BE} &= {\pi \Gamma(1+\epsilon)  g_2^4 \over (4\pi)^{2-\epsilon} }  \bigg\{
\bigg[ m_W^{-3-2\epsilon} {\cal C}_W + { m_Z^{-3-2\epsilon} {\cal C}_Z \over 2 c_W^4} 
\big[ c_V^{(D)2}+c_A^{(D)2} \big]  \bigg]  \bigg[ 
{1\over 3} + \left( {11\over 9} -{2\over 3}\log 2  \right) \epsilon 
\bigg]
\nl
&\quad
 + \delta_{Db}  { m_W^{-3-2\epsilon} {\cal C}_W \over 2}  
\bigg[ {(3x_t +2) \over 3(x_t+1)^3} - \frac23
+ \bigg(
{ 2 x_t(7x_t^2 -3) \over 3(x_t^2-1)^3 }\log x_t 
- {2(3x_t+2)\over 3(x_t+1)^3}\log 2 
\nl
&\quad 
- {2( 25x_t^2-2x_t -11) \over 9(x_t^2-1)^2(x_t+1) } - {22 \over 9} + {4 \over 3}\log 2
\bigg)\epsilon \bigg]  
+ \order(\epsilon^2) 
\bigg\}\, ,
\end{align}
where $x_t = m_t/m_W$, and the Kronecker delta, $\delta_{Db}$, is equal to unity for $D=b$ and vanishes for $D=d,s$. We obtain the pure triplet (doublet) result upon setting ${\cal
C}_W=2$ and ${\cal C}_Z=0$ (${\cal C}_W=1/2$ and ${\cal C}_Z=1/4$) in
(\ref{eq:2BEquarkpureCO}).

\subsubsection{Singlet-doublet admixture} 

For the singlet-doublet case
the amplitudes for the different types of two-boson exchange are 
\begin{align}
i{\cal M}_q^{ZZ} &= {g_2^4 \over 64 c_W^4} c^2_{\rho\over 2} \,
\bar{u}_q(p)
\bigg[  \big[ c_V^{(q)2} + c_A^{(q)2} \big] \slash{v} \big[ \slash{J}(p,m_Z,0,\delta_0^{(0)}) + 
\slash{p}J(m_Z,0,\delta_0^{(0)})  \big] \slash{v} 
\nl
&\quad 
 + m_q \big[ c_V^{(q) 2} - c_A^{(q)2} \big] J(m_Z,0,\delta_0^{(0)})  \bigg] 
u_q(p) \, ,
\nl
i{\cal M}_U^{WW} &= 
{g_2^4 \over 16} c^2_{\rho\over 2} \,
\bar{u}_U(p)
\slash{v} \big[ \slash{J}(p,m_W,0,\delta_0^{(0)}) + \slash{p} J(m_W,0,\delta_0^{(0)}) \big] \slash{v}
u_U(p)  \,,
\nl
i{\cal M}_D^{WW} &=  \sum_U {g_2^4 \over 16} c^2_{\rho\over 2}  |V_{UD}|^2 \,
\bar{u}_D(p) 
\slash{v} \big[ \slash{J}(p,m_W,m_U,\delta_0^{(0)}) + \slash{p} J(m_W,m_U,\delta_0^{(0)}) \big] \slash{v} 
u_D(p) \,, 
\nl
i{\cal M}_q^{Z\phi_Z} &= -{g_2^3 a \over 16 c_W^2} s_\rho {m_q\over m_W} \bar{u}_q(p) \big[ v\cdot J_-(m_Z,0,\delta_0^{(0)}) \big] u_q(p) \,,
\nl
i{\cal M}_U^{W\phi_W} &= - {g_2^3 a \over 8 } s_\rho  {m_U\over m_W}
\bar{u}_U(p) \big[ v\cdot J_-(m_W,0,\delta_0^{(0)}) \big]
u_U(p) \,, 
\nl
i{\cal M}_D^{W\phi_W} &=  \sum_U  {g_2^3 a \over 8 } s_\rho   |V_{UD}|^2 
\bar{u}_D(p) \bigg[  
- {m_D\over m_W} v\cdot J_-(m_W,m_U,\delta_0^{(0)})
+ \slash{v} {m_U^2\over m_W} J_-(p,m_W,m_U,\delta_0^{(0)}) 
 \bigg]
u_D(p) \,, 
\nl
i{\cal M}_D^{\phi_W\phi_W} &= {g_2^2 a^2 \over 4 } s^2_{\rho\over 2} {m_t^2 \over m_W^2} |V_{tD}|^2 \,
\bar{u}_D(p) 
\big[ -m_D J(m_W,m_t,\delta_0^{(0)}) + \slash{J}(p,m_W,m_t,\delta_0^{(0)}) \big] u_D(p)\,,
\nl
i{\cal M}_U^{\phi_W\phi_W} &= 0 \, , \quad 
i{\cal M}_q^{\phi_Z\phi_Z} = 0 \,, \quad
i{\cal M}_q^{hh} = 0 \, .
\end{align}
The amplitudes ${\cal M}_q^{hh}$, ${\cal M}_q^{\phi_Z \phi_Z}$, and
${\cal M}_U^{\phi_W \phi_W}$ are suppressed by light quark
masses. Comparing each amplitude above with (\ref{eq:2BEquark}), we
find the contributions to spin-0 and spin-2 quark matching
coefficients,
\begin{align}\label{eq:c2BESD}
c_{q}^{(0)} {}^{ZZ} &={g_2^4 \over 64 c_W^4} c_{\rho \over 2}^2 \Bigg\{  \big[ c_V^{(q)2} - c_A^{(q)2} \big] J(m_Z,0,\delta_0^{(0)}) 
\nl
&\quad + \big[ c_V^{(q)2} + c_A^{(q)2 }\big] \left[-J(m_Z,0,\delta_0^{(0)}) - J_2(m_Z,0,\delta_0^{(0)}) +\frac1d {\hat J}(m_Z,0,\delta_0^{(0)})  \right]   \Bigg\} \,,
\nl
c_{q}^{(2)} {}^{ZZ} &=  {g_2^4 \over 64 c_W^4} c_{\rho \over 2}^2 \big[ c_V^{(q) 2} + c_A^{(q) 2}\big] \hat{J}(m_Z,0,\delta_0^{(0)}) \, ,
\nl
c_{U}^{(0)} {}^{WW} &= {g_2^4 \over 16} c_{\rho \over 2}^2  \left[ -J(m_W,0,\delta_0^{(0)}) -J_2(m_W,0,\delta_0^{(0)})  + \frac1d \hat{J}(m_W,0,\delta_0^{(0)})  \right]\,,
\nl
c_{U}^{(2)} {}^{WW} &= {g_2^4 \over 16} c_{\rho \over 2}^2  \hat{J}(m_W,0,\delta_0^{(0)})  \,,
\nl
c_{D}^{(0)} {}^{WW} &= \sum_U {g_2^4 \over 16} c_{\rho \over 2}^2 |V_{UD}|^2 \left[ -J(m_W,m_U,\delta_0^{(0)}) -J_2(m_W,m_U,\delta_0^{(0)}) +\frac1d \hat{J}(m_W,m_U,\delta_0^{(0)})  \right] \,,
\nl
c_{D}^{(2)} {}^{WW} &= \sum_U {g_2^4 \over 16} c_{\rho \over 2}^2 |V_{UD}|^2 \hat{J}(m_W,m_U,\delta_0^{(0)})\,,
\nl
c_{q}^{(0)} {}^{Z\phi_Z} &= -{g_2^3 a \over 16 c_W^2 } {s_\rho \over m_W} J_{1-}(m_Z,0,\delta_0^{(0)})  \,,
\nl
c_{q}^{(2)} {}^{Z\phi_Z} &= 0 \,,
\nl
c_{U}^{(0)} {}^{W \phi_W} &= -{g_2^3 a \over 8} {s_\rho \over m_W} J_{1-}(m_W,0,\delta_0^{(0)})\,,
\nl
c_{U}^{(2)} {}^{W\phi_W} &= 0\,,
\nl
c_{D}^{(0)} {}^{W\phi_W} &= \sum_U {g_2^3 a \over 8} s_\rho |V_{UD}|^2 \left[- {1 \over m_W} J_{1-}(m_W,m_U,\delta_0^{(0)}) + \frac1d {m_U^2 \over m_W} J_-(m_W,m_U,\delta_0^{(0)}) \right]\,,
\nl
c_{D}^{(2)} {}^{W\phi_W} &= \sum_U {g_2^3 a \over 8} s_\rho |V_{UD}|^2 {m_U^2 \over m_W} J_-(m_W,m_U,\delta_0^{(0)}) \,,
\nl
c_{D}^{(0)} {}^{\phi_W \phi_W} &= {g_2^2 a^2 \over 4} s_{\rho \over 2}^2 {m_t^2 \over m_W^2} | V_{tD}|^2 \left[ J_2(m_W,m_t,\delta_0^{(0)}) - J(m_W,m_t,\delta_0^{(0)}) + \frac 1d J_1(m_W,m_t,\delta_0^{(0)}) \right] \,,
\nl
c_{D}^{(2)} {}^{\phi_W \phi_W} &= {g_2^2 a^2 \over 4} s_{\rho \over 2}^2 {m_t^2 \over m_W^2} | V_{tD}|^2 J_1 (m_W,m_t,\delta_0^{(0)})\,,
\end{align}
where we have defined
\be \label{eq:Jhat}
\hat{J}(m_x,m_y,\delta_z) \equiv J_1(m_x,m_y,\delta_z) + 2J_2(m_x,m_y,\delta_z) + 2J(m_x,m_y,\delta_z).
\ee
The integrals $J(m_V,M,\delta)$, $J_1(m_V,M,\delta)$, $J_2(m_V,M,\delta)$, $J_-(m_V,M,\delta)$ and $J_{1-}(m_V,M,\delta)$ are given in
Appendix~\ref{sec:box}. The matching coefficients $c_{q}^{(0)} {}_{\rm
2BE}$ and $c_{q}^{(2)} {}_{\rm 2BE}$ for a given quark $q$ are
obtained by summing the nonvanishing contributions above,
\begin{align} \label{eq:sumquark2BE} c_{U}^{(0)} {}_{\rm 2BE} &=
c_{U}^{(0)} {}^{ZZ} + c_{U}^{(0)} {}^{WW} + c_{U}^{(0)} {}^{Z\phi_Z} +
c_{U}^{(0)} {}^{W\phi_W} \, , \nl c_{D}^{(0)} {}_{\rm 2BE} &=
c_{D}^{(0)} {}^{ZZ} + c_{D}^{(0)} {}^{WW} + c_{D}^{(0)} {}^{Z\phi_Z} +
c_{D}^{(0)} {}^{W\phi_W} + c_{D}^{(0)} {}^{\phi_W\phi_W} \, , \nl
c_{U}^{(2)} {}_{\rm 2BE} &= c_{U}^{(2)} {}^{ZZ} + c_{U}^{(2)} {}^{WW}
\, , \nl c_{D}^{(2)} {}_{\rm 2BE} &= c_{D}^{(2)} {}^{ZZ} + c_{D}^{(2)}
{}^{WW} + c_{D}^{(2)} {}^{W\phi_W} + c_{D}^{(2)} {}^{\phi_W\phi_W} .
\end{align}
Upon taking the pure-case limits described in Sec.~\ref{sec:limits},
we recover the results (\ref{eq:2BEquarkpureAMP}) and
(\ref{eq:2BEquarkpureCO}) for a pure doublet. In the pure singlet
limit, the two-boson exchange amplitudes vanish.

\subsubsection{Triplet-doublet admixture} 

We may similarly compute the two-boson exchange amplitudes for the
triplet-doublet system, and upon comparing with (\ref{eq:2BEquark}),
we find the following contributions to spin-0 and spin-2 quark
matching coefficients,
\begin{align}\label{eq:c2BETD}
c_{q}^{(0)} {}^{ZZ} &={g_2^4 \over 64 c_W^4} c_{\rho \over 2}^2 \Bigg\{  \big[ c_V^{(q)2} - c_A^{(q)2} \big] J(m_Z,0,\delta_0^{(0)}) 
\nl
&\quad + \big[ c_V^{(q)2} + c_A^{(q)2} \big] \left(-J(m_Z,0,\delta_0^{(0)}) - J_2(m_Z,0,\delta_0^{(0)}) +\frac1d {\hat J}(m_Z,0,\delta_0^{(0)})  \right)   \Bigg\} \,,
\nl
c_{q}^{(2)} {}^{ZZ} &=  {g_2^4 \over 64 c_W^4} c_{\rho \over 2}^2 \big[ c_V^{(q)2} + c_A^{(q)2} \big] \hat{J}(m_Z,0,\delta_0^{(0)}) \, ,
\nl
c_{U}^{(0)} {}^{WW} &= {g_2^4 \over 16} \Bigg\{ (1+s_{\rho \over 2}^2)^2  \left[ -J(m_W,0,\delta_0^{(-)}) -J_2(m_W,0,\delta_0^{(-)})  + \frac1d \hat{J}(m_W,0,\delta_0^{(-)})  \right]
\nl
&\quad +\frac14 s_{\rho}^2  \left[ -J(m_W,0,\delta_0^{(+)}) -J_2(m_W,0,\delta_0^{(+)})  + \frac1d \hat{J}(m_W,0,\delta_0^{(+)})  \right] \Bigg\}\,,
\nl
c_{U}^{(2)} {}^{WW} &= {g_2^4 \over 16} \left[ (1+s_{\rho \over 2}^2)^2  \hat{J}(m_W,0,\delta_0^{(-)}) + \frac14 s_{\rho}^2  \hat{J}(m_W,0,\delta_0^{(+)})  \right] \,,
\nl
c_{D}^{(0)} {}^{WW} &= \sum_U {g_2^4 \over 16} |V_{UD}|^2 \Bigg\{ (1+s_{\rho \over 2}^2)^2 \left[ -J(m_W,m_U,\delta_0^{(-)}) -J_2(m_W,m_U,\delta_0^{(-)}) +\frac1d \hat{J}(m_W,m_U,\delta_0^{(-)}) \right]  
\nl
& \quad + \frac14 s_{\rho}^2\left[ -J(m_W,m_U,\delta_0^{(+)}) -J_2(m_W,m_U,\delta_0^{(+)}) +\frac1d \hat{J}(m_W,m_U,\delta_0^{(+)}) \right] \Bigg\} \,,
\nl
c_{D}^{(2)} {}^{WW} &= \sum_U {g_2^4 \over 16} |V_{UD}|^2 \left[ (1+s_{\rho \over 2}^2)^2 \hat{J}(m_W,m_U,\delta_0^{(-)}) + \frac14 s_{\rho}^2 \hat{J}(m_W,m_U,\delta_0^{(+)}) \right] \,,
\nl
c_{q}^{(0)} {}^{Z\phi_Z} &= -{g_2^3 a \over 16 c_W^2 } {s_\rho \over m_W} J_{1-}(m_Z,0,\delta_0^{(0)})  \,,
\nl
c_{q}^{(2)} {}^{Z\phi_Z} &= 0 \,,
\nl
c_{U}^{(0)} {}^{W \phi_W} &= -{g_2^3 a \over 8} {s_\rho \over m_W} J_{1-}(m_W,0,\delta_0^{(+)})\,,
\nl
c_{U}^{(2)} {}^{W\phi_W} &= 0\,,
\nl
c_{D}^{(0)} {}^{W\phi_W} &= \sum_U {g_2^3 a \over 8} s_\rho |V_{UD}|^2 \left[- {1 \over m_W} J_{1-}(m_W,m_U,\delta_0^{(+)}) + \frac1d {m_U^2 \over m_W} J_-(m_W,m_U,\delta_0^{(+)}) \right]\,,
\nl
c_{D}^{(2)} {}^{W\phi_W} &= \sum_U {g_2^3 a \over 8} s_\rho |V_{UD}|^2 {m_U^2 \over m_W} J_-(m_W,m_U,\delta_0^{(+)}) \,,
\nl
c_{D}^{(0)} {}^{\phi_W \phi_W} &= {g_2^2 a^2 \over 4}  {m_t^2 \over m_W^2} | V_{tD}|^2 \left[ J_2(m_W,m_t,\delta_0^{(+)}) - J(m_W,m_t,\delta_0^{(+)}) + \frac 1d J_1(m_W,m_t,\delta_0^{(+)}) \right] \,,
\nl
c_{D}^{(2)} {}^{\phi_W \phi_W} &= {g_2^2 a^2 \over 4}  {m_t^2 \over m_W^2} | V_{tD}|^2 J_1 (m_W,m_t,\delta_0^{(+)})\,,
\end{align}
where $\hat{J}(m_x, m_y, \delta_z)$
is given in (\ref{eq:Jhat}), and the relevant integrals can be found in Appendix~\ref{sec:box}. The total matching
coefficients $c_{q}^{(0)} {}_{\rm 2BE}$ and $c_{q}^{(2)} {}_{\rm 2BE}$
are obtained by summing the contributions above as in
(\ref{eq:sumquark2BE}). Upon taking the pure-case limits described in
Sec.~\ref{sec:limits}, we recover the results
(\ref{eq:2BEquarkpureAMP}) and (\ref{eq:2BEquarkpureCO}) for both pure
triplet and pure doublet.

\subsection{Gluon matching: two-boson exchange}

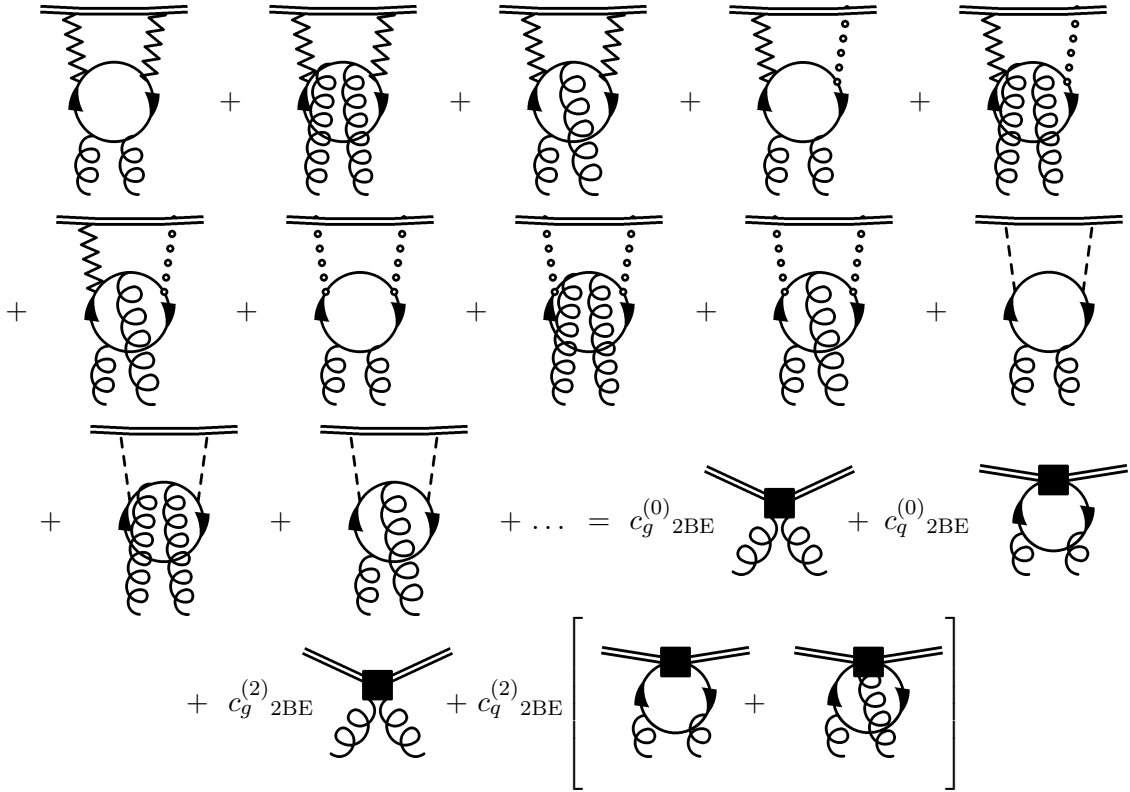
\begin{figure}[htb]
\begin{center}
\parbox{25mm}{
\begin{fmfgraph*}(70,70)
  \fmfleftn{l}{3}
  \fmfrightn{r}{3}
  \fmftopn{t}{4}
  \fmfbottomn{b}{4}
  \fmf{double}{l3,t2,t3,r3}
  \fmf{phantom,tag=3}{l1,r1}
  \fmf{phantom,tag=4}{l3,r3}
  \fmffreeze
  \fmf{phantom,tension=3}{l2,v1}
  \fmf{phantom,tension=3}{r2,v2}
  \fmf{phantom,left,tag=1}{v1,v2}
  \fmf{phantom,left,tag=2}{v2,v1}
  \fmfposition
  \fmffreeze
  \fmfipath{p[]}
  \fmfiset{p1}{vpath1(__v1,__v2)}
  \fmfiset{p2}{vpath2(__v2,__v1)}
  \fmfiset{p3}{vpath3(__l1,__r1)}
  \fmfiset{p4}{vpath4(__l3,__r3)}
  \fmfi{fermion}{subpath (length(p1)/2, 3length(p1)/2) of (p1 & p2)}
  \fmfi{fermion}{subpath (length(p1)/2, 3length(p1)/2) of (p2 & p1)}
  \fmfi{gluon}{point length(p2)/3 of p2 -- point 2length(p3)/3 of p3 }
  \fmfi{gluon}{point 2length(p2)/3 of p2 -- point length(p3)/3 of p3 }
  \fmfset{zigzag_width}{1.5thick}
  \fmfi{zigzag}{point length(p1)/6 of p1 -- point length(p4)/5 of p4 } 
  \fmfi{zigzag}{point 5length(p1)/6 of p1 -- point 4length(p4)/5 of p4 } 
\end{fmfgraph*}
}
+
\parbox{25mm}{
\begin{fmfgraph*}(70,70)
  \fmfleftn{l}{3}
  \fmfrightn{r}{3}
  \fmftopn{t}{4}
  \fmfbottomn{b}{4}
  \fmf{double}{l3,t2,t3,r3}
  \fmf{phantom,tag=3}{l1,r1}
  \fmf{phantom,tag=4}{l3,r3}
  \fmffreeze
  \fmf{phantom,tension=3}{l2,v1}
  \fmf{phantom,tension=3}{r2,v2}
  \fmf{phantom,left,tag=1}{v1,v2}
  \fmf{phantom,left,tag=2}{v2,v1}
  \fmfposition
  \fmffreeze
  \fmfipath{p[]}
  \fmfiset{p1}{vpath1(__v1,__v2)}
  \fmfiset{p2}{vpath2(__v2,__v1)}
  \fmfiset{p3}{vpath3(__l1,__r1)}
  \fmfiset{p4}{vpath4(__l3,__r3)}
  \fmfi{fermion}{subpath (length(p1)/2, 3length(p1)/2) of (p1 & p2)}
  \fmfi{fermion}{subpath (length(p1)/2, 3length(p1)/2) of (p2 & p1)}
  \fmfset{curly_len}{2.5mm}
  \fmfi{gluon}{point 2length(p2)/5 of p1 -- point length(p3)/3 of p3 }
  \fmfi{gluon}{point 3length(p2)/5 of p1 -- point 2length(p3)/3 of p3 }
  \fmfset{zigzag_width}{1.5thick}
  \fmfi{zigzag}{point length(p1)/6 of p1 -- point length(p4)/5 of p4 } 
  \fmfi{zigzag}{point 5length(p1)/6 of p1 -- point 4length(p4)/5 of p4 } 
\end{fmfgraph*}
}
+ 
\parbox{25mm}{
\begin{fmfgraph*}(70,70)
  \fmfleftn{l}{3}
  \fmfrightn{r}{3}
  \fmftopn{t}{4}
  \fmfbottomn{b}{4}
  \fmf{double}{l3,t2,t3,r3}
  \fmf{phantom,tag=3}{l1,r1}
  \fmf{phantom,tag=4}{l3,r3}
  \fmffreeze
  \fmf{phantom,tension=3}{l2,v1}
  \fmf{phantom,tension=3}{r2,v2}
  \fmf{phantom,left,tag=1}{v1,v2}
  \fmf{phantom,left,tag=2}{v2,v1}
  \fmfposition
  \fmffreeze
  \fmfipath{p[]}
  \fmfiset{p1}{vpath1(__v1,__v2)}
  \fmfiset{p2}{vpath2(__v2,__v1)}
  \fmfiset{p3}{vpath3(__l1,__r1)}
  \fmfiset{p4}{vpath4(__l3,__r3)}
  \fmfi{fermion}{subpath (length(p1)/2, 3length(p1)/2) of (p1 & p2)}
  \fmfi{fermion}{subpath (length(p1)/2, 3length(p1)/2) of (p2 & p1)}
  \fmfi{gluon}{point 2length(p2)/3 of p2 -- point length(p3)/3 of p3 }
  \fmfi{gluon}{point length(p1)/2 of p1 -- point 2length(p3)/3 of p3 }
  \fmfset{zigzag_width}{1.5thick}
  \fmfi{zigzag}{point length(p1)/6 of p1 -- point length(p4)/5 of p4 } 
  \fmfi{zigzag}{point 5length(p1)/6 of p1 -- point 4length(p4)/5 of p4 } 
\end{fmfgraph*}
}
+ 
\parbox{25mm}{
\begin{fmfgraph*}(70,70)
  \fmfleftn{l}{3}
  \fmfrightn{r}{3}
  \fmftopn{t}{4}
  \fmfbottomn{b}{4}
  \fmf{double}{l3,t2,t3,r3}
  \fmf{phantom,tag=3}{l1,r1}
  \fmf{phantom,tag=4}{l3,r3}
  \fmffreeze
  \fmf{phantom,tension=3}{l2,v1}
  \fmf{phantom,tension=3}{r2,v2}
  \fmf{phantom,left,tag=1}{v1,v2}
  \fmf{phantom,left,tag=2}{v2,v1}
  \fmfposition
  \fmffreeze
  \fmfipath{p[]}
  \fmfiset{p1}{vpath1(__v1,__v2)}
  \fmfiset{p2}{vpath2(__v2,__v1)}
  \fmfiset{p3}{vpath3(__l1,__r1)}
  \fmfiset{p4}{vpath4(__l3,__r3)}
  \fmfi{fermion}{subpath (length(p1)/2, 3length(p1)/2) of (p1 & p2)}
  \fmfi{fermion}{subpath (length(p1)/2, 3length(p1)/2) of (p2 & p1)}
  \fmfi{gluon}{point length(p2)/3 of p2 -- point 2length(p3)/3 of p3 }
  \fmfi{gluon}{point 2length(p2)/3 of p2 -- point length(p3)/3 of p3 }
  \fmfset{zigzag_width}{1.5thick}
  \fmfi{zigzag}{point length(p1)/6 of p1 -- point length(p4)/5 of p4 } 
  \fmfi{dbl_dots}{point 5length(p1)/6 of p1 -- point 4length(p4)/5 of p4 } 
\end{fmfgraph*}
}
+
\parbox{25mm}{
\begin{fmfgraph*}(70,70)
  \fmfleftn{l}{3}
  \fmfrightn{r}{3}
  \fmftopn{t}{4}
  \fmfbottomn{b}{4}
  \fmf{double}{l3,t2,t3,r3}
  \fmf{phantom,tag=3}{l1,r1}
  \fmf{phantom,tag=4}{l3,r3}
  \fmffreeze
  \fmf{phantom,tension=3}{l2,v1}
  \fmf{phantom,tension=3}{r2,v2}
  \fmf{phantom,left,tag=1}{v1,v2}
  \fmf{phantom,left,tag=2}{v2,v1}
  \fmfposition
  \fmffreeze
  \fmfipath{p[]}
  \fmfiset{p1}{vpath1(__v1,__v2)}
  \fmfiset{p2}{vpath2(__v2,__v1)}
  \fmfiset{p3}{vpath3(__l1,__r1)}
  \fmfiset{p4}{vpath4(__l3,__r3)}
  \fmfi{fermion}{subpath (length(p1)/2, 3length(p1)/2) of (p1 & p2)}
  \fmfi{fermion}{subpath (length(p1)/2, 3length(p1)/2) of (p2 & p1)}
  \fmfset{curly_len}{2.5mm}
  \fmfi{gluon}{point 2length(p2)/5 of p1 -- point length(p3)/3 of p3 }
  \fmfi{gluon}{point 3length(p2)/5 of p1 -- point 2length(p3)/3 of p3 }
  \fmfset{zigzag_width}{1.5thick}
  \fmfi{zigzag}{point length(p1)/6 of p1 -- point length(p4)/5 of p4 } 
  \fmfi{dbl_dots}{point 5length(p1)/6 of p1 -- point 4length(p4)/5 of p4 } 
\end{fmfgraph*}
}
\nl
\vspace{3mm}
+
\parbox{25mm}{
\begin{fmfgraph*}(70,70)
  \fmfleftn{l}{3}
  \fmfrightn{r}{3}
  \fmftopn{t}{4}
  \fmfbottomn{b}{4}
  \fmf{double}{l3,t2,t3,r3}
  \fmf{phantom,tag=3}{l1,r1}
  \fmf{phantom,tag=4}{l3,r3}
  \fmffreeze
  \fmf{phantom,tension=3}{l2,v1}
  \fmf{phantom,tension=3}{r2,v2}
  \fmf{phantom,left,tag=1}{v1,v2}
  \fmf{phantom,left,tag=2}{v2,v1}
  \fmfposition
  \fmffreeze
  \fmfipath{p[]}
  \fmfiset{p1}{vpath1(__v1,__v2)}
  \fmfiset{p2}{vpath2(__v2,__v1)}
  \fmfiset{p3}{vpath3(__l1,__r1)}
  \fmfiset{p4}{vpath4(__l3,__r3)}
  \fmfi{fermion}{subpath (length(p1)/2, 3length(p1)/2) of (p1 & p2)}
  \fmfi{fermion}{subpath (length(p1)/2, 3length(p1)/2) of (p2 & p1)}
  \fmfi{gluon}{point 2length(p2)/3 of p2 -- point length(p3)/3 of p3 }
  \fmfi{gluon}{point length(p1)/2 of p1 -- point 2length(p3)/3 of p3 }
  \fmfset{zigzag_width}{1.5thick}
  \fmfi{zigzag}{point length(p1)/6 of p1 -- point length(p4)/5 of p4 } 
  \fmfi{dbl_dots}{point 5length(p1)/6 of p1 -- point 4length(p4)/5 of p4 } 
\end{fmfgraph*}
}
+
\parbox{25mm}{
\begin{fmfgraph*}(70,70)
  \fmfleftn{l}{3}
  \fmfrightn{r}{3}
  \fmftopn{t}{4}
  \fmfbottomn{b}{4}
  \fmf{double}{l3,t2,t3,r3}
  \fmf{phantom,tag=3}{l1,r1}
  \fmf{phantom,tag=4}{l3,r3}
  \fmffreeze
  \fmf{phantom,tension=3}{l2,v1}
  \fmf{phantom,tension=3}{r2,v2}
  \fmf{phantom,left,tag=1}{v1,v2}
  \fmf{phantom,left,tag=2}{v2,v1}
  \fmfposition
  \fmffreeze
  \fmfipath{p[]}
  \fmfiset{p1}{vpath1(__v1,__v2)}
  \fmfiset{p2}{vpath2(__v2,__v1)}
  \fmfiset{p3}{vpath3(__l1,__r1)}
  \fmfiset{p4}{vpath4(__l3,__r3)}
  \fmfi{fermion}{subpath (length(p1)/2, 3length(p1)/2) of (p1 & p2)}
  \fmfi{fermion}{subpath (length(p1)/2, 3length(p1)/2) of (p2 & p1)}
  \fmfi{gluon}{point length(p2)/3 of p2 -- point 2length(p3)/3 of p3 }
  \fmfi{gluon}{point 2length(p2)/3 of p2 -- point length(p3)/3 of p3 }
  \fmfset{zigzag_width}{1.5thick}
  \fmfi{dbl_dots}{point length(p1)/6 of p1 -- point length(p4)/5 of p4 } 
  \fmfi{dbl_dots}{point 5length(p1)/6 of p1 -- point 4length(p4)/5 of p4 } 
\end{fmfgraph*}
}
+
\parbox{25mm}{
\begin{fmfgraph*}(70,70)
  \fmfleftn{l}{3}
  \fmfrightn{r}{3}
  \fmftopn{t}{4}
  \fmfbottomn{b}{4}
  \fmf{double}{l3,t2,t3,r3}
  \fmf{phantom,tag=3}{l1,r1}
  \fmf{phantom,tag=4}{l3,r3}
  \fmffreeze
  \fmf{phantom,tension=3}{l2,v1}
  \fmf{phantom,tension=3}{r2,v2}
  \fmf{phantom,left,tag=1}{v1,v2}
  \fmf{phantom,left,tag=2}{v2,v1}
  \fmfposition
  \fmffreeze
  \fmfipath{p[]}
  \fmfiset{p1}{vpath1(__v1,__v2)}
  \fmfiset{p2}{vpath2(__v2,__v1)}
  \fmfiset{p3}{vpath3(__l1,__r1)}
  \fmfiset{p4}{vpath4(__l3,__r3)}
  \fmfi{fermion}{subpath (length(p1)/2, 3length(p1)/2) of (p1 & p2)}
  \fmfi{fermion}{subpath (length(p1)/2, 3length(p1)/2) of (p2 & p1)}
  \fmfset{curly_len}{2.5mm}
  \fmfi{gluon}{point 2length(p2)/5 of p1 -- point length(p3)/3 of p3 }
  \fmfi{gluon}{point 3length(p2)/5 of p1 -- point 2length(p3)/3 of p3 }
  \fmfset{zigzag_width}{1.5thick}
  \fmfi{dbl_dots}{point length(p1)/6 of p1 -- point length(p4)/5 of p4 } 
  \fmfi{dbl_dots}{point 5length(p1)/6 of p1 -- point 4length(p4)/5 of p4 } 
\end{fmfgraph*}
}
+ 
\parbox{25mm}{
\begin{fmfgraph*}(70,70)
  \fmfleftn{l}{3}
  \fmfrightn{r}{3}
  \fmftopn{t}{4}
  \fmfbottomn{b}{4}
  \fmf{double}{l3,t2,t3,r3}
  \fmf{phantom,tag=3}{l1,r1}
  \fmf{phantom,tag=4}{l3,r3}
  \fmffreeze
  \fmf{phantom,tension=3}{l2,v1}
  \fmf{phantom,tension=3}{r2,v2}
  \fmf{phantom,left,tag=1}{v1,v2}
  \fmf{phantom,left,tag=2}{v2,v1}
  \fmfposition
  \fmffreeze
  \fmfipath{p[]}
  \fmfiset{p1}{vpath1(__v1,__v2)}
  \fmfiset{p2}{vpath2(__v2,__v1)}
  \fmfiset{p3}{vpath3(__l1,__r1)}
  \fmfiset{p4}{vpath4(__l3,__r3)}
  \fmfi{fermion}{subpath (length(p1)/2, 3length(p1)/2) of (p1 & p2)}
  \fmfi{fermion}{subpath (length(p1)/2, 3length(p1)/2) of (p2 & p1)}
  \fmfi{gluon}{point 2length(p2)/3 of p2 -- point length(p3)/3 of p3 }
  \fmfi{gluon}{point length(p1)/2 of p1 -- point 2length(p3)/3 of p3 }
  \fmfset{zigzag_width}{1.5thick}
  \fmfi{dbl_dots}{point length(p1)/6 of p1 -- point length(p4)/5 of p4 } 
  \fmfi{dbl_dots}{point 5length(p1)/6 of p1 -- point 4length(p4)/5 of p4 } 
\end{fmfgraph*}
}
+
\parbox{25mm}{
\begin{fmfgraph*}(70,70)
  \fmfleftn{l}{3}
  \fmfrightn{r}{3}
  \fmftopn{t}{4}
  \fmfbottomn{b}{4}
  \fmf{double}{l3,t2,t3,r3}
  \fmf{phantom,tag=3}{l1,r1}
  \fmf{phantom,tag=4}{l3,r3}
  \fmffreeze
  \fmf{phantom,tension=3}{l2,v1}
  \fmf{phantom,tension=3}{r2,v2}
  \fmf{phantom,left,tag=1}{v1,v2}
  \fmf{phantom,left,tag=2}{v2,v1}
  \fmfposition
  \fmffreeze
  \fmfipath{p[]}
  \fmfiset{p1}{vpath1(__v1,__v2)}
  \fmfiset{p2}{vpath2(__v2,__v1)}
  \fmfiset{p3}{vpath3(__l1,__r1)}
  \fmfiset{p4}{vpath4(__l3,__r3)}
  \fmfi{fermion}{subpath (length(p1)/2, 3length(p1)/2) of (p1 & p2)}
  \fmfi{fermion}{subpath (length(p1)/2, 3length(p1)/2) of (p2 & p1)}
  \fmfi{gluon}{point length(p2)/3 of p2 -- point 2length(p3)/3 of p3 }
  \fmfi{gluon}{point 2length(p2)/3 of p2 -- point length(p3)/3 of p3 }
  \fmfset{zigzag_width}{1.5thick}
  \fmfi{dashes}{point length(p1)/6 of p1 -- point length(p4)/5 of p4 } 
  \fmfi{dashes}{point 5length(p1)/6 of p1 -- point 4length(p4)/5 of p4 } 
\end{fmfgraph*}
}
\nl
\vspace{3mm}
+
\parbox{25mm}{
\begin{fmfgraph*}(70,70)
  \fmfleftn{l}{3}
  \fmfrightn{r}{3}
  \fmftopn{t}{4}
  \fmfbottomn{b}{4}
  \fmf{double}{l3,t2,t3,r3}
  \fmf{phantom,tag=3}{l1,r1}
  \fmf{phantom,tag=4}{l3,r3}
  \fmffreeze
  \fmf{phantom,tension=3}{l2,v1}
  \fmf{phantom,tension=3}{r2,v2}
  \fmf{phantom,left,tag=1}{v1,v2}
  \fmf{phantom,left,tag=2}{v2,v1}
  \fmfposition
  \fmffreeze
  \fmfipath{p[]}
  \fmfiset{p1}{vpath1(__v1,__v2)}
  \fmfiset{p2}{vpath2(__v2,__v1)}
  \fmfiset{p3}{vpath3(__l1,__r1)}
  \fmfiset{p4}{vpath4(__l3,__r3)}
  \fmfi{fermion}{subpath (length(p1)/2, 3length(p1)/2) of (p1 & p2)}
  \fmfi{fermion}{subpath (length(p1)/2, 3length(p1)/2) of (p2 & p1)}
  \fmfset{curly_len}{2.5mm}
  \fmfi{gluon}{point 2length(p2)/5 of p1 -- point length(p3)/3 of p3 }
  \fmfi{gluon}{point 3length(p2)/5 of p1 -- point 2length(p3)/3 of p3 }
  \fmfset{zigzag_width}{1.5thick}
  \fmfi{dashes}{point length(p1)/6 of p1 -- point length(p4)/5 of p4 } 
  \fmfi{dashes}{point 5length(p1)/6 of p1 -- point 4length(p4)/5 of p4 } 
\end{fmfgraph*}
}
+ 
\parbox{25mm}{
\begin{fmfgraph*}(70,70)
  \fmfleftn{l}{3}
  \fmfrightn{r}{3}
  \fmftopn{t}{4}
  \fmfbottomn{b}{4}
  \fmf{double}{l3,t2,t3,r3}
  \fmf{phantom,tag=3}{l1,r1}
  \fmf{phantom,tag=4}{l3,r3}
  \fmffreeze
  \fmf{phantom,tension=3}{l2,v1}
  \fmf{phantom,tension=3}{r2,v2}
  \fmf{phantom,left,tag=1}{v1,v2}
  \fmf{phantom,left,tag=2}{v2,v1}
  \fmfposition
  \fmffreeze
  \fmfipath{p[]}
  \fmfiset{p1}{vpath1(__v1,__v2)}
  \fmfiset{p2}{vpath2(__v2,__v1)}
  \fmfiset{p3}{vpath3(__l1,__r1)}
  \fmfiset{p4}{vpath4(__l3,__r3)}
  \fmfi{fermion}{subpath (length(p1)/2, 3length(p1)/2) of (p1 & p2)}
  \fmfi{fermion}{subpath (length(p1)/2, 3length(p1)/2) of (p2 & p1)}
  \fmfi{gluon}{point 2length(p2)/3 of p2 -- point length(p3)/3 of p3 }
  \fmfi{gluon}{point length(p1)/2 of p1 -- point 2length(p3)/3 of p3 }
  \fmfset{zigzag_width}{1.5thick}
  \fmfi{dashes}{point length(p1)/6 of p1 -- point length(p4)/5 of p4 } 
  \fmfi{dashes}{point 5length(p1)/6 of p1 -- point 4length(p4)/5 of p4 } 
\end{fmfgraph*}
}
+
\dots \,
= 
$\,\,c_{g}^{(0)} {}_{\rm 2BE}\!\!\!\!\!\!\!\!$ 
\parbox{20mm}{
\begin{fmfgraph*}(70,40)
  \fmfleftn{l}{2}
  \fmfrightn{r}{2}
  \fmftopn{t}{4}
  \fmfbottomn{b}{5}
  \fmf{double}{l2,v,r2}
  \fmf{gluon,tension=0.5}{b2,v,b4}
  \fmffreeze
  \fmfv{decor.shape=square,decor.size=0.15w}{v}
\end{fmfgraph*} 
} 
+
$\, c_{q}^{(0)} {}_{\rm 2BE}\!\!\!\!$ 
\parbox{20mm}{
\begin{fmfgraph*}(70,40)
  \fmfleftn{l}{3}
  \fmfrightn{r}{3}
  \fmftopn{t}{3}
  \fmfbottomn{b}{3}
  \fmf{phantom}{l3,t2,r3}
  \fmf{phantom}{l1,b2,r1}
  \fmf{phantom,tag=3}{l1,r1}
  \fmf{phantom,tension=5}{t2,b2}
  \fmf{phantom,tension=10}{t2,v1}
  \fmf{phantom,tension=6}{b2,v2}
  \fmf{plain,left,tag=1}{v1,v2}
  \fmf{plain,left,tag=2}{v2,v1}
  \fmf{phantom,tag=4}{l3,v1}  
  \fmf{phantom,tag=5}{v1,r3}  
  \fmfposition
  \fmffreeze
   \fmfipath{p[]}
   \fmfiset{p1}{vpath1(__v1,__v2)}
   \fmfiset{p2}{vpath2(__v2,__v1)}
   \fmfiset{p3}{vpath3(__l1,__r1)}
   \fmfiset{p4}{vpath4(__l3,__v1)}
   \fmfiset{p5}{vpath5(__v1,__r3)}
   \fmfi{fermion}{subpath (0,length(p1)) of p1}
   \fmfi{fermion}{subpath (0,length(p2)) of p2}
   \fmfi{gluon}{point 2length(p1)/3 of p1 -- point 2length(p3)/3 of p3 }
   \fmfi{gluon}{point length(p2)/3 of p2 -- point length(p3)/3 of p3 }
   \fmfi{double}{subpath (0,length(p4)) of p4}
   \fmfi{double}{subpath (0,length(p5)) of p5}
  \fmfv{decor.shape=square,decor.size=0.15w}{v1}
\end{fmfgraph*}
}
\nl
+
$\,\,c_{g}^{(2)} {}_{\rm 2BE}\!\!\!\!\!\!\!\!$ 
\parbox{20mm}{
\begin{fmfgraph*}(70,40)
  \fmfleftn{l}{2}
  \fmfrightn{r}{2}
  \fmftopn{t}{4}
  \fmfbottomn{b}{5}
  \fmf{double}{l2,v,r2}
  \fmf{gluon,tension=0.5}{b2,v,b4}
  \fmffreeze
  \fmfv{decor.shape=square,decor.size=0.15w}{v}
\end{fmfgraph*} 
} 
+$\, \, c_{q}^{(2)} {}_{\rm 2BE} \Vast[\!\!$ 
\parbox{20mm}{
\begin{fmfgraph*}(70,40)
  \fmfleftn{l}{3}
  \fmfrightn{r}{3}
  \fmftopn{t}{3}
  \fmfbottomn{b}{3}
  \fmf{phantom}{l3,t2,r3}
  \fmf{phantom}{l1,b2,r1}
  \fmf{phantom,tag=3}{l1,r1}
  \fmf{phantom,tension=5}{t2,b2}
  \fmf{phantom,tension=10}{t2,v1}
  \fmf{phantom,tension=6}{b2,v2}
  \fmf{plain,left,tag=1}{v1,v2}
  \fmf{plain,left,tag=2}{v2,v1}
  \fmf{phantom,tag=4}{l3,v1}  
  \fmf{phantom,tag=5}{v1,r3}  
  \fmfposition
  \fmffreeze
   \fmfipath{p[]}
   \fmfiset{p1}{vpath1(__v1,__v2)}
   \fmfiset{p2}{vpath2(__v2,__v1)}
   \fmfiset{p3}{vpath3(__l1,__r1)}
   \fmfiset{p4}{vpath4(__l3,__v1)}
   \fmfiset{p5}{vpath5(__v1,__r3)}
   \fmfi{fermion}{subpath (0,length(p1)) of p1}
   \fmfi{fermion}{subpath (0,length(p2)) of p2}
   \fmfi{gluon}{point 2length(p1)/3 of p1 -- point 2length(p3)/3 of p3 }
   \fmfi{gluon}{point length(p2)/3 of p2 -- point length(p3)/3 of p3 }
   \fmfi{double}{subpath (0,length(p4)) of p4}
   \fmfi{double}{subpath (0,length(p5)) of p5}
  \fmfv{decor.shape=square,decor.size=0.15w}{v1}
\end{fmfgraph*}
}
+
\parbox{20mm}{
\begin{fmfgraph*}(70,40)
  \fmfleftn{l}{3}
  \fmfrightn{r}{3}
  \fmftopn{t}{3}
  \fmfbottomn{b}{3}
  \fmf{phantom}{l3,t2,r3}
  \fmf{phantom}{l1,b2,r1}
  \fmf{phantom,tag=3}{l1,r1}
  \fmf{phantom,tension=5}{t2,b2}
  \fmf{phantom,tension=10}{t2,v1}
  \fmf{phantom,tension=6}{b2,v2}
  \fmf{plain,left,tag=1}{v1,v2}
  \fmf{plain,left,tag=2}{v2,v1}
  \fmf{phantom,tag=4}{l3,v1}  
  \fmf{phantom,tag=5}{v1,r3}  
  \fmfposition
  \fmffreeze
   \fmfipath{p[]}
   \fmfiset{p1}{vpath1(__v1,__v2)}
   \fmfiset{p2}{vpath2(__v2,__v1)}
   \fmfiset{p3}{vpath3(__l1,__r1)}
   \fmfiset{p4}{vpath4(__l3,__v1)}
   \fmfiset{p5}{vpath5(__v1,__r3)}
   \fmfi{fermion}{subpath (0,length(p1)) of p1}
   \fmfi{fermion}{subpath (0,length(p2)) of p2}
   \fmfi{gluon}{point length(p2)/3 of p2 -- point length(p3)/3 of p3 }
   \fmfi{gluon}{point 0 of p1 -- point 2length(p3)/3 of p3 }
   \fmfi{double}{subpath (0,length(p4)) of p4}
   \fmfi{double}{subpath (0,length(p5)) of p5}
  \fmfv{decor.shape=square,decor.size=0.15w}{v1}
\end{fmfgraph*}
}
$\ \Vast]$
\caption{\label{fig:gluon2match} Matching condition for two-boson
exchange contributions to gluon operators. The notation for the
different lines and vertices is as in Fig.~\ref{fig:quark1match}. The diagrams with a quark loop are obtained by closing the external legs of the
corresponding diagrams in Fig.~\ref{fig:quark2match}, and considering
the possible attachments of two external gluons. All active quark
flavors, such as the top quark in the full theory, are included in the
loops.
}
\end{center}
\end{figure}

The gluon matching condition for two-boson exchange is pictured in
Fig.~\ref{fig:gluon2match}. If we consider the external gluons as a
background field~\cite{Novikov:1983gd}, we may express the full theory
diagrams in terms of electroweak polarization tensors induced by a
loop of quarks. For example, using the Feynman rules for the WIMP-$Z^0$ coupling 
from (\ref{eq:heavy2}), the contributions from exchanging two
$Z^0$ bosons may be written as
\be\label{eq:ZZexchange}
{\cal M}^{ZZ}\sim \int (dL) 
{1 \over  - v \cdot L - \delta +i0}  
{ 1 \over (L^2- m_Z^2 +i0)^2} v_\mu v_\nu i \Pi^{\mu \nu}_{(ZZ)} (L)\, ,
\ee
where $(dL) = d^dL / (2\pi)^d$ (this shorthand notation is used
throughout this work), $\delta$ is a residual mass depending on the 
intermediate WIMP state, and $ \Pi^{\mu\nu}_{(ZZ)}(L)$ is the two-gluon part of the $Z^0$ boson polarization
tensor in a background gluon field. The amplitudes with exchange
of one gauge and one Goldstone boson, two Goldstone bosons, or two
Higgs bosons, have the same structure but with vector and scalar
electroweak polarization tensors appearing. 
A similar analysis of gluon contributions to DM-nucleon scattering in
Ref.~\cite{Hisano:2010ct} focused on the spin-0 operator. Here we perform  a complete matching for both spin-0 and spin-2 gluon
operators, and consider new contributions appearing in the case
of mixed states.

The background field method presents the following strategy for
evaluating the two-loop diagrams of the full theory. First, we
determine the two-gluon part of the relevant polarization
tensors. These amplitudes depend only on SM parameters, and can be
used for gluon matching in general DM scenarios; in
particular, this part of the computation is independent of whether the
heavy-particle expansion is employed. Second, we insert the
polarization tensors into the boson loop and perform the remaining
integrals. We illustrate this second part by identifying a basis
of heavy-particle integrals to compute the universal heavy WIMP limit.

In our evaluation we neglect subleading corrections of $\order(m_q/m_W)$ for light quarks ($q\ne t$).
The two-loop diagrams in the full theory (cf. Fig.~\ref{fig:gluon2match}) 
are UV finite, and may be evaluated in $d=4$.  
However, we regulate the effective theory with dimensional regularization, 
and in performing the effective theory subtractions to determine Wilson coefficients 
it is convenient to also use dimensional regularization as IR regulator.   
Thus we choose to implement dimensional regularization as IR regulator also in the 
full theory.  When considering only those terms contributing to the scalar operators 
appearing in (\ref{eq:LWIMPSM}), the relevant amplitudes do not involve 
$\gamma_5$ or $\epsilon^{\mu\nu\alpha\beta}$.    
In particular, the specification of $\gamma_5$ for $d\ne 4$ is unnecessary. 
Further discussion of effective theory contributions can be found in Sec.~\ref{sec:EFTsubtract}.

\subsubsection{Electroweak polarization tensors in a background gluon field} 

Let us isolate the two-gluon amplitude of the relevant electroweak 
polarization tensors in a background gluon field. The generalized 
polarization tensors appearing in two-boson exchange contributions are
\begin{align} \label{eq:POL} 
i\Pi_{(W^+W^+)}^{\nu\mu}(L) &=  \qquad
\parbox{35mm}{
\begin{fmfgraph*}(60,30)
  \fmfleftn{l}{3}
  \fmfrightn{r}{3}
  \fmfset{zigzag_width}{1.5thick}
  \fmf{zigzag}{l2,v}
  \fmf{zigzag}{w,r2}
  \fmfset{arrow_len}{3mm}
  \fmf{plain_arrow,left,tension=0.7,label=$U$}{v,w}
  \fmf{plain_arrow,left,tension=0.7,label=$D$}{w,v}
  \fmflabel{$\mu$}{l2}
  \fmflabel{$\nu$}{r2} 
\end{fmfgraph*}
}
\nl
&= 
- \sum_{U,D} 
{g_2^2  |V_{UD}|^2 \over 8 }
\int d^dx \, e^{iL\cdot x} \langle T\{ \bar{D}(x) \gamma^\nu (1-\gamma_5) U(x) 
\bar{U}(0) \gamma^\mu (1-\gamma_5) D(0) \}\rangle  
\,,
\nl
i\Pi_{(ZZ)}^{\nu\mu}(L) &= 
\qquad
\parbox{25mm}{
\begin{fmfgraph*}(60,30)
  \fmfleftn{l}{3}
  \fmfrightn{r}{3}
  \fmf{zigzag}{l2,v}
  \fmf{zigzag}{w,r2}
  \fmfset{arrow_len}{3mm}
  \fmf{plain_arrow,left,tension=0.7,label=$q$}{v,w}
  \fmf{plain_arrow,left,tension=0.7, label=$q$}{w,v}
  \fmflabel{$\mu$}{l2}
  \fmflabel{$\nu$}{r2} 
\end{fmfgraph*}
}
\nl 
&=
- \sum_q 
{ g_2^2\over 16 c_W^2} 
\int d^dx \, e^{iL\cdot x} \langle T\{ \bar{q}(x) \gamma^\nu (c_V^{(q)}+c_A^{(q)}\gamma_5) q(x) 
\bar{q}(0) \gamma^\mu (c_V^{(q)} + c_A^{(q)}\gamma_5 ) q(0) \} \rangle 
\,,
\nl
i\Pi_{(W^+\phi^+_{W})}^{\mu}(L) &= 
\qquad
\parbox{25mm}{
\begin{fmfgraph*}(60,30)
  \fmfleftn{l}{3}
  \fmfrightn{r}{3}
  \fmf{zigzag}{l2,v}
  \fmf{dbl_dots}{w,r2}
    \fmfset{arrow_len}{3mm}
  \fmf{plain_arrow,left,tension=0.7,label=$U$}{v,w}
  \fmf{plain_arrow,left,tension=0.7, label=$D$}{w,v}
  \fmflabel{$\mu$}{l2}
\end{fmfgraph*}
}
\nl 
&=
\sum_{U,D} {g_2^2  |V_{UD}|^2\over 8 m_W}  
\int d^dx \, e^{iL\cdot x} \langle T\{ \bar{D}(x) \big[ -(m_U-m_D) -(m_U+m_D)\gamma_5 \big] U(x) 
\nl
&\qquad
\bar{U}(0) \gamma^\mu ( 1 - \gamma_5 ) D(0) \} \rangle 
\,,
\nl
i\Pi_{(Z\phi_Z)}^{\mu}(L) 
 &= 
\qquad
\parbox{25mm}{
\begin{fmfgraph*}(60,30)
  \fmfleftn{l}{3}
  \fmfrightn{r}{3}
  \fmf{zigzag}{l2,v}
  \fmf{dbl_dots}{w,r2}
    \fmfset{arrow_len}{3mm}
  \fmf{plain_arrow,left,tension=0.7,label=$q$}{v,w}
  \fmf{plain_arrow,left,tension=0.7, label=$q$}{w,v}
  \fmflabel{$\mu$}{l2}
\end{fmfgraph*}
}
\nl 
&= \sum_q { ig_2^2  m_q  \over 8 c_W m_W} 
\int d^dx \, e^{iL\cdot x} \langle T\{ \bar{q}(x) c_A^{(q)} \gamma_5 q(x) 
\bar{q}(0) \gamma^\mu (c_V^{(q)} + c_A^{(q)}\gamma_5 ) q(0) \} \rangle 
\,,
\nl
i\Pi_{(\phi^+_W \phi^+_W)}(L) 
 &= 
 \quad
\parbox{25mm}{
\begin{fmfgraph*}(60,30)
  \fmfleftn{l}{3}
  \fmfrightn{r}{3}
  \fmf{dbl_dots}{l2,v}
  \fmf{dbl_dots}{w,r2}
      \fmfset{arrow_len}{3mm}
  \fmf{plain_arrow,left,tension=0.7,label=$U$}{v,w}
  \fmf{plain_arrow,left,tension=0.7, label=$D$}{w,v}
\end{fmfgraph*}
}
\nl 
&= - \sum_{UD} { g_2^2 |V_{UD}|^2 \over 8 m_W^2}   
\int d^dx \, e^{iL\cdot x} \langle T\{ \bar{D}(x) \big[ -(m_U-m_D) -(m_U+m_D)\gamma_5 \big] U(x) 
\nl
&\qquad 
\bar{U}(0) \big[ -(m_U-m_D) + (m_U+m_D)\gamma_5 \big] D(0) \} \rangle \,,
\nl
i\Pi_{(\phi_Z\phi_Z)}(L)
 &= 
 \quad
\parbox{20mm}{
\begin{fmfgraph*}(60,30)
  \fmfleftn{l}{3}
  \fmfrightn{r}{3}
  \fmf{dbl_dots}{l2,v}
  \fmf{dbl_dots}{w,r2}
      \fmfset{arrow_len}{3mm}
  \fmf{plain_arrow,left,tension=0.7,label=$q$}{v,w}
  \fmf{plain_arrow,left,tension=0.7, label=$q$}{w,v}
\end{fmfgraph*}
}
\nl
&= \sum_q {g_2^2 m_q^2 \over 4 m_W^2}  
\int d^dx \, e^{iL\cdot x} \langle T\{ \bar{q}(x)\gamma_5 q(x) 
\bar{q}(0) \gamma_5 q(0) \} \rangle 
\,,
\nl
i\Pi_{(hh)}(L) 
 &= 
 \quad
\parbox{20mm}{
\begin{fmfgraph*}(60,30)
  \fmfleftn{l}{3}
  \fmfrightn{r}{3}
  \fmf{dashes}{l2,v}
  \fmf{dashes}{w,r2}
      \fmfset{arrow_len}{3mm}
  \fmf{plain_arrow,left,tension=0.7,label=$q$}{v,w}
  \fmf{plain_arrow,left,tension=0.7, label=$q$}{w,v}
\end{fmfgraph*}
}
\nl
&= - \sum_q { g_2^2 m_q^2 \over 4 m_W^2} 
\int d^dx \, e^{iL\cdot x} \langle T\{ \bar{q}(x) q(x) 
\bar{q}(0) q(0) \} \rangle 
\, ,
\end{align}
where the momentum $L$ is flowing from left to right in the above diagrams. We also require the polarization tensors 
\be\label{eq:otherPOL} 
\Pi_{(W^- W^-)}^{\mu \nu}(L)\, , \quad 
\Pi_{(W^- \phi^{-}_W)}^{\mu}(L) \, , \quad \Pi_{(\phi^{\pm}_W
W^{\pm})}^{\mu}(L)\, , \quad  \Pi_{(\phi_Z Z)}^{\mu}(L)  
 \, , \quad \Pi_{(\phi^-_W \phi^-_W)}(L)\, ,
\ee
which are related to those we have specified in (\ref{eq:POL}) through the 
identities in (\ref{eq:identities}).

Let us now focus on the object,
\begin{align} \label{eq:PItilde} 
i {\tilde \Pi}(L) \equiv \int d^dx \,
e^{iL\cdot x} \langle T\{ \bar{q}^\prime(x) \Gamma q(x)  \bar{q}(0)
\Gamma^\prime q^\prime(0) \} \rangle \, ,
\end{align}
where $\Gamma$ and $\Gamma^\prime$ denote the possible Dirac 
structures whose indices we here suppress. The sum over quark mass 
eigenstates and other prefactors appearing in (\ref{eq:POL}) are included in 
the final result for the polarization tensors. Let us write ${\tilde \Pi}$ in terms 
of momentum-space propagators in a background field,
\begin{align} \label{eq:fourier1}
i {\tilde \Pi}(L) 
= 
- \int d^dx \, e^{iL\cdot x} {\rm Tr}\big[ 
\Gamma iS^{(q)}(x,0) \Gamma^\prime iS^{(q^\prime)}(0,x) \big] 
= \int (dp)  {\rm Tr}\big[ 
\Gamma S^{(q)}(p) \Gamma^\prime \tilde{S}^{(q^\prime)}(p-L) \big] \,,
\end{align}
where 
\be
iS^{(q)}(x,y) = \langle T\{ q(x)\bar{q}(y) \} \rangle 
\ee
and we have used
\begin{align}
S^{(q)}(p) &\equiv \int d^d x\, e^{ip\cdot x} S^{(q)}(x,0) \,,
\quad 
\tilde{S}^{(q)}(p) \equiv \int d^d x\, e^{-ip\cdot x} S^{(q)}(0,x) \,.
\end{align}
We may expand the background field propagators at weak coupling,
\begin{align} \label{eq:FSprop}
iS(p) &= {i\over \slash{p} - m} + g \int (dq) {i\over \slash{p} - m} i\slash{A}(q) 
{i\over \slash{p} - \slash{q} - m} 
\nl
&\quad
+ g^2 \int (dq_1)(dq_2) {i\over \slash{p}-m}i\slash{A}(q_1){i\over \slash{p}-\slash{q}_1 -m}
i\slash{A}(q_2){i\over \slash{p}-\slash{q}_1-\slash{q}_2 - m} + \dots \,,
\nl
i\tilde{S}(p) &= 
{i\over \slash{p} - m} + g \int (dq) {i\over \slash{p}+\slash{q} - m} i\slash{A}(q) 
{i\over \slash{p} - m} 
\nl
&\quad
+ g^2 \int (dq_1)(dq_2) {i\over \slash{p}+\slash{q}_1+\slash{q}_2 - m}
i\slash{A}(q_1){i\over \slash{p}+\slash{q}_2 -m}
i\slash{A}(q_2){i\over \slash{p} - m} + \dots \,,
\end{align}
and upon insertion of these expressions into (\ref{eq:fourier1}), the terms with two gluon fields are readily identified. Furthermore, in Fock-Schwinger gauge the gluon field can be written as
\begin{align}
\slash{A}(q) &= t^a \gamma^\alpha \int d^d x \, e^{iq\cdot x} A^a_\alpha(x) 
\nl
&= t^a \gamma^\alpha \left[ 
{-i\over 2} {\partial \over \partial q_\rho} G^a_{\rho\alpha}(0) (2\pi)^d \delta^d(q) + \dots \right]
\,,
\end{align} 
where the ellipsis denotes terms with derivatives acting on $G^a_{\mu\nu}$. 
Thus the amplitudes with gluon emission are given directly in terms of field-strengths, and intermediate steps involving gauge-variant combinations can 
be avoided.

In isolating the two-gluon amplitude, we may separately consider three 
cases depending on where the gluons are attached. Contributions with both 
gluons attached to the upper quark line in (\ref{eq:POL}) are referred to as``$a$-type", those with both gluons attached to the lower quark line in (\ref{eq:POL}) are referred to as ``$b$-type", and those with one gluon 
attached to each of the upper and lower quark lines are referred to as ``$c$-type". We thus have  
\be \label{eq:sumabc} 
{\tilde \Pi} (L)=
{\tilde \Pi}_a(L) + {\tilde \Pi}_b (L) + {\tilde \Pi}_c(L) \,,  
\ee 
with
\begin{align} \label{eq:abcglue}
i{\tilde \Pi}_a(L)
&= 
{-g^2 \over 4} 
{\rm Tr}(t^a t^b) G^a_{\rho\alpha}(0) G^b_{\sigma\tau}(0) 
\int (dp) {\partial \over \partial q_\rho} {\partial \over \partial q^\prime_\sigma}  
\nl
&\quad
{\rm Tr}\bigg[ 
\Gamma 
{1\over \slash{p}-\mymu} \gamma^\alpha {1\over \slash{p}-\slash{q}-\mymu}
\gamma^\tau {1\over \slash{p}-\slash{q}-\slash{q}^\prime - \mymu} 
\Gamma^\prime
{1\over \slash{p}-\slash{L} - \mymd} 
\bigg]_{q=q^\prime =0} \,,
\nl
i{\tilde \Pi}_b(L) 
&= 
{-g^2 \over 4} 
{\rm Tr}(t^a t^b) G^a_{\rho\alpha}(0) G^b_{\sigma\tau}(0) 
\int (dp) {\partial \over \partial q_\rho} {\partial \over \partial q^\prime_\sigma}  
\nl
&\quad 
{\rm Tr}\bigg[
\Gamma 
{1\over \slash{p}-\mymu}
\Gamma^\prime
{1\over \slash{p}-\slash{L}+\slash{q}
+\slash{q}^\prime - \mymd} \gamma^\alpha {1\over \slash{p}-\slash{L}+\slash{q}^\prime - \mymd} 
\gamma^\tau {1\over \slash{p}-\slash{L} - \mymd} \bigg]_{q=q^\prime=0} \,,
\nl
i{\tilde \Pi}_c(L)
&= 
{-g^2 \over 4} 
{\rm Tr}(t^a t^b) G^a_{\rho\alpha}(0) G^b_{\sigma\tau}(0) 
\int (dp) {\partial \over \partial q_\rho} {\partial \over \partial q^\prime_\sigma}  
\nl
&\quad
{\rm Tr}\bigg[ 
\Gamma
{1\over \slash{p}-\mymu} \gamma^\alpha {1\over \slash{p}-\slash{q} - \mymu}
\Gamma^\prime
{1\over \slash{p}-\slash{L}+\slash{q}^\prime - \mymd} \gamma^\tau 
{1\over \slash{p}-\slash{L} - \mymd} \bigg]_{q=q^\prime=0} \,,
\end{align}
where $\mymu$ and $\mymd$ are the masses of the quarks in the upper 
and lower lines in (\ref{eq:POL}), respectively. To project these onto the 
spin-0 and spin-2 QCD gluon operators, ${\cal O}_g^{(0)}$ and ${\cal O}_g^{(2)}$ in (\ref{eq:ops}), consider the four-index tensor $T_{\alpha\rho\gamma\delta} = G^A_{\alpha\rho}G^A_{\gamma\delta}$ with index 
symmetries $T_{\alpha\rho\gamma\delta} = T_{\gamma\delta\alpha\rho} = -T_{\rho\alpha\gamma\delta}$. We can decompose $T$ into components $T = T^{(0)} + T^{(2)} + \Delta T $, where 
\begin{align}\label{eq:tensordecomp}
T^{(0)}_{\alpha\rho\gamma\delta} &= {1 \over d(d-1)} {\cal O}^{(0)}_g ( g_{\alpha\gamma}g_{\rho\delta} 
- g_{\alpha\delta} g_{\rho\gamma} ) \,,
\nl
T^{(2)}_{\alpha\rho\gamma\delta} &= 
{1 \over d-2} \left( - g_{\alpha\gamma} {\cal O}^{(2)}_{g \, \rho\delta} + g_{\alpha\delta} {\cal O}^{(2)}_{g \, \rho\gamma}
- g_{\rho\delta} {\cal O}^{(2)}_{g \, \alpha\gamma} + g_{\rho\gamma} {\cal O}^{(2)}_{g \, \alpha\delta}  \right) \,,
\end{align} 
and $\Delta T$, satisfying 
\be 
g^{\alpha\gamma}g^{\rho\delta} (\Delta T)_{\alpha \rho \gamma
\delta} =  v^\alpha v^\gamma g^{\rho\delta} (\Delta T)_{\alpha \rho
\gamma \delta} =0 \, , 
\ee
is not needed for the present analysis. The proportionality constants in $T^{(0)}$ and $T^{(2)}$ were obtained by contraction with $g^{\alpha\gamma}g^{\rho\delta}$ or $v^\alpha v^\gamma g^{\rho\delta}$. Upon 
applying the above decomposition to the expressions in (\ref{eq:abcglue}), 
we obtain
\be \label{eq:POLspin}
 i{\tilde \Pi}_k (L) \equiv 
{-g^2 \over 8} \left[{1 \over d(d-1)} {\cal O}^{(0)}_g I_k^{(0)}(L) + {1 \over d-2}{\cal O}_g^{(2)\mu \nu} 
I^{(2)}_{k \, \mu\nu}(L) + \dots \right]\,,
\ee
where $k=a,b,c$ and the ellipsis denotes irrelevant $\Delta T$ 
contributions. 

Let us now determine $I_k^{(0)}(L)$ and $I_{k \, \mu \nu}^{(2)}(L)$ for the 
different cases of two-boson exchange. The trace and derivatives with 
respect to momenta $q$ and $q^\prime$ in (\ref{eq:abcglue}) are 
straightforward to evaluate, and the result is projected onto gluon operators 
of definite spin using (\ref{eq:tensordecomp}). The quark-loop integral over 
momentum $p$ is computed using standard methods, leaving an integral 
over a Feynman parameter, $x$, which will be evaluated after performing 
the boson-loop integral over momentum $L$. We may express the results in 
the form
\begin{align}\label{eq:Ndef}
I_k^{(S)}(L) &\equiv   {i \Gamma[1+\epsilon] \over (4\pi)^{2-\epsilon}} \int_0^1 dx \ u_k(x) N_k^{(S)}(L) \,, \quad u_a(x) = { (1-x)^3 \over 3!}  \, , \quad u_b(x) = { x^3 \over 3!} \, , \quad u_c(x) = x(1-x) \, ,
\end{align}
where $S=0,2$ and for $S=2$ the Lorentz indices are suppressed. Let us also introduce the parameters
\be\label{eq:zDelta}
z_n \equiv {(-1)^n  \over  2^{3-n} }{ \Gamma[n+\epsilon] \over \Gamma[1+\epsilon] } \, , \quad 
\Delta \equiv (1-x)\mymu^2 + x \mymd^2 -x(1-x)L^2 -i0  \, ,
\ee
which appear in the expressions for $N_k^{(S)}(L)$ given below. 

For the operators of interest in (\ref{eq:LWIMPSM}), the relevant projections of the $a$- and $b$-type amplitudes in (\ref{eq:abcglue}) and (\ref{eq:tensordecomp}) are related by CP transformation. This condition can be stated in terms of $N_k^{(S)}(L)$ as 
\be \label{eq:abrelation} 
N_b^{(S)}(L) = N_a^{(S)}(L) \bigg|_{x \,\leftrightarrow \, 1-x, \, \mymu \,  \leftrightarrow \, \mymd } \, .
\ee
In the case of flavor-diagonal currents ($Z^0,\phi_Z^0,h$) where we set 
$\mymu=\mymd=m_q$ in $\tilde{\Pi}_k(L)$, the above relation implies 
$I_b^{(S)}(L) = I_a^{(S)}(L)$. For flavor-changing currents ($W^\pm,\phi_W^\pm$) we set the down-type quark mass $m_2=m_D=0$ but keep 
the up-type quark mass $m_1=m_U \ne 0$ to accommodate the top quark. 
This asymmetry in treating the masses does not allow us to systematically 
recover $N_b^{(S)}(L)$ from $N_a^{(S)}(L)$ using the relation 
(\ref{eq:abrelation}). Below we provide $N_b^{(S)}(L)$ explicitly for 
flavor-changing currents.

To illustrate the explicit implementation of this program, we again focus on 
the heavy WIMP limit, retaining the leading order (in $1/M$) WIMP-SM 
couplings as in (\ref{eq:ZZexchange}). Anticipating the insertion of 
polarization tensors into the boson loop with leading order heavy-particle 
Feynman rules, we thus contract the free Lorentz indices of $\Gamma$ and 
$\Gamma^{\prime}$ in (\ref{eq:POL}), (\ref{eq:PItilde}) with $v_\mu$'s from 
the WIMP-vector boson vertices. It is straightforward to analyze the 
remaining components of $\Pi^{\mu\nu} (L)$ by the same methods. 
The following results are labelled by the bosons in the corresponding
electroweak polarization tensor. For $N_k^{(0)}(L)$ we find,
\begin{align} \label{eq:Nzero}
N_{a}^{(0)}{}_{(W^+W^+)} &= 64 (3-2\epsilon) m_U^2 \bigg\{ 
2(1-\epsilon)  {  z_2 \over \Delta^{2+ \epsilon } } + x(1-x) \left(  2 (v\cdot L)^2  -L^2 \right) {z_3 \over \Delta^{3+\epsilon} }
\bigg\}\, ,
\nl
N_b^{(0)}{}_{(W^+W^+)} &= 0 \, ,
\nl
N_c^{(0)}{}_{(W^+W^+)} &= 64(1-\epsilon)\bigg\{ - 2(1+\epsilon)(3- 2\epsilon ) {z_1 \over \Delta^{1+\epsilon}} 
\nl
&\quad
+ x(1-x) \big[ 
2(1-2\epsilon) (v\cdot L)^2
+ (1+2\epsilon)  L^2 
 \big] {z_2 \over \Delta^{2+\epsilon}}
\bigg\}\, ,
\nl
N_a^{(0)}{}_{(ZZ)} &= 32(3-2\epsilon)m_q^2 \bigg\{ \big[ c_V^{(q)2}+c_A^{(q)2} \big]  \bigg[ 2(1-\epsilon) {z_2 \over \Delta^{2+\epsilon}} + x(1-x)(  2 (v\cdot L)^2 - L^2 ) {z_3 \over \Delta^{3+\epsilon} } \bigg] 
\nl
&\quad
- \big[ c_V^{(q)2} - c_A^{(q)2} \big]  \bigg[ 2(2-\epsilon) {z_2 \over \Delta^{2+\epsilon} } + x^2 L^2 {z_3 \over \Delta^{3+\epsilon} } \bigg]
\bigg\} \, ,
\nl
N_c^{(0)}{}_{(ZZ)} &= 32 \bigg\{ \big[ c_V^{(q)2}+c_A^{(q)2} \big] (1-\epsilon) \bigg[ - 2(3 - 2\epsilon) (1+\epsilon) {z_1\over \Delta^{1+\epsilon}} + x(1-x) \big[ 2(1-2\epsilon) (v \cdot L)^2  
\nl
&\quad 
+(1+2\epsilon) L^2 \big] {z_2 \over \Delta^{2+\epsilon} } \bigg]
+ \big[ c_V^{(q)2} - c_A^{(q)2} \big] \epsilon(3-2\epsilon) m_q^2 {z_2 \over \Delta^{2+\epsilon} } 
\bigg\} 
\,,
\nl
N_a^{(0)}{}_{(W^+\phi_W^+)} &= - 64 (3-2\epsilon) m_U^2 v \cdot L \bigg[ 2\big[ 2 - 3x  - \epsilon (1-x) \big] {z_2 \over \Delta^{2+\epsilon}} + x^2 (1-x) L^2 {z_3 \over \Delta^{3+\epsilon}} \bigg]   \,,
\nl
N_b^{(0)}{}_{(W^+\phi_W^+)} &= 0 \, ,
\nl
N_c^{(0)}{}_{(W^+\phi_W^+)} &= - 64(3-2\epsilon)(1 - \epsilon)(1-x) m_U^2 v\cdot L {z_2 \over \Delta^{2+\epsilon} }\,,
\nl
N_a^{(0)}{}_{(Z\phi_Z)} &= - 32(3 - 2\epsilon)  c_A^{(q)2} m_q v\cdot L \bigg[ 2\big[2 - 3x -\epsilon(1-x) \big] {z_2 \over \Delta^{2+\epsilon} } + x \big[ m_q^2 + x(1-x) L^2 \big] {z_3 \over \Delta^{3+\epsilon} } \bigg] \,,
\nl
N_c^{(0)}{}_{(Z\phi_Z)} &= - 32 (3-2\epsilon)(1 - \epsilon)  c_A^{(q)2} m_q v\cdot L {z_2 \over \Delta^{2+\epsilon}} \,, 
\nl
N_a^{(0)}{}_{(\phi_W^+ \phi_W^+)} &= 64 (3-2\epsilon) m_U^4 \big[ - 2(2- \epsilon) {z_2 \over \Delta^{2+\epsilon}} +x(1-x) L^2 {z_3 \over \Delta^{3+\epsilon}} \big] \,,
\nl
N_b^{(0)}{}_{(\phi_W^+ \phi_W^+)} &= 0 \, ,
\nl
N_c^{(0)}{}_{(\phi_W^+ \phi_W^+)} &= 64 (1-\epsilon)(3-2\epsilon) m_U^2 \big[- 2(2- \epsilon) {z_1 \over \Delta^{1+\epsilon}}  + x(1-x) L^2 {z_2 \over \Delta^{2+\epsilon}} \big] \,,
\nl
N_a^{(0)}{}_{(\phi_Z \phi_Z)} &= - 32(3- 2\epsilon) x m_q^2 L^2 {z_3 \over \Delta^{3+\epsilon} }\,,
\nl
N_c^{(0)}{}_{(\phi_Z \phi_Z)} &= 32(3-2\epsilon)\bigg[ 2(1-\epsilon)(2-\epsilon) {z_1 \over \Delta^{1+\epsilon}} - \big[ (2- \epsilon) m_q^2 + (1-\epsilon) x(1-x) L^2 \big] {z_2 \over \Delta^{2+\epsilon}} \bigg] \,,
\nl
N_a^{(0)}{}_{(hh)} &= 32(3-2\epsilon) m_q^2 \bigg[- 4(2- \epsilon) {z_2 \over \Delta^{2+\epsilon} } + x(1-2x) L^2 {z_3 \over \Delta^{3+\epsilon}}  \bigg] \,,
\nl
N_c^{(0)}{}_{(hh)} &= 32(3-2\epsilon) \bigg[ - 2(1-\epsilon)(2- \epsilon) {z_1 \over \Delta^{1+\epsilon} }  +  \big[ (1-\epsilon) x(1-x) L^2 - (2- \epsilon) m_q^2  \big] {z_2 \over \Delta^{2+\epsilon}} \bigg] \,.
\end{align}
For $N_{k \, \mu \nu}^{(2)}(L)$ the open indices are to be contracted with 
${\cal O}_g^{(2) \, \mu \nu}$, which is symmetric in $\mu$ and $\nu$ and 
satisfies $g_{\mu\nu} {\cal O}_g^{(2)\,\mu\nu}=0$. The results are
\begin{align} \label{eq:Ntwo}
N_{a \, \mu \nu}^{(2)}{}_{(W^+W^+)} &= 128(1-\epsilon) \bigg\{ 
- 4(2- \epsilon) v_\mu v_\nu {z_1 \over \Delta^{1+\epsilon} }
\nl
& \quad
+2\bigg[( m_U^2 - x^2 L^2 ) v_\mu v_\nu 
+2 (2-\epsilon) x (1-x)  v\cdot L v_{\mu} L_{\nu  } 
- x ( 2- x-\epsilon)  L_\mu L_\nu  
\bigg] {z_2 \over \Delta^{2+ \epsilon} } 
\nl
& \quad 
+x(1-x)\bigg[ (m_U^2 - 2x^2 (v \cdot L)^2) L_\mu L_\nu - 2 (m_U^2 - x^2 L^2) v \cdot L v_{ \mu} L_{ \nu  } \bigg] {z_3 \over \Delta^{3+\epsilon}}
\bigg\} 
\,, 
\nl
N_{b \, \mu \nu}^{(2)}{}_{(W^+W^+)} &= 128(1-\epsilon) \bigg\{ 
- 4(2 - \epsilon ) v_\mu v_\nu {z_1 \over \Delta^{1+\epsilon} }
\nl
& \quad
-2(1-x) \bigg[ (1-x) L^2  v_\mu v_\nu 
-2 (2-\epsilon) x  v\cdot L v_{\mu} L_{\nu  } 
+ ( 1 + x - \epsilon )  L_\mu L_\nu  
\bigg] {z_2 \over \Delta^{2+ \epsilon} } 
\nl
& \quad 
+2x(1-x)^3 \bigg[ - (v \cdot L)^2 L_\mu L_\nu +  L^2 v \cdot L v_{ \mu} L_{ \nu  } \bigg] {z_3 \over \Delta^{3+\epsilon}}
\bigg\} 
\,, 
\nl
N_{c \, \mu \nu}^{(2)}{}_{(W^+W^+)} &= 128 \bigg\{ 2 (1-\epsilon) (1-2\epsilon)  v_\mu v_\nu {z_1 \over \Delta^{1 + \epsilon} } 
\nl
&\quad 
+ x(1-x) \big[ \epsilon L_\mu L_\nu  + 2 (1-2\epsilon)  v \cdot L v_{ \mu} L_{\nu  } - (1 - 2\epsilon)  L^2 v_\mu v_\nu \big] {z_2 \over \Delta^{2+\epsilon}}  \bigg\} \,,
\nl
N_{a\, \mu \nu}^{(2)}{}_{(ZZ)} &= 64 (1-\epsilon) \bigg\{ \big[ c_V^{(q)2} - c_A^{(q)2} \big] x^2 m_q^2 L_\mu L_\nu {z_3 \over \Delta^{3+\epsilon}}  + \big[ c_V^{(q)2}+c_A^{(q)2} \big]  \bigg[ - 4(2- \epsilon)v_\mu v_\nu {z_1 \over \Delta^{1+\epsilon}} 
\nl
&\quad
 +2 \big[  (m_q^2 - x^2 L^2 ) v_\mu v_\nu + 2(2-\epsilon) x (1-x) v \cdot L v_{ \mu} L_{ \nu } + x(2- x- \epsilon )L_\mu L_\nu \big] {z_2 \over \Delta^{2+\epsilon} }  
 \nl
 & \quad 
+ x(1-x) \big[ (m_q^2 -2x^2 (v\cdot L)^2 L_\mu L_\nu - 2 (m_q^2 - x^2 L^2 ) v \cdot L v_{  \mu} L_{ \nu  } \big]  {z_3 \over \Delta^{3+ \epsilon} }
  \bigg] \bigg\}\, ,
\nl
N_{c \, \mu \nu}^{(2)}{}_{(ZZ)} &= 64 \bigg\{  - \big[ c_V^{(q)2} - c_A^{(q)2} \big] 2(1- \epsilon)  m_q^2 v_\mu v_\nu {z_2 \over \Delta^{2+\epsilon}}  +   \big[ c_V^{(q)2}+c_A^{(q)2} \big] \bigg[ 2(1-\epsilon)(1-2 \epsilon) v_\mu v_\nu { z_1 \over \Delta^{1 + \epsilon} }
\nl
&\quad 
+x(1-x) \big[ \epsilon L_\mu L_\nu + 2 (1-2\epsilon) v \cdot L v_{ \mu } L_{ \nu } - (1- 2 \epsilon ) L^2 v_\mu v_\nu  \big]   {z_2 \over \Delta^{2+\epsilon}} \bigg] \bigg\} \, ,
\nl
N_{a\, \mu \nu}^{(2)}{}_{(W^+\phi_W^+)} &= - 128 (1- \epsilon ) x m_U^2 \bigg\{ 2 v_{ \mu } L_{ \nu } {z_2 \over \Delta^{2+\epsilon} } - x(1-x) v \cdot L L_\mu L_\nu {z_3 \over \Delta^{3+\epsilon} } \bigg\} \, ,
\nl
N_{b\, \mu \nu}^{(2)}{}_{(W^+\phi_W^+)} &= - 128 (1- \epsilon ) (1-x) m_U^2 \bigg \{ 2(2-\epsilon) v_{ \mu } L_{ \nu  } {z_2 \over \Delta^{2+\epsilon} } +(1-x)^2 \big[ L^2v_{\mu } L_{ \nu  } - v \cdot L L_\mu L_\nu  \big] {z_3 \over \Delta^{3+\epsilon} } \bigg\} \, ,
\nl
N_{c\, \mu \nu}^{(2)}{}_{(W^+\phi_W^+)} &= - 128 (1- \epsilon ) (1-x) m_U^2 v_{ \mu } L_{ \nu  } {z_2 \over \Delta^{2+\epsilon} } \, ,
\nl
N_{a\, \mu \nu}^{(2)}{}_{(Z\phi_Z)} &= \big[ c_A^{(q)2} \big] 64 (1-\epsilon)  x  m_q \bigg \{ - 2(3- \epsilon ) v_{ \mu } L_{ \nu  } {z_2 \over \Delta^{2+\epsilon} } 
\nl
&\quad
+ \big[ (m_q^2 -x^2 L^2) v_{ \mu } L_{ \nu  } +x v \cdot L L_\mu L_\nu  \big]  {z_3 \over \Delta^{3+\epsilon } } \bigg \} \, ,
\nl
N_{c\, \mu \nu}^{(2)}{}_{(Z\phi_Z)} &= -  \big[ c_A^{(q)2} \big] 64 (1- \epsilon ) m_q  v_{  \mu } L_{ \nu  } { z_2 \over \Delta^{2+\epsilon}} \, ,
\nl
N_{a\, \mu \nu}^{(2)}{}_{(\phi_W^+ \phi_W^+)} &= 128 (1-\epsilon ) x m_U^2 L_\mu L_\nu \bigg\{ 2(2-\epsilon) {z_2 \over \Delta^{2+\epsilon} } -(1-x) m_U^2 {z_3 \over \Delta^{3+\epsilon} }  \bigg\} \, ,
\nl
N_{b\, \mu \nu}^{(2)}{}_{(\phi_W^+ \phi_W^+)} &= 256 (1-\epsilon)(2-\epsilon)(1-x) m_U^2 L_\mu L_\nu {z_2 \over \Delta^{2+\epsilon} }  \, ,
\nl
N_{c\, \mu \nu}^{(2)}{}_{(\phi_W^+ \phi_W^+)} &= 128 (1-\epsilon) x (1-x) m_U^2 L_\mu L_\nu {z_2 \over \Delta^{2+\epsilon} } \, ,
\nl
N_{a\, \mu \nu}^{(2)}{}_{(\phi_Z \phi_Z)} &= 64 (1-\epsilon) x L_\mu L_\nu \bigg\{- 2(2- \epsilon) {z_2 \over \Delta^{2+\epsilon}} + m_q^2 {z_3 \over \Delta^{3+\epsilon} } \bigg\}  \, ,
\nl
N_{c\, \mu \nu}^{(2)}{}_{(\phi_Z \phi_Z)} &= - 64 (1- \epsilon ) x (1-x) L_\mu L_\nu {z_2 \over \Delta^{2+\epsilon}} \, ,
\nl
N_{a \, \mu \nu }^{(2)}{}_{(hh)} &= 64 (1-\epsilon ) x L_\mu L_\nu \bigg\{ 2 (2-\epsilon) {z_2 \over \Delta^{2+\epsilon}} - (1- 2x) m_q^2 {z_3 \over \Delta^{3+\epsilon}}  \bigg\} \, ,
\nl
N_{c \, \mu \nu}^{(2)}{}_{(hh)} &= 64(1-\epsilon) x(1-x) L_\mu L_\nu {z_2 \over \Delta^{2+\epsilon} }.
\end{align}

The results for $N_k^{(S)}(L)$ in (\ref{eq:Nzero}) and (\ref{eq:Ntwo}) specify 
$I_a^{(S)}(L)$ through (\ref{eq:Ndef}), and hence ${\tilde \Pi}_k(L)$ through 
(\ref{eq:POLspin}), and ${\tilde \Pi}(L)$ through (\ref{eq:sumabc}). 
This completes our determination of the polarization tensors in (\ref{eq:POL}). 
The polarization tensors in (\ref{eq:otherPOL}) are obtained through the following relations
\begin{align} \label{eq:identities}
\Pi_{(W^- W^-)}^{\mu \nu}(L) &= \Pi_{(W^+ W^+)}^{\mu \nu}(-L) \, , 
\quad
\Pi_{(\phi_Z Z)}^{\mu}(L) =  \Pi_{(Z \phi_Z)}^{\mu}(-L) \, ,
\quad
\Pi_{(\phi^-_W \phi^-_W)}(L) = \Pi_{(\phi^+_W \phi^+_W)}( - L) \, ,
\nl 
\Pi_{(\phi^{-}_W W^{-})}^{\mu}(L) &= \Pi_{( W^+ \phi^{+}_W)}^{\mu}(-L) \, , \quad
\Pi_{(\phi^{+}_W W^{+})}^{\mu}(L) = \Pi_{( W^- \phi^{-}_W)}^{\mu}(- L) \, ,
\nl
\Pi_{(W^- \phi_W^- )}^{\mu}( L) &= \Pi_{( W^+ \phi^{+}_W)}^{\mu}(-L)\, .
\end{align}
The identities in the first two lines are consequences of reversing the 
direction of momentum $L$ in the diagrams in (\ref{eq:POL}). The last 
relation follows from Hermitian conjugation and the identification 
$\overline{S(p)} \equiv \gamma^0 S(p)^\dagger \gamma^0 = \tilde{S}(p)$. 
We note that polarization tensors with one gauge and one Goldstone boson 
are odd in $L$, while all others are even in $L$. This property also holds for 
the corresponding $N_k^{(S)}(L)$, and we use it in the next section to 
systematically reduce the boson loop integrals  into a convenient basis.

\subsubsection{Basis reduction of the full theory boson loop}

Having determined the generalized polarization tensors, we now proceed
with the reduction of the remaining boson loop integrals. 
Upon insertion of the polarization tensors into the
boson loop, we find the required set of basic loop integrals
\begin{align}\label{eq:bosonloop}
 \int (dL) \bigg[ {1 \over v \cdot L - \delta +i 0} + {1 \over -v \cdot L - \delta +i 0} \bigg]{1 \over (L^2 - m_V^2 +i0)^2} N_k^{(S)}(L) \equiv {\cal I}_{\rm even}(\delta, m_V) N_k^{(S)}(L) \, ,
 \nl
 \int (dL) \bigg[ {1 \over v \cdot L- \delta  +i 0} - {1 \over -v \cdot L - \delta +i 0} \bigg]{1 \over (L^2 - m_V^2 +i0)^2} N_k^{(S)}(L)  \equiv {\cal I}_{\rm odd} (\delta, m_V) N_k^{(S)}(L) \, ,
\end{align}
where $\delta$ is the residual mass of the intermediate 
WIMP state, and $m_V$ is the mass of the exchanged bosons. We
suppress the arguments, $(\delta, m_V)$, of these integral operators
when making generic statements below. The integral operator ${\cal
I}_{\rm even}$ requires that $N_k^{(S)}(L)$ be even in $L$ as in those
for polarization tensors with a single type of boson, while the
integral operator ${\cal I}_{\rm odd}$ requires that $N_k^{(S)}(L)$ be
odd in $L$ as in those for polarization tensors with one gauge and one
Goldstone boson. Let us denote even $N_k^{(S)}(L)$ by $N_{k\, {\rm
even}}^{(S)}(L)$ and odd $N_k^{(S)}(L)$ by $N_{k \, {\rm odd}}^{(S)}(L)$. The
subscripts even and odd may be dropped if we mean either type, or if
the exchanged bosons are specified.

To reduce (\ref{eq:bosonloop}) to a set of basis integrals for
evaluation, we begin by replacing factors of $L^2$ in $N_k^{(S)}(L)$
with
\begin{align}\label{eq:replaceL}
L^2 = -{\Delta \over x(1-x) } + {\mymu^2 \over x}  + {\mymd^2 \over (1-x)} \, ,
\end{align}
which follows from the definition of $\Delta$ in (\ref{eq:zDelta}). The
$N_k^{(S)}(L)$ of (\ref{eq:Nzero}) and (\ref{eq:Ntwo}) may then be written in terms of
$\Delta$ and the vectors $v_\mu$ and $L_\mu$. 
In $N_{k\, {\rm even}}^{(S)}(L)$ each term must have two or zero $v_\mu$'s, while in $N_{k \, {\rm odd} }^{(S)}(L)$ each term must have one $v_\mu$. Organizing the result in powers of $(v \cdot L)$, we obtain
\begin{align}\label{eq:NexpandvL}
N_{k\, {\rm even}}^{(0)} (L)&=  (v \cdot L)^0  \sum_n a^{(1)}_n \Delta^{-n-\epsilon}  + (v\cdot L)^2  \sum_{n} a_n^{(2)} \Delta^{-n-\epsilon}   \, ,
\nl
N_{k\, {\rm odd}}^{(0)}  (L)&=  (v\cdot L)^1  \sum_n a_n^{(3)} \Delta^{-n-\epsilon} \, ,
\nl
 N_{k \, {\rm even}}^{(2) \, \mu \nu}  (L)&= (v \cdot L)^0 \sum_n \bigg[ v^\mu v^\nu  a^{(4)}_n \Delta^{-n-\epsilon}  + L^\mu L^\nu   a^{(5)}_n \Delta^{-n-\epsilon} \bigg] + (v\cdot L)^1 \sum_n v^\mu L^\nu  a^{(6)}_n \Delta^{-n-\epsilon} 
 \nl
 &\quad + (v\cdot L)^2 \sum_n L^\mu L^\nu   a_n^{(7)} \Delta^{-n-\epsilon}    \, ,
 \nl
 N_{k \, {\rm odd}}^{(2) \, \mu \nu} (L) &=   (v \cdot L)^0  \sum_n v^\mu L^\nu  a^{(8)}_n \Delta^{-n-\epsilon} + (v\cdot L)^1 \sum_n L^\mu L^\nu  a^{(9)}_n \Delta^{-n-\epsilon}    \, ,
\end{align}
where the sums run over $n=1,2, \dots \,$, and the coefficients
$a_n^{(i)}$ are functions of $x$ and $\epsilon$. The above $N_k^{(S)} (L)$
structures require the set of integrals
\begin{align}\label{eq:tildebasis}
H(n) &= {\cal I}_{\rm even} \Delta^{-n-\epsilon} \, ,
\quad
H^{\mu}(n) =  {\cal I}_{\rm odd} \Delta^{-n-\epsilon} L^\mu \,, 
\quad 
H^{\mu \nu}(n)  = {\cal I}_{\rm even} \Delta^{-n-\epsilon} L^\mu L^\nu \,, 
\nl
F(n) &= \int (dL) {1 \over (L^2 - m_V^2 +i0)^2} \Delta^{-n-\epsilon} \,. 
\quad 
\end{align}
The integrals $H^\mu$ and $H^{\mu \nu}$ may be expressed in terms of 
$H(n)$ and $F(n)$ through standard reduction methods and the relation
\begin{align}\label{eq:relationPM} \bigg[ {1 \over v \cdot L - \delta
+i 0} \pm {1 \over -v \cdot L - \delta +i 0} \bigg] v \cdot L & =
\delta \bigg[ {1 \over v \cdot L - \delta +i 0} \mp {1 \over -v \cdot
L - \delta +i 0} \bigg]  + 1 \mp 1 \,.
\end{align}
Furthermore, recursion relations in $n$ may be
derived by taking derivatives of parameters. A detailed discussion of
these relations, as well as the evaluation of the above integrals, can
be found in Appendix~\ref{sec:appC}. Note that the $(v \cdot L)^2$ term in
$N_{k \, {\rm even} }^{(2)\, \mu \nu} (L)$ also requires the integral 
\be \int (dL) {1 \over (L^2 - m_V^2 +i0)^2} \Delta^{-n-\epsilon} L^\mu
L^\nu \sim g^{\mu \nu} \, , \ee
however this does not contribute since it vanishes upon contraction with the
traceless spin-2 gluon operator, $O_g^{(2) \mu \nu}$. Upon feeding the
general expressions for $N_k^{(S)} (L)$ in (\ref{eq:NexpandvL}) into the
integrals in (\ref{eq:bosonloop}), we find the following decomposition
in terms of basis integrals,
\begin{align}\label{eq:HFreduction}
{\cal I}_{\rm even} N_{k\, {\rm even}}^{(0)}(L) 
&=  \sum_n \bigg[ a^{(1)}_n H(n)  + a^{(2)}_n \big[ \delta^2 H(n) + 2 \delta F(n) \big]   \bigg] \, ,
\nl
{\cal I}_{\rm odd} N_{k\, {\rm odd}}^{(0)}(L)
&=   \sum_n a_n^{(3)} \big[ \delta H(n) + 2 F(n) \big]  \, ,
\nl
{\cal I}_{\rm even} N_{k \, {\rm even} }^{(2) \, \mu \nu}(L)
&= v^\mu v^\nu \sum_n \bigg[  a^{(4)}_n H(n)  +    a^{(5)}_n H_1(n) +   a^{(6)}_n \big[ \delta^2 H(n) +2 \delta F(n) \big] +  a_n^{(7)} \delta^2 H_1(n)   \bigg] \, ,
\nl
{\cal I}_{\rm odd} N_{k \, {\rm odd} }^{(2)\, \mu \nu}(L)
&= v^\mu v^\nu \sum_n \bigg[  a^{(8)}_n \big[ \delta H(n) +2 F(n) \big]  +      a^{(9)}_n \delta H_1(n)   \bigg] \, ,
\end{align}
where
\begin{align}
H_1(n) = {1 \over 3-2\epsilon} \bigg\{ 
(4-2\epsilon)\big[ \delta^2 H(n) + 2 \delta F(n) \big]  + {H(n-1) \over x(1-x)}   - \bigg[ {\mymu^2 \over x} + {\mymd^2 \over 1-x} \bigg] H (n) \bigg\} \, .
\end{align}
The above results apply generally to both pure and mixed
states. Comparing with the explicit expressions for $N_k^{(S)}(L)$ in
(\ref{eq:Nzero}) and (\ref{eq:Ntwo}), we find that $H(n)$ for
$n=1,2,3$ and $F(n)$ for $n=2,3$ are required.

For pure states there is no residual mass, and ${\cal I}_{\rm odd}$ is
irrelevant since the only contributions are from exchanges of $W^\pm$
and $Z^0$, involving $N_{k \, {\rm even}}^{(S)}(L)$. The vanishing of
certain contributions in ${\cal I}_{\rm even} N_{k\, {\rm
even}}^{(S)}(L)$ at $\delta =0$ can be traced to the identity in
(\ref{eq:relationPM}).%
\footnote{In particular, this can be used to demonstrate gauge
invariance for the electroweak part of the amplitudes since in a
general $R_\xi$ gauge the $\xi$-dependent terms carry a factor of $(v
\cdot L)$.}
Setting $\delta=0$ in ${\cal I}_{\rm even} N_{k \, {\rm
even}}^{(S)}(L)$ above and using the explicit expressions for $N_k^{(S)}(L)$
in (\ref{eq:Nzero}) and (\ref{eq:Ntwo}), we find pure-state results
that depend on $H(n)$ only,
\begin{align}\label{eq:purebasis}
{\cal I}(0, m_W) N_a^{(0)} {}_{(W^+W^+)}  &= 64  (1+\epsilon) (3- 2\epsilon) m_U^2 \bigg\{ (2+\epsilon)(1-x) m_U^2 H(3) - (1+2\epsilon) H(2)   \bigg\} \, , 
\nl
{\cal I}(0, m_W) N_c^{(0)} {}_{(W^+W^+)}  &= 32 (1-\epsilon^2) \bigg\{  (1+2\epsilon)(1-x) m_U^2 H(2) + 2(1-2\epsilon) H(1)  \bigg\} \, , 
\nl
{\cal I}(0, m_Z) N_a^{(0)} {}_{(ZZ)}  &= { 32(1+\epsilon)(3-2\epsilon) m_q^2 \over 1-x}  \bigg\{ \big[ c_V^{(q)2}+c_A^{(q)2} \big] (1-x) \big[ (2+\epsilon) m_q^2 H(3) - (1+2\epsilon)  H(2) \big]  
\nl &\quad
+ \big[ c_V^{(q)2} - c_A^{(q)2} \big] \big[ (2+\epsilon) x m_q^2 H(3) - (2  - \epsilon + 2\epsilon x ) H(2) \big] \bigg\} \, , 
\nl
{\cal I}(0, m_Z) N_c^{(0)} {}_{(ZZ)}  &= 16 (1+\epsilon) \bigg\{ 
\big[ c_V^{(q)2}+c_A^{(q)2} \big] (1-\epsilon)   \big[ (1+2 \epsilon) m_q^2 H(2) + 2 (1-2\epsilon)  H(1) \big]  
\nl &\quad
+ \big[ c_V^{(q)2} - c_A^{(q)2} \big] \epsilon (3-2 \epsilon ) m_q^2  H(2) \bigg\} \, ,
\nl
{\cal I}(0, m_W) N_{a \, \mu \nu}^{(2)} {}_{(W^+W^+)}  &= { 128 (1-\epsilon ) v_\mu v_\nu \over (3-2\epsilon) (1-x) } \bigg \{ (2-\epsilon) (2 - x - 3\epsilon +4\epsilon x) H(1) 
\nl &\quad
+  (1+\epsilon) (1-x) m_U^2 \big[ (2+\epsilon)(1-x) m_U^2 H(3) + (3-4x -4\epsilon + 2 \epsilon x) H(2) \big]   \bigg \}  \, , 
\nl
{\cal I}(0, m_W) N_{c\, \mu \nu}^{(2)} {}_{(W^+W^+)}  &= {64 (1-\epsilon) v_\mu v_\nu \over (3-2\epsilon) } \bigg\{ - (3- \epsilon - 4\epsilon^2) (1-x) m_U^2 H(2)  + \epsilon (7 - 8\epsilon) H(1) \bigg\} \, , 
\nl
{\cal I}(0, m_Z) N_{a \, \mu \nu}^{(2)} {}_{(ZZ)}  &= { 64 (1-\epsilon)  v_\mu v_\nu \over (3-2\epsilon)(1-x) } \bigg\{ \big[ c_V^{(q)2}+c_A^{(q)2} \big] \bigg[ (2- \epsilon ) (2-x-3\epsilon +4\epsilon x )  H(1) 
\nl &\quad
+ m_q^2 (1+\epsilon) \big[ (2+\epsilon) (1-x) m_q^2 H(3) +( 3 - 5 x - 4 \epsilon + 5 \epsilon x  )H(2)   \big]  \bigg]  
\nl &\quad
+\big[ c_V^{(q)2} - c_A^{(q)2} \big] (1+\epsilon) (2+\epsilon) x  m_q^2 \big[ m_q^2 H(3) - H(2) \big] \bigg\} \, ,
\nl
{\cal I}(0, m_Z) N_{c \, \mu \nu}^{(2)} {}_{(ZZ)} &= {32 (1-\epsilon) v_\mu v_\nu \over (3-2\epsilon) } \bigg\{ \big[ c_V^{(q)2}+c_A^{(q)2} \big] \big[ - (1+\epsilon) (3- 4\epsilon) m_q^2 H(2) + \epsilon (7-8 \epsilon) H(1) \big] 
\nl &\quad
- \big[ c_V^{(q)2} - c_A^{(q)2} \big] 2(1+\epsilon) (3-2\epsilon)  m_q^2 H(2) \bigg\} \, ,
\end{align}
where the subscript on ${\cal I}_{\rm even}$ has been suppressed. The
reduction for admixtures, where there are nonzero residual masses and
the integral ${\cal I}_{\rm odd}$ is relevant, is also straightforward
to obtain. 

We collect in Appendix~\ref{sec:appC} useful results for the remaining
task of integrating over Feynman parameters. The singularity structure
and evaluation of integrals can be classified into three cases
corresponding to zero, one, or two heavy fermions contributing to the
electroweak polarization tensor. 
The case of zero heavy fermions is
for polarization tensors with no top quark in the loop. 
With subleading powers of light quark masses neglected, only polarization tensors of $W^\pm$ and $Z^0$ 
bosons are relevant in this case.  The case of one heavy
fermion is for polarization tensors of flavor-changing 
currents with one top quark and one down-type quark. The case of two
heavy fermions is for polarization tensors of flavor-diagonal
currents with a top quark loop. 

\subsubsection{Full theory contributions and matching coefficients for pure states}

Let us now determine the full theory contributions to the matching
using the generalized electroweak polarization tensors and the
reduction method for the boson loop integral. For pure states, the
total amplitude receives two-boson exchange contributions from $W^\pm$
and $Z^0$ bosons,
\begin{align}
{\cal M} = {\cal M}^{WW}  +  {\cal M}^{ZZ} \, ,
\end{align}
which may be written in terms of electroweak polarization tensors in a background field as
\begin{align} 
i{\cal M}^{WW} &={ i g_2^2 {\cal C}_W \over 2} \int (dL)
{1 \over - v \cdot L + i0 } {1 \over (L^2 - m_W^2 + i0)^2} v_\mu v_\nu
\bigg[ i\Pi^{\mu \nu}_{(W^+W^+)} (L) + i\Pi^{\mu \nu}_{(W^-W^-)} (L)
\bigg] \, , \nl i{\cal M}^{ZZ} &={ i g_2^2 {\cal C}_Z \over c_W^2}
\int (dL) {1 \over - v \cdot L + i0 } {1 \over (L^2 - m_Z^2 + i0)^2}
v_\mu v_\nu  i\Pi^{\mu \nu}_{(ZZ)} (L)  \, ,
\end{align}
with ${\cal C}_W$ and ${\cal C}_Z$ given in (\ref{eq:WZcoeff}). 
The parity of the polarization tensors under $L \to -L$ and the identities in (\ref{eq:identities}) allow us to write the above amplitudes in terms of the integrals defined in (\ref{eq:bosonloop}),
\begin{align}
i {\cal M}^{WW} &= {i g_2^2  {\cal C}_W \over 2} 
{\cal I}_{\rm even}(0, m_W) v_\mu v_\nu  i\Pi^{\mu \nu}_{(W^+W^+)} (L) \, ,
\nl
i {\cal M}^{ZZ} &= {i g_2^2  {\cal C}_Z \over 2 c_W^2} 
{\cal I}_{\rm even}(0,m_Z) v_\mu v_\nu  i\Pi^{\mu \nu}_{(ZZ)} (L) \, .
\end{align}
Upon inserting the explicit polarization tensors from (\ref{eq:POL})
into the expressions above, we may employ the reduction of integrals
given in (\ref{eq:purebasis}) and write each contribution in terms of
the gluon operators of definite spin,
\begin{align}\label{eq:gluonamplitudespin} {\cal M}^{B B^\prime} &=
{\cal M}^{{B B^\prime} (0)} {\cal O}_g^{(0)} +  {\cal M}^{{B B^\prime}
(2)} \, v_\mu v_\nu {\cal O}_g^{(2)\mu\nu}  \, ,
\end{align}
where the superscript ${B B^\prime}$ denotes the different types of
two-boson exchange. From the expression in
(\ref{eq:gluonamplitudespin}), we readily identify the contribution of
each amplitude to $c_g^{(0)}{}_{\rm 2BE}$ and $c_g^{(2)}{}_{\rm 2BE}$
as ${\cal M}^{BB^\prime (0)}$ and ${\cal M}^{BB^\prime  (2)}$,
respectively. Let us decompose ${\cal M}^{WW(S)}$, for $S=0,2$, into
contributions from each up-type quark flavor, and the $a$-, $b$-,
and $c$-type gluon attachments,
\be\label{eq:ampcharge} {\cal M}^{WW(S)} =   -{[\Gamma (1+\epsilon)]^2
\over (4\pi)^d }  { \pi g^2 g_2^4 \over m_W^{3+4\epsilon} }  {{\cal
C}_W \over 16}  \sum_{U=u,c,t} \sum_{\ k={a,b,c}} {\cal M}_{U,k}^{WW(S)}
\,.  \ee
Similarly, we decompose ${\cal M}^{ZZ(S)}$ into contributions from
each quark flavor, and the $a$-, $b$-, and $c$-type gluon
attachments,
\be\label{eq:ampneutral}
{\cal M}^{ZZ(S)} =  -    { [\Gamma (1+\epsilon)]^2   \over (4\pi)^d }  { \pi g^2 g_2^4   \over m_Z^{3+4\epsilon} } 
{ {\cal C}_Z \over 64 c_W^4}  \sum_{q=u,c,t,d,s,b} \sum_{\ k={a,b,c}} {\cal M}_{q,k}^{ZZ(S)} \,. 
\ee
The results for $W^\pm$ exchange are as follows. The amplitudes with one top quark are
\begin{align} 
{\cal M}_{t,a}^{WW(0)} &= 
{4 x_t^2}\log{x_t+1\over x_t} - {2x_t ( 6x_t^2 + 9 x_t + 2 )\over 3 (x_t+1)^2 }  \, , 
\nl
{\cal M}_{t,b}^{WW(0)} &= 0 \,,
\nl
{\cal M}_{t,c}^{WW(0)} &= 
-4 x_t^2 \log{x_t+1 \over x_t} + {2 (6 x_t^3 + 9 x_t^2 + 2 x_t - 2)\over 3(x_t+1)^2}  \,,
\nl
 {\cal M}^{WW(2)}_{t,a} &= 
{16 (30 x_t^4 -3 x_t^2 -4) \over 9}\log{x_t+1\over x_t}
- { 8(60 x_t^5 + 90 x_t^4 + 14 x_t^3 -14 x_t^2 -8x_t -9)\over 9(x_t+1)^2 } \, ,
\nl
{\cal M}^{WW(2)}_{t,b} &= 
 {16(3x_t+2)\over 9(x_t+1)^3} {1\over \epsilon} 
+{32 x_t ( 15 x_t^9 - 48 x_t^7 + 52 x_t^5 - 15 x_t^3 + 14 x_t^2 - 6 ) \over 9 (x_t^2-1)^3} \log x_t 
\nl
&\quad
- {32 ( 15 x_t^{10} - 48 x_t^8 + 52 x_t^6 - 12 x_t^4 + 3 x_t^2 - 2) \over 9 (x_t^2-1)^3} \log(x_t+1) 
\nl
&\quad 
- {32 ( 3 x_t^4 -21 x_t^3 + 3 x_t^2 + 9 x_t - 2) \over 9 (x_t^2-1)^3}\log 2 
\nl
&\quad 
+ {8 (180 x_t^8 + 90 x_t^7 - 426 x_t^6 - 183 x_t^5 + 285 x_t^4 + 111 x_t^3 - 220 x_t^2 -4 x_t + 71)\over
27 (x_t^2 -1)^2 (x_t+1) } \, ,
\nl
 {\cal M}^{WW(2)}_{t,c} &=
- 48 x_t^2 \log{x_t+1\over x_t} + {8x_t (6x_t^2 + 9x_t + 2) \over (x_t+1)^2}
\,,
\end{align}
where $x_t= m_t/m_W$. The amplitudes with only light quarks are
\begin{align}
{\cal M}_{U,a}^{WW(0)}&= 
{\cal M}_{U,b}^{WW(0)}
= 
0 \,, 
\quad 
{\cal M}_{U,c}^{WW(0)} = 
-\frac43 \, ,
\nl
{\cal M}^{WW(2)}_{U,a}
&= 
 {\cal M}^{WW(2)}_{U,b}
=   {32\over 9 \epsilon} + {568\over 27} - {64\over 9}\log{2} \,,
\quad 
 {\cal M}^{WW (2)}_{U,c}
=  0 \,,
\end{align}
for $U=u,c$. The results for $Z^0$ exchange with a top quark loop are
\begin{align}\label{eq:pureZamptop}
{\cal M}_{t,a}^{ZZ(0)} &= {\cal M}_{t,b}^{ZZ(0)}
= \big[ c_V^{(t)2}+c_A^{(t)2} \big] \bigg[ 
{ 4y_t^2 (32 y_t^6 -28y_t^4 +14 y_t^2 -1 )\over (4y_t^2-1)^{7/2} } 
\arctan\big(\sqrt{\smash[b]{4y_t^2-1}} \big)
 - {\pi y_t \over 2} 
\nl
&\quad   
+ {4y_t^2(y_t^2-1)(24y_t^2-1)\over 3(4y_t^2-1)^3 } \bigg]
+ \big[ c_V^{(t)2} - c_A^{(t)2} \big] \bigg[ 
{16 y_t^4(24y_t^4-21y_t^2+5) \over (4y_t^2-1)^{7/2} }
\arctan\big(\sqrt{\smash[b]{4y_t^2-1}} \big)
\nl
&\quad 
+{2(144 y_t^6-70y_t^4 + 9y_t^2 -2) \over 3(4y_t^2-1)^3} 
- {3\pi y_t\over 2} 
\bigg] 
\,,
\nl
{\cal M}_{t,c}^{ZZ(0)} &= \big[ c_V^{(t)2}+c_A^{(t)2} \big] \bigg[ 
- { 8y_t^2 (8y_t^2-1)(2y_t^2-1) \over (4y_t^2-1)^{5/2} }
 \arctan\big(\sqrt{\smash[b]{4y_t^2-1}} \big) -{4(24y_t^4-7y_t^2+1)\over 3(4y_t^2-1)^2} + 2\pi y_t 
\bigg] \, ,
\nl
{\cal M}^{ZZ(2)}_{t,a} &= 
 {\cal M}^{ZZ(2)}_{t,b}  =
\big[ c_V^{(t)2}+c_A^{(t)2} \big] \bigg[ {16(480 y_t^8 -420y_t^6 +214 y_t^4-47 y_t^2 + 4)\over 9 (4y_t^2-1)^{7/2}}
\arctan\big(\sqrt{\smash[b]{4y_t^2-1}} \big)
\nl
&\quad 
+ {8(240y_t^6-314y_t^4+92y_t^2-9)\over 9(4y_t^2-1)^3} 
- {10\pi y_t\over 3} 
\bigg]
+ \big[ c_V^{(t)2} - c_A^{(t)2} \big] \bigg[ - {8y_t^2(48y_t^4-34y_t^2+13)\over 9 (4y_t^2-1)^3} 
\nl
&\quad
-  {32y_t^2(16y_t^6-14y_t^4+4y_t^2-1 )\over 3 (4y_t^2-1)^{7/2} } 
\arctan\big(\sqrt{\smash[b]{4y_t^2-1}} \big)
 + {2\pi y_t\over 3} 
\bigg] \, ,
\nl
{\cal M}^{ZZ(2)}_{t,c}&= 
\big\{ \big[ c_V^{(t)2}+c_A^{(t)2} \big]  + 2 \big[ c_V^{(t)2} - c_A^{(t)2} \big] \big\}
 \bigg[ - {32 y_t^2 (16y_t^4 -10 y_t^2 + 3) \over (4y_t^2-1)^{5/2}  } 
\arctan\big(\sqrt{\smash[b]{4y_t^2-1}} \big)
\nl
&\quad
- {16 y_t^2 (8y_t^2-5)\over (4y_t^2-1)^2 } + 8 \pi y_t 
\bigg] \,,
\end{align}
where $y_t = m_t/m_Z$. The amplitudes for $Z^0$ exchange with a light quark loop are
\begin{align}\label{eq:pureZamplight}
{\cal M}_{q,a}^{ZZ(0)} &= {\cal M}_{q,b}^{ZZ(0)}= 0 \,, \quad
{\cal M}_{q,c}^{ZZ(0)} =
 \big[ c_V^{(q)2}+c_A^{(q)2} \big] \bigg[ -\frac43 \bigg] 
\,,
\nl
{\cal M}^{ZZ(2)}_{q,a}
&= 
{\cal M}^{ZZ(2)}_{q,b}
=   \big[ c_V^{(q)2}+c_A^{(q)2} \big] \left[ {32\over 9 \epsilon} + {568\over 27} - {64\over 9}\log{2} \right]\,,
\quad  {\cal M}^{ZZ(2)}_{q,c}=0
\,,
\end{align}
where $q=u,d,s,c,b$. The $\frac1{\epsilon}$ pieces in the above
amplitudes are IR divergences that cancel upon subtraction of the
effective theory contributions, $ {\cal M}_{\rm EFT}^{(S)}$, discussed in Sec.~\ref{sec:EFTsubtract}.
The bare coefficients are then given
by
\begin{align}
c_g^{(S)}{}_{\rm 2BE} &= {\cal M}^{WW(S)}+ {\cal M}^{ZZ(S)} - {\cal M}_{\rm EFT}^{(S)} \, ,
\end{align}
where the remaining $\frac1{\epsilon}$ pieces are UV divergences. 

\subsubsection{Full theory contributions and matching coefficients for admixtures}

For admixtures, the total amplitude receives contributions from other
types of two-boson exchange beyond $WW$ and $ZZ$,
\be {\cal M} = {\cal M}^{WW} + {\cal M}^{ZZ} +  {\cal M}^{\phi_W
\phi_W} + {\cal M}^{\phi_Z \phi_Z}+ {\cal M}^{hh} + {\cal M}^{Z
\phi_Z} +  {\cal M}^{W \phi_W} .  \ee
Let us first consider the singlet-doublet case. In terms of the
electroweak polarization tensors, we find integrals involving nonzero
residual masses,
\begin{align}
i{\cal M}^{WW} &={ i g_2^2  \over  4} c_{\rho \over 2}^2 \int (dL) {1 \over - v \cdot L -\delta_0^{(0)} + i0 } {1 \over (L^2 - m_W^2 + i0)^2} v_\mu v_\nu \bigg[ i\Pi^{\mu \nu}_{(W^+W^+)} (L) + i\Pi^{\mu \nu}_{(W^-W^-)} (L) \bigg] \, ,
\nl
i{\cal M}^{ZZ} &={ i g_2^2  \over 4 c_W^2} c_{\rho \over 2}^2 \int (dL) {1 \over - v \cdot L -\delta_0^{(0)} + i0} {1 \over (L^2 - m_Z^2 +i0)^2} v_\mu v_\nu  i\Pi^{\mu \nu}_{(ZZ)} (L)  \, ,
\nl
i{\cal M}^{hh} &= i a^2   \int (dL) \left[ {s_\rho^2 \over - v \cdot L -\delta_0^{(-)} + i0 } + {c_\rho^2 \over - v \cdot L -\delta_0^{(+)} + i0 } \right] {1 \over (L^2 - m_h^2 + i0)^2}  i\Pi_{(hh)} (L)  \, ,
\nl
i{\cal M}^{\phi_Z \phi_Z} &=i  a^2  s_{\rho \over 2}^2 \int (dL) {1 \over - v \cdot L -\delta_0^{(0)} + i0 } {1 \over (L^2 - m_Z^2 +i0)^2} i\Pi_{(\phi_Z \phi_Z)} (L) \, ,
\nl
i{\cal M}^{\phi_W \phi_W} &=i  a^2  s_{\rho \over 2}^2 \int (dL) {1 \over - v \cdot L -\delta_0^{(0)} + i0 } {1 \over (L^2 - m_W^2 +i0)^2} \bigg[  i\Pi_{(\phi_W^+ \phi_W^+)} (L) +  i\Pi_{(\phi_W^- \phi_W^-)} (L)   \bigg]\, ,
 \nl
 i{\cal M}^{Z \phi_Z} &={ g_2 a \over 4 c_W}  s_{\rho } \int (dL) {1 \over - v \cdot L -\delta_0^{(0)} + i0 } {1 \over (L^2 - m_Z^2 +i0)^2} v_\mu \bigg[  i\Pi^\mu_{(Z \phi_Z)} (L) -  i\Pi^\mu_{(\phi_Z Z)} (L)  \bigg]\, ,
\nl
i{\cal M}^{W \phi_W} &= { i g_2 a \over 4}  s_{\rho } \int (dL) {1 \over - v \cdot L -\delta_0^{(0)} + i0 } {1 \over (L^2 - m_W^2 +i0)^2} v_\mu \bigg[  i\Pi^\mu_{(W^+ \phi_W^+)} (L) -  i\Pi^\mu_{(W^- \phi_W^-)} (L) 
\nl
& \quad 
 + i\Pi^\mu_{(\phi_W^+ W^+)} (L) -  i\Pi^\mu_{(\phi_W^- W^-)} (L) \bigg]\, .
\end{align}
Using the behavior of the polarization tensors under $L \to -L$ and the
identities in (\ref{eq:identities}), we may write these amplitudes in
terms of the integrals defined in (\ref{eq:bosonloop}),
\begin{align}\label{eq:2BEglueSD}
i {\cal M}^{WW} &={  i g_2^2   \over 4} c_{\rho \over 2}^2
\, {\cal I}_{\rm even}(\delta_0^{(0)}, m_W) v_\mu v_\nu  i\Pi^{\mu \nu}_{(W^+W^+)} (L) \, ,
\nl
i {\cal M}^{ZZ} &= {i g_2^2  \over 8 c_W^2} c_{\rho \over 2}^2
\, {\cal I}_{\rm even}(\delta_0^{(0)},m_Z) v_\mu v_\nu  i\Pi^{\mu \nu}_{(ZZ)} (L) \, ,
\nl
i {\cal M}^{hh} &= {i a^2\over 2} s_\rho^2 
\, {\cal I}_{\rm even}(\delta_0^{(-)}, m_h)   i\Pi_{(hh)} (L) 
+ {i a^2 \over 2} c_\rho^2 
\, {\cal I}_{\rm even}(\delta_0^{(+)}, m_h)   i\Pi_{(hh)} (L) \, ,
\nl
i {\cal M}^{\phi_Z \phi_Z} &= {i a^2 \over 2} s_{\rho \over 2}^2
\, {\cal I}_{\rm even}(\delta_0^{(0)},m_Z)   i\Pi_{(\phi_Z\phi_Z)} (L) \, ,
\nl
i {\cal M}^{\phi_W \phi_W} &= i a^2  s_{\rho \over 2}^2
\, {\cal I}_{\rm even}(\delta_0^{(0)},m_W)   i\Pi_{(\phi_W^+\phi_W^+)} (L) \, ,
\nl
i {\cal M}^{Z \phi_Z} &= {g_2 a \over 4 c_W} s_{\rho}
\, {\cal I}_{\rm odd}(\delta_0^{(0)},m_Z)   v_\mu i\Pi^\mu_{(Z\phi_Z)} (L) \, ,
\nl
i {\cal M}^{W \phi_W} &= {ig_2 a \over 2} s_{\rho}
\, {\cal I}_{\rm odd}(\delta_0^{(0)},m_W)   v_\mu i\Pi^\mu_{(W^+ \phi_W^+)} (L) \, .
\end{align}
The required polarization tensors are specified in (\ref{eq:POL}),
and, in particular, the complete set of functions $N_k^{(S)}(L)$ are
explicitly given in (\ref{eq:Nzero}) and (\ref{eq:Ntwo}). Thus, the
general result in (\ref{eq:HFreduction}) for reducing these integrals
may be applied. Each amplitude may be written in the form of
(\ref{eq:gluonamplitudespin}), i.e., in terms of its contributions to
the gluon operators of definite spin. The bare coefficients are then
given by
\begin{align}\label{eq:2BEgluesum}
c_g^{(S)}{}_{\rm 2BE} &= {\cal M}^{(S)}{}^{WW} 
+ {\cal M}^{(S)}{}^{ZZ} 
+ {\cal M}^{(S)}{}^{hh} 
+ {\cal M}^{(S)}{}^{\phi_Z \phi_Z} 
\nl
& \quad 
+ {\cal M}^{(S)}{}^{\phi_W \phi_W} 
+ {\cal M}^{(S)}{}^{Z \phi_Z} 
+ {\cal M}^{(S)}{}^{W \phi_W} - {\cal M}^{(S)}_{\rm EFT}  \, ,
\end{align}
where the remaining $\frac1{\epsilon}$ pieces are UV divergences. We may 
again organize each contribution in the previous equation in terms of the 
quark flavors in the loop, and the $a$-,
$b$-, and $c$-type gluon attachments, as we have done in
(\ref{eq:ampcharge}) and (\ref{eq:ampneutral}).

For the triplet-doublet case we find, 
\begin{align}
i {\cal M}^{WW} &={  i g_2^2   \over 16} s_{\rho}^2
\, {\cal I}_{\rm even}(\delta_0^{(+)}, m_W) v_\mu v_\nu  i\Pi^{\mu \nu}_{(W^+W^+)} (L)
\nl
&\quad
+ {  i g_2^2   \over 4} \big( 1 + s_{\rho \over 2}^2 \big)^2
\, {\cal I}_{\rm even}(\delta_0^{(-)}, m_W) v_\mu v_\nu  i\Pi^{\mu \nu}_{(W^+W^+)} (L) \, ,
\nl
i {\cal M}^{ZZ} &= {i g_2^2  \over 8 c_W^2} c_{\rho \over 2}^2
\, {\cal I}_{\rm even}(\delta_0^{(0)},m_Z) v_\mu v_\nu  i\Pi^{\mu \nu}_{(ZZ)} (L) \, ,
\nl
i {\cal M}^{hh} &= {i a^2\over 2} s_\rho^2 
\, {\cal I}_{\rm even}(\delta_0^{(-)}, m_h)   i\Pi_{(hh)} (L) 
+ {i a^2 \over 2} c_\rho^2 
\, {\cal I}_{\rm even}(\delta_0^{(+)}, m_h)   i\Pi_{(hh)} (L) \, ,
\nl
i {\cal M}^{\phi_Z \phi_Z} &= {i a^2 \over 2} s_{\rho \over 2}^2
\, {\cal I}_{\rm even}(\delta_0^{(0)},m_Z)   i\Pi_{(\phi_Z\phi_Z)} (L) \, ,
\nl
i {\cal M}^{\phi_W \phi_W} &= i a^2  
\, {\cal I}_{\rm even}(\delta_0^{(+)},m_W)   i\Pi_{(\phi_W^+\phi_W^+)} (L) \, ,
\nl
i {\cal M}^{Z \phi_Z} &= {g_2 a \over 4 c_W} s_{\rho}
\, {\cal I}_{\rm odd}(\delta_0^{(0)},m_Z)   v_\mu i\Pi^\mu_{(Z\phi_Z)} (L) \, ,
\nl
i {\cal M}^{W \phi_W} &= {ig_2 a \over 2} s_{\rho}
\, {\cal I}_{\rm odd}(\delta_0^{(+)},m_W)   v_\mu i\Pi^\mu_{(W^+ \phi_W^+)} (L) \, .
\end{align}
The rest of the analysis proceeds as above, using the same
polarization tensors and integral reduction method. We check for both
types of admixtures that the expected results are recovered 
upon taking the pure-case limits described in Sec.~\ref{sec:limits}.

\subsection{Effective theory amplitudes and infrared regulator}\label{sec:EFTsubtract}

In the computation of both pure- and mixed-case amplitudes above, we have neglected subleading 
corrections of $\order(m_q/m_W)$ by Taylor expanding integrands about vanishing light quark masses.%
\footnote{For matching onto quark operators, we of course include the leading $m_q$ 
factor appearing in ${\cal O}_q^{(0)}$ and ${\cal O}_q^{(2)}$.
For matching onto gluon operators we may neglect light quark masses.} 
This requires a regulator to control IR divergences (the full theory 
diagrams in Figs.~\ref{fig:gluon1match} and~\ref{fig:gluon2match} are UV finite but 
the projection onto the spin-2 operator ${\cal O}_g^{(2)}$ is IR divergent).
 
It is technically simplest to compute the full and effective theory amplitudes 
using dimensional regularization as IR regulator.
Effective theory loop diagrams on the right hand sides of Figs.~\ref{fig:gluon1match} and~\ref{fig:gluon2match}
then result in dimensionfull but scaleless integrals that are required to vanish. 
Upon subtracting the effective theory amplitude, remaining $1/\epsilon$ pieces
in matching coefficients are identified as UV divergences. 

We have obtained identical renormalized matching coefficients by retaining
light quark masses, $m_q \ne 0$, as an alternative IR regulator. In this
scheme, the effective theory loop diagrams on the right-hand
side of Figs.~\ref{fig:gluon1match} and~\ref{fig:gluon2match} yield nonvanishing contributions. The full theory diagrams on the left-hand side
are correspondingly modified so that, upon subtracting the effective theory amplitude, consistent results are obtained.

\section{Results for matching coefficients \label{sec:pheno}}

We may now collect the results of the preceding analysis of quark and
gluon matching to present the bare coefficients of the effective
theory at the weak scale. We have analyzed the Wilson coefficients 
of the effective theory described by (\ref{eq:LWIMPSM})
in terms of contributions from exchanges of
one or two electroweak bosons, as expressed in (\ref{eq:cBE}). The
results for one-boson exchange matching to quark and gluon operators
are given by (\ref{eq:1BEquark}) and (\ref{eq:1BEglue}),
respectively. The results for two-boson exchange matching to quark and
gluon operators are given by summing contributions of the form
(\ref{eq:2BEquark}) and (\ref{eq:gluonamplitudespin}), respectively.

For pure cases, the results for the bare matching coefficients are as follows,
\begin{align}\label{eq:purebare}
c_{U}^{(0)} &= {\pi \Gamma(1+\epsilon) g_2^4 \over (4\pi)^{2-\epsilon} } \Bigg\{ 
- {m_W^{-3-2\epsilon}\over 2 x_h^2}\left[ {\cal C}_W + {{\cal C}_Z \over c_W^3}\right]  
+  {m_Z^{-3-2\epsilon} {\cal C}_Z \over 8 c_W^4} \big[ c_V^{(U)2} - c_A^{(U)2} \big]
+ \order(\epsilon) 
\Bigg\} 
\,,
\nl
c_{D}^{(0)} &= {\pi \Gamma(1+\epsilon) g_2^4 \over (4\pi)^{2-\epsilon} } \Bigg\{ 
- {m_W^{-3-2\epsilon}\over 2 x_h^2}\left[ {\cal C}_W + {{\cal C}_Z \over c_W^3}\right]  
+  {m_Z^{-3-2\epsilon} {\cal C}_Z \over 8 c_W^4} \big[ c_V^{(D)2} - c_A^{(D)2} \big]
\nl
&\quad
-  \delta_{Db} \, m_W^{-3-2\epsilon} {\cal C}_W    {x_t \over 8(x_t+1)^3}
+ \order(\epsilon) 
\Bigg\} 
\,,
\nl
c_{g}^{(0)} &= {\pi [\Gamma(1+\epsilon)]^2 g_2^4 g^2 \over (4\pi)^{4-2\epsilon}}\Bigg\{
{m_W^{-3-4\epsilon} \over 2} \Bigg[ 
{1\over 3 x_h^2}\left[ {\cal C}_W + { {\cal C}_Z \over  c_W^3}\right]   
+ {\cal C}_W \left[ \frac13 + {1 \over 6 (x_t+1)^2 }   \right]
\Bigg] 
\nl
&\quad
+  {m_Z^{-3-4\epsilon} {\cal C}_Z\over 64 c_W^4}
\Bigg[
 4 \big[ c_V^{(D)2} + c_A^{(D)2} \big] 
 + \big[ c_V^{(U)2} + c_A^{(U)2} \big]
\bigg[ \frac83
+  {32 y_t^6 (8 y_t^2 -7) \over (4y_t^2-1)^{7/2} }\arctan\big(\sqrt{\smash[b]{4y_t^2-1}} \big) 
\nl
&\quad
- \pi y_t 
+{4(48y_t^6-2y_t^4+9y_t^2-1) \over 3(4y_t^2-1)^3} 
\bigg] 
+ \big[ c_V^{(U)2} - c_A^{(U)2} \big]
\bigg[ 3\pi y_t 
- {4(144y_t^6 -70y_t^4+9y_t^2-2)\over 3(4y_t^2-1)^3} 
\nl
&\quad
- {32 y_t^4(24y_t^4-21 y_t^2+5) \over (4y_t^2-1)^{7/2} }\arctan\big(\sqrt{\smash[b]{4y_t^2-1}} \big)
\bigg] 
\Bigg]
+ \order(\epsilon) 
\Bigg\} \, ,
\nl
c_{U}^{(2)} &= {\pi \Gamma(1+\epsilon)  g_2^4 \over (4\pi)^{2-\epsilon} }  \Bigg\{
\bigg[ m_W^{-3-2\epsilon} {\cal C}_W + { m_Z^{-3-2\epsilon} {\cal C}_Z \over 2 c_W^4} 
\big[ c_V^{(U)2}+c_A^{(U)2} \big]  \bigg]  \bigg[ 
{1\over 3} + \left( {11\over 9} -{2\over 3}\log 2  \right) \epsilon 
\bigg]+ \order(\epsilon^2) 
\Bigg\} \, ,
\nl
c_{D}^{(2)} &= {\pi \Gamma(1+\epsilon)  g_2^4 \over (4\pi)^{2-\epsilon} }  \Bigg\{
\bigg[ m_W^{-3-2\epsilon} {\cal C}_W + { m_Z^{-3-2\epsilon} {\cal C}_Z \over 2 c_W^4} 
\big[ c_V^{(D)2}+c_A^{(D)2} \big]  \bigg]  \bigg[ 
{1\over 3} + \left( {11\over 9} -{2\over 3}\log 2  \right) \epsilon 
\bigg]
\nl
&\quad
+ \delta_{Db} \, { m_W^{-3-2\epsilon} {\cal C}_W \over 2}  
  \bigg[ {(3x_t +2) \over 3(x_t+1)^3} - \frac23
+ \bigg(
{ 2 x_t(7x_t^2 -3) \over 3(x_t^2-1)^3 }\log x_t 
- {2(3x_t+2)\over 3(x_t+1)^3}\log 2
\nl
&\quad 
- {2( 25x_t^2-2x_t -11) \over 9(x_t^2-1)^2(x_t+1) } 
- {22 \over 9} + {4 \over 3}\log 2
\bigg)\epsilon \bigg]  
+ \order(\epsilon^2) 
\Bigg\} \, ,
\nl
c_g^{(2)} &= 
{\pi [\Gamma(1+\epsilon)]^2 g_2^4 g^2 \over (4\pi)^{4-2\epsilon}}\Bigg\{
{m_W^{-3-4\epsilon} {\cal C}_W \over 2} 
\bigg[
 -{16 \over 9 \epsilon} -{284 \over 27} +{32 \over 9} \log2
- {2 (3x_t + 2)\over 9(x_t+1)^3} {1\over \epsilon} 
\nl
&\quad 
+ {8 ( 6 x_t^8 - 18 x_t^6 + 21 x_t^4 - 3x_t^2 -2 )\over 9 (x_t^2-1)^3}\log(x_t+1) 
+ {4 (3 x_t^4 -21 x_t^3 + 3 x_t^2 + 9x_t - 2)\over 9(x_t^2-1)^3}\log 2 
\nl
&\quad
- {4 ( 12 x_t^8 - 36 x_t^6 + 39 x_t^4 + 14 x_t^3- 9 x_t^2 - 6x_t - 2 )\over 9(x_t^2-1)^3}\log x_t 
\nl
&\quad
- {144 x_t^6 + 72 x_t^5 -312 x_t^4 - 105 x_t^3 - 40 x_t^2 + 47 x_t + 98 \over 27 (x_t^2-1)^2(x_t+1)} 
\bigg]
\nl
&\quad 
+  {m_Z^{-3-4\epsilon} {\cal C}_Z \over 64 c_W^4}
\Bigg[ 
\bigg[ 8 
\big[  c_V^{(U)2} + c_A^{(U)2} \big]
+ 12 \big[ c_V^{(D)2} + c_A^{(D)2} \big]
\bigg] 
\bigg[
-{16\over 9\epsilon} - {284\over 27} + {32\over 9}\log{2} 
\bigg]
\nl
&\quad 
+ \big[ c_V^{(U)2}+c_A^{(U)2}\big]  \bigg[ 
{ 128 ( 24 y_t^8 -21 y_t^6 -4y_t^4 + 5y_t^2 -1  )  \over 9 (4y_t^2-1)^{7/2} } \arctan\big(\sqrt{\smash[b]{4y_t^2-1}} \big)
- {4 \pi y_t \over 3}
\nl
&\quad 
+ {16 ( 48 y_t^6 +62 y_t^4 -47 y_t^2 +9 )\over 9(4y_t^2-1)^3 } 
 \bigg]
 + \big[ c_V^{(U)2}-c_A^{(U)2} \big] \bigg[ 
{16 y_t^2 ( 624 y_t^4 - 538 y_t^2 +103 ) \over 9 (4y_t^2-1)^3} 
- {52 \pi y_t \over 3} 
\nl
&\quad 
+{ 128 y_t^2 (104 y_t^6 -91 y_t^4 + 35 y_t^2 - 5) \over 3 (4y_t^2-1)^{7/2} }
\arctan\big(\sqrt{\smash[b]{4y_t^2-1}} \big)
\bigg] 
\Bigg]
+ \order(\epsilon) 
\Bigg\} \,,
\end{align}
where, as before, $x_t= m_t/m_W$ and $y_t=m_t/m_Z$. Above, the Kronecker delta, $\delta_{Db}$, is equal to unity for $D=b$, and vanishes for $D=d,s$.
The pure triplet
(doublet) results are given by setting ${\cal C}_W = 2$ and ${\cal
C}_Z =0$ (${\cal C}_W = 1/2$ and ${\cal C}_Z =1/4$). 
The renormalization of the theory involving these bare coefficients
will be detailed in a forthcoming paper~\cite{part2}. In particular,
the relation between the bare coefficient $c_g^{(2)}$ given above and
the renormalized coefficient $c_g^{(2)}(\mu)$ involves a
nontrivial subtraction requiring the $\order(\epsilon)$ part of
$c^{(2)}_{q}$ which we have retained.

\begin{figure}[t]
\centering
\includegraphics[scale=0.45]{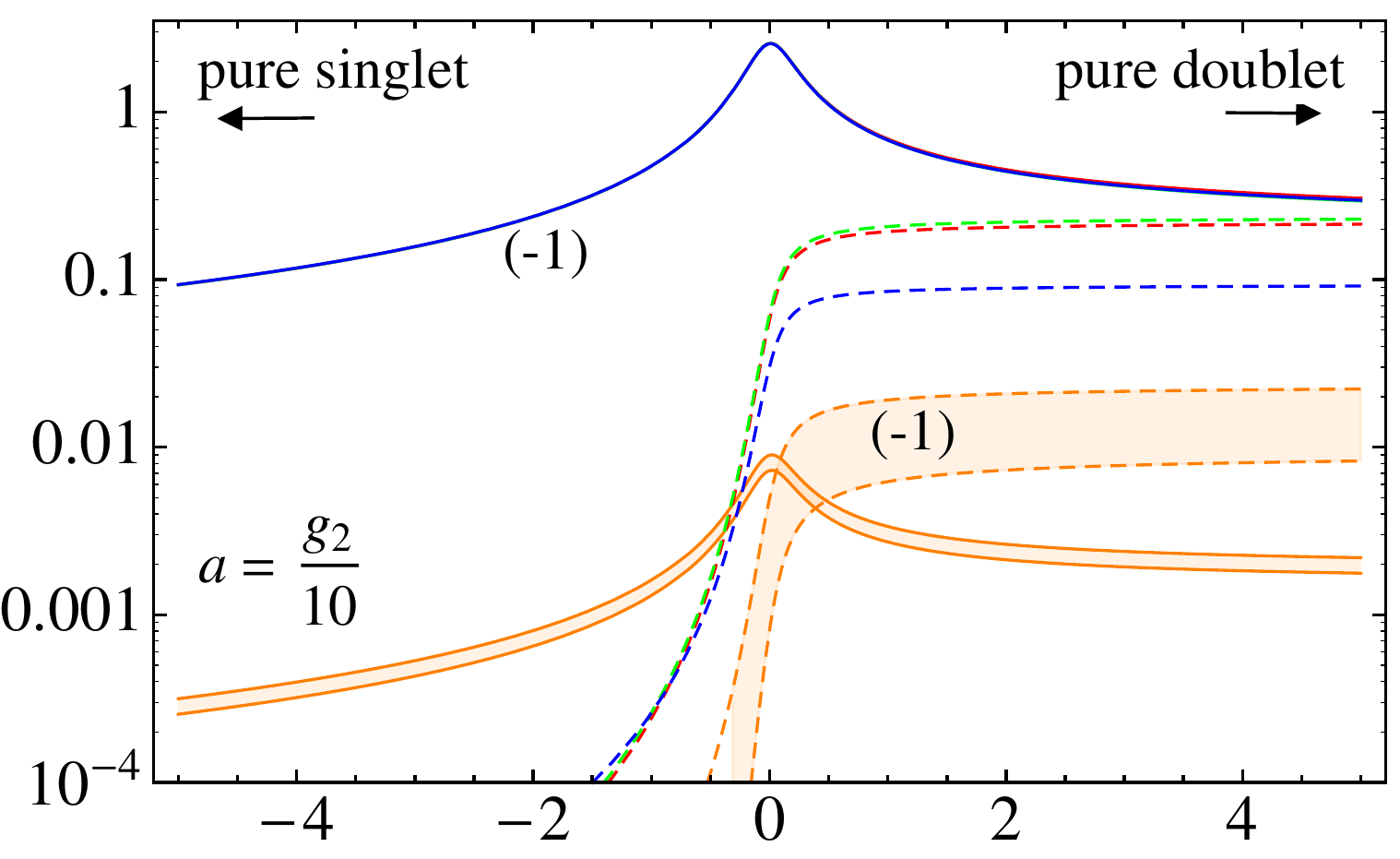}
\hspace*{12mm}\includegraphics[scale=0.45]{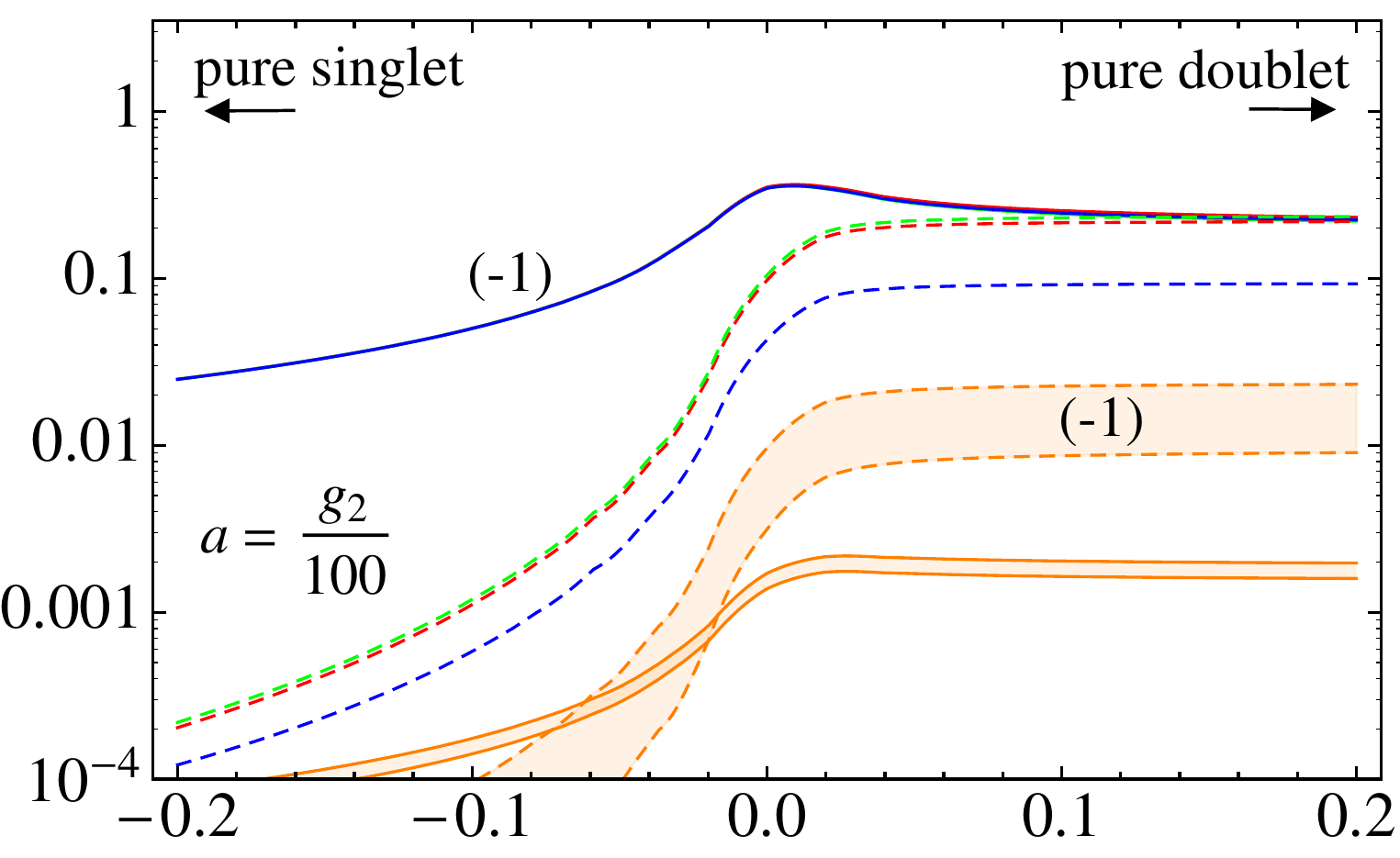}
\centering
\begin{minipage}{0cm}
\vspace*{0.5cm}\hspace*{-3.9cm}$\Delta/m_W$\hspace*{-9.5cm}$\Delta/m_W$
\end{minipage}
\vspace*{0.5cm}
\begin{minipage}{0cm}
\vspace*{-4.6cm}\hspace*{-16.1cm}\rotatebox{90}{$ \pm c \, {\pi \alpha_2^2 \over m_W^3} $}
\hspace*{7.45cm}\rotatebox{90}{$\pm c \, {\pi \alpha_2^2 \over m_W^3} $}
\end{minipage}
\centering
\includegraphics[scale=0.45]{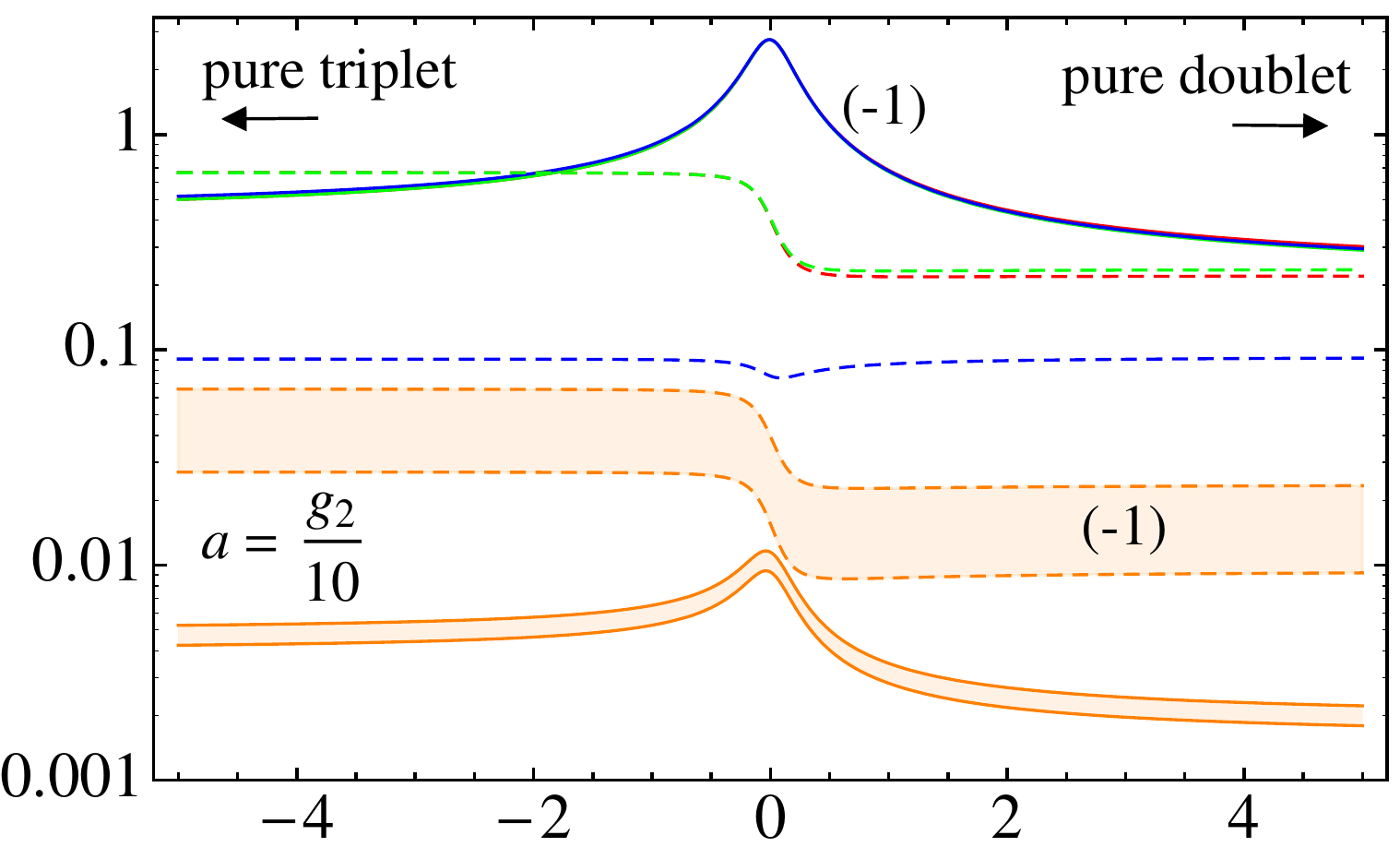}
\hspace*{12mm}\includegraphics[scale=0.45]{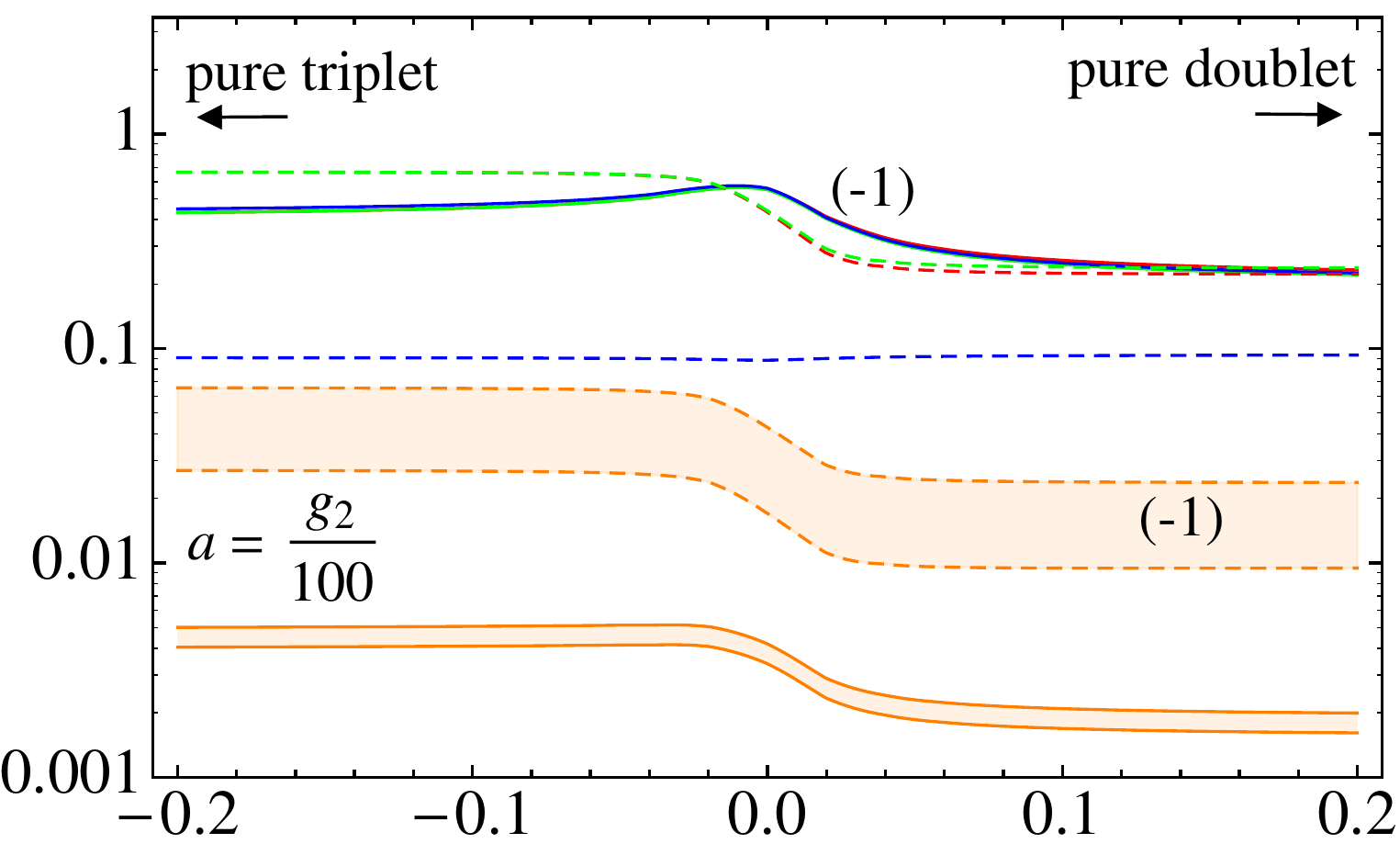}
\centering
\begin{minipage}{0cm}
\vspace*{0.5cm}\hspace*{-3.9cm}$\Delta/m_W$\hspace*{-9.5cm}$\Delta/m_W$
\end{minipage}
\begin{minipage}{0cm}
\vspace*{-4.6cm}\hspace*{-16.1cm}\rotatebox{90}{$ \pm c \, {\pi \alpha_2^2 \over m_W^3} $}
\hspace*{7.45cm}\rotatebox{90}{$\pm c \, {\pi \alpha_2^2 \over m_W^3} $}
\end{minipage}
\caption{\label{fig:cplots} \emph{Renormalized} coefficients (with
$\pi \alpha_2^2 /m_W^3$ extracted) for the singlet-doublet (upper
panels) and triplet-doublet (lower panels) mixtures as a function of
the respective mass splittings $\Delta = (M_S - M_D)/2$ and $\Delta =
(M_T - M_D)/2$, in units of $m_W$. The panels on the left (right) use
$a = g_2/10$ ($a=g_2/100$). The negative
coefficients $c_q^{(0)}$ and $c_g^{(2)}$ are presented with opposite sign,
as indicated by $(-1)$. The
solid red, green, and blue lines are respectively for
$-c_{U=u,c}^{(0)}$, $-c_{D=d,s}^{(0)}$, and $-c_b^{(0)}$. The dashed
red, green, and blue lines are respectively for $c_{U=u,c}^{(2)}$,
$c_{D=d,s}^{(2)}$, and $c_b^{(2)}$. Some quark matching coefficients 
appear degenerate. The
orange band with solid borders is $c_g^{(0)}$, and the orange band
with dashed borders is $-c_g^{(2)}$. The band thickness represents
renormalization scale variation, taking $m_W^2 /2 < \mu_t^2 < 2m_t^2$ \cite{Hill:2011be}. We indicate the
pure-case limits at large $| \Delta |$. }
\end{figure}

The results for admixtures are similarly obtained by collecting
contributions to the coefficients specified in (\ref{eq:cBE}). For
example, the amplitudes in (\ref{eq:1BEquarkSDAMP}) for a
singlet-doublet admixture, combined with the integrals defined in
Appendix~\ref{sec:self}, specify $c_{q}^{(0)}{}_{\rm 1BE}$ through
(\ref{eq:1BEquark}), and $c_g^{(0)}{}_{\rm 1BE}$ through
(\ref{eq:1BEglue}). The coefficients $c_{q}^{(S)}{}_{\rm 2BE}$ are
specified in (\ref{eq:sumquark2BE}) in terms of the results in
(\ref{eq:c2BESD}), which require the integrals in
Appendix~\ref{sec:box}. Finally, $c_g^{(S)}{}_{\rm 2BE}$ is specified in
(\ref{eq:2BEgluesum}) in terms of the amplitudes in
(\ref{eq:2BEglueSD}) which require the polarization tensors in
(\ref{eq:POL}), the basis reduction in (\ref{eq:HFreduction}), and the
integrals in Appendix~\ref{sec:appC}. 

The matching coefficients for admixtures are functions of the mass
splitting $\Delta$ and the coupling $a$, as defined in
(\ref{eq:SDparameters}) for the singlet-doublet mixture. We illustrate
numerical values in Fig.~\ref{fig:cplots} for both the singlet-doublet
and triplet-doublet mixtures. Numerical inputs are collected in
Table~\ref {tab:inputs} of Appendix~\ref{app:inputs}. Depending on the
value of $a$, the $\order(\alpha_2^1)$ tree-level Higgs exchange
contribution to the spin-0 coefficients may dominate near $\Delta
=0$. When $m_W/\Delta$ suppression is significant, the ${\cal
O}(\alpha_2^2)$ loop contributions dominate. The curves
approach the correct pure-case values upon taking the limits described
in Sec.~\ref{sec:limits}. In particular, the coefficients vanish in
the pure singlet limit.

The contributions of these coefficients to scattering cross sections
depend on the detailed mapping onto the low-energy $n_f=3$ flavor
theory through renormalization group running and heavy quark threshold
matching, and on the evaluation of nucleon matrix elements at a low
scale, $\mu \sim 1\, {\rm GeV}$. These effects enhance the
contribution from certain coefficients, upsetting the $\alpha_s$
counting reflected in the relative magnitudes of the high-scale
coefficients. One example is the enhancement of the spin-0 gluon
contribution due both to 
a large anomalous dimension in the RG running, and to the large nucleon matrix  
element of the scalar gluon operator~\cite{Shifman:1978zn}. 
Another example is the enhanced
impact of numerically subleading contributions due to a partial cancellation
at leading order. The relative signs between high-scale coefficients in
Fig.~\ref{fig:cplots}, combined with details of the mapping onto
low-energy coefficients and evaluation of matrix elements, lead to a
cancellation between the spin-0 and spin-2 amplitude
contributions\cite{Hisano:2011cs,Hill:2011be}. Therefore,
a robust determination of DM-nucleon scattering cross sections demands
a careful analysis of the complete set of leading operators in
(\ref{eq:ops}).

The coefficient $c_g^{(2)}$ has been omitted in previous works~\cite{Hisano:2010ct,Hisano:2011cs}. Due to a cancellation between spin-0 and spin-2 amplitude contributions to cross sections, the effect of neglecting $c_g^{(2)}$ ranges from a factor of a few to an order of magnitude difference in cross sections. For the pure-doublet and pure-triplet states, neglecting $c_g^{(2)}$ leads to an $\order(10-20 \%)$ shift in the spin-2 amplitude, depending on the choice of renormalization scale, and an underestimation of its perturbative uncertainty by $\order(70 \%)$. For comparison, neglecting $c_q^{(2)}$ for $q=b,c,s,d,u$ shifts the spin-2 amplitude by $\order(1 \%)$, $\order(10 \%)$, $\order(10 \%)$, $\order(30 \%)$, and $\order(50 \%)$, respectively.

\section{Summary \label{sec:summary}}

The present analysis focused on weak-scale matching conditions 
necessary for robustly computing WIMP-nucleon interactions, both in 
specified UV completions involving electroweak-charged DM, 
and in the model-independent heavy WIMP limit. 
Careful computation of competing Standard
Model contributions is necessary to estimate the correct order of
magnitude of scattering cross sections in many simple and motivated
models of DM.  For example, a simple dimensional estimate of
the cross section for spin-independent, low-velocity scattering of a 
pure-state WIMP on a nucleon yields%
\footnote{Cross sections of this magnitude were
obtained in previous 
estimates that missed the cancellation
between spin-0 and spin-2 amplitude contributions (and ignored gluon
contributions)~\cite{Cirelli:2005uq}.} 
\be
\sigma_{\rm SI} \sim  { \alpha_2^4 m_N^4 \over m_W^2 } \bigg( {1 \over m_W^2} , \, {1 \over m_h^2} \bigg)^2 \sim 10^{-45} {\rm \, cm}^2 \, .
\ee
Cross sections of this order of magnitude are currently being probed by 
direct detection searches (e.g., see Refs.~\cite{Cheung:2012qy} for detection prospects computed using tree-level cross sections). 
However, a cancellation
between spin-0 and spin-2 amplitude contributions leads to much
smaller cross section values for motivated candidates such as the pure
wino ($\sigma_{\rm SI} \sim 10^{-47} {\rm \, cm}^2$) and the pure
higgsino ($\sigma_{\rm SI} \lesssim 10^{-48} {\rm \, cm}^2$) of
supersymmetric SM extensions. This cancellation demands a careful
analysis of perturbative contributions from weak-scale matching
amplitudes presented here, e.g., the inclusion of the spin-2 gluon
contribution, and of remaining theoretical and input uncertainties, which will
be discussed in a companion paper~\cite{part2}. Robust predictions for
the cross sections of the pure triplet, pure doublet, singlet-doublet
admixture, and triplet-doublet admixture can be found in
Refs.~\cite{Hill:2011be}. Given the matching coefficients
in (\ref{eq:purebare}), the cross sections for pure states with
arbitrary electroweak quantum numbers can also be computed.

Although we find that cancellations are generic, their severity
depends on SM parameters and on properties of DM such as its
electroweak quantum numbers. The presence of additional low-lying
states could also have impact, and the formalism for weak-scale
matching presented here can be readily extended to investigate such
scenarios. For example, including a second Higgs doublet in the
pure-state analysis simply requires modification of the vertices in
the amplitudes computed in Figs.~\ref{fig:quark1match} and
\ref{fig:gluon1match}. An extra Higgs boson modifies the spin-0
amplitude, and could potentially weaken 
the cancellation between spin-0 and spin-2
amplitudes. The case where the second Higgs-like doublet itself plays the role of DM (e.g., ``inert Higgs DM"~\cite{Klasen:2013btp}) is related to the pure-doublet case in the heavy WIMP limit by heavy particle universality.

While we have focused here on the case of a heavy, self-conjugate
WIMP, deriving from one or two electroweak multiplets, much of the
formalism applies more generally. The construction of the heavy
particle effective theories in Sec.~\ref{sec:lagrangians} could be
straightforwardly extended to include power corrections, and other
light states within the context of specific UV completions. 
The generalized electroweak polarization tensors obtained
through background field techniques depend only on SM parameters, and
hence can be applied for gluon operator matching in general
DM scenarios. Within the context of heavy particle effective
theories, the new integral basis evaluated here may be applied to
other processes such as low-energy lepton-nucleon scattering~\cite{Hill:2012rh}.

Separating improvable SM uncertainties from DM model dependence demands
precise theoretical formalism. The focus of the present paper is on
the systematic treatment of weak-scale matching calculations, with
particular attention paid to the heavy-particle limit. The remaining
analysis below the weak scale may be applied to a broader class of
theories, and is the subject of a forthcoming
paper~\cite{part2}. There, we present the necessary ingredients to systematically map
high-scale matching coefficients onto the low-energy theory ($n_f=3$
or $n_f=4$ flavor QCD plus interactions with DM) where hadronic matrix
elements are evaluated.

\vskip 0.2in
\noindent
{\bf Acknowledgements}
\vskip 0.1in
\noindent We acknowledge useful discussions with C.E.M.~Wagner. M.S. also thanks T.~Cohen and J.~Kearney for useful discussions.
This work was supported by the United States Department of Energy under Grant
No.~DE-FG02-13ER41958.
M.S. is supported by a Bloomenthal Fellowship.

\begin{appendix}

\section{The singlet-doublet mixture \label{sec:SDfull}}

The heavy-particle lagrangians in Sec.~\ref{sec:lagrangians} may be
obtained from a manifestly relativistic lagrangian by performing field
redefinitions at tree level. Consider the case of a singlet-doublet
mixture (see also~\cite{Cohen:2011ec}),
\be
{\cal L} = {\cal L}_{\rm SM} + \frac12 \bar{b} (i\dslash - M_1 ) b 
 + \bar{\psi} ( i\Dslash - M_2 )\psi 
- ( y \, \bar{b} P_L H^\dagger \psi  + y^\prime \, \bar{b} P_L H^T \psi^c + {\rm h.c.} )\, ,
\ee 
where $b$ is a gauge singlet (Majorana) fermion represented as a Dirac spinor with $b^c=b$, and $\psi$ is a Dirac fermion in
the $(\bm{2}, 1/2)$ representation of $SU(2)_W \times U(1)_Y$. In the above equation, $P_{\rm R,L} = (1 \pm \gamma_5)/2$, and we
have included all renormalizable gauge-invariant interactions
involving the SM Higgs field. Expressing the result in terms of
Majorana combinations,
\be\label{eq:lambdai}
\lambda_{1} = {1\over \sqrt{2}} \left( \psi + \psi^c \right) \,,
\quad 
\lambda_{2} = {i\over \sqrt{2}} \left( \psi - \psi^c \right) \,,
\ee
and collecting the fermions in the column vector $\lambda = (b,
\lambda_{1}, \lambda_{2})$, we may write the interactions with the
Higgs field as
\begin{align}
{\cal L}_{H \bar{\lambda} \lambda} &= -{1\over\sqrt{2}}\bar{b}{1-\gamma_5\over 2} 
\left[ ( y H^\dagger + y^\prime H^T)\lambda_{1} - i(y H^\dagger - y^\prime H^T)\lambda_{2} 
\right]
+ {\rm h.c.}
\nl
& \equiv -\frac12 \bar{\lambda} \bigg[ f(H) + i\gamma_5 g(H) \bigg] \lambda \, ,
\end{align}
with
\begin{align}
f(H) &= 
{a_1 \over \sqrt{2}}\left(\begin{array}{ccc} 
0 & H^\dagger + H^T & i(H^T-H^\dagger) \\
H+H^* & \mathbb{0}_2 & \mathbb{0}_2 \\
i(H-H^*) & \mathbb{0}_2 & \mathbb{0}_2 
\end{array}
\right)
+ {a_2 \over \sqrt{2}}\left(\begin{array}{ccc} 
0 & -i(H^T - H^\dagger) & H^T+H^\dagger \\
-i(H-H^*) & \mathbb{0}_2 & \mathbb{0}_2 \\
H+H^* & \mathbb{0}_2 & \mathbb{0}_2 
\end{array}
\right) \,,
\nl
g(H) &= 
{b_1 \over \sqrt{2}}\left(\begin{array}{ccc} 
0 & -i(H^T-H^\dagger) & H^T+H^\dagger \\
-i(H-H^*) & \mathbb{0}_2 & \mathbb{0}_2 \\
H+H^* & \mathbb{0}_2 & \mathbb{0}_2 
\end{array}
\right)
+ {b_2 \over \sqrt{2}}\left(\begin{array}{ccc} 
0 & H^\dagger + H^T & i(H^T-H^\dagger) \\
H+H^* & \mathbb{0}_2 & \mathbb{0}_2 \\
i(H-H^*) & \mathbb{0}_2 & \mathbb{0}_2 
\end{array}
\right)\,.
\end{align} 
The real parameters $a_i$ and $b_i$ are given by 
\be
a_1 = \frac12 {\rm Re}(y+y^\prime) \,, \quad
a_2 = \frac12 {\rm Im}(y-y^\prime) \,, \quad
b_1 = \frac12 {\rm Re}(y-y^\prime) \,, \quad
b_2 = -\frac12 {\rm Im}(y+y^\prime) \,.
\ee
We employ phase redefinitions of $b$, $\psi_L$ and $\psi_R$ to ensure
that $M_1$ and $M_2$ are real and positive.%
\footnote{An additional
phase redefinition could be used to eliminate $a_1$, $a_2$, $b_1$ or $b_2$.} 
The gauge generators will be those
given in (\ref{eq:higgsinogauge}), extended trivially to include the
singlet. Upon performing the tree-level field redefinition
\be 
\lambda = \sqrt{2} e^{-i(M-\delta M) v\cdot x} ( h_v + H_v ) \,,
\ee
where the fields $h_v$ and $H_v$ obey $\vslash h_v =h_v$ and $\vslash H_v = -H_v$, we obtain the heavy-particle lagrangian in (\ref{eq:heavy}). It follows from $\lambda^c = \lambda$ that the resulting lagrangian is invariant under the simultaneous transformations in (\ref{eq:vparity}). Note that $f(H)$ is the only term surviving the projection from the condition $\vslash h_v = h_v$. The remaining analysis follows that of Sec.~\ref{sec:SDmatrices}.

\section{Self energy integrals and Standard Model two-point functions \label{sec:self}}

Here and in the following sections we use the notation
\be
[c_\epsilon] = { i \Gamma(1+\epsilon) \over (4\pi)^{2-\epsilon} } \, , \quad (dL) = {d^dL \over (2\pi)^d}.
\ee
The self-energies in Sec.~\ref{sec:renscheme} and the $h
\bar{\chi}\chi$ three-point functions in Sec.~\ref{sec:1BEquark}
require the following integrals,
\begin{align} 
I_1(\delta,m) &= \int (dL) {1\over v\cdot L - \delta + i0} {1\over (L^2 -m^2 + i0)^2 } 
\nl
&= {\partial \over \partial m^2} I_3(\delta,m) 
\nl
&= [c_\epsilon] m^{-2\epsilon} \bigg\{ {2\over \sqrt{m^2-\delta^2-i0} }\bigg[ \arctan\left( \delta \over\sqrt{m^2-\delta^2-i0}\right) - {\pi\over 2} 
\bigg] + \order(\epsilon) \bigg\} \,,
\nl
I_2(\delta,m) &= \int (dL) v\cdot L {1\over v\cdot L - \delta + i0} {1\over (L^2-m^2+i0)^2 }
\nl
&= \delta I_1(\delta,m) + {i\over (4\pi)^2} B_0(0,m,m) 
\nl
&= [c_\epsilon] m^{-2\epsilon} \bigg\{ {1\over \epsilon} 
+ {2\delta \over \sqrt{m^2-\delta^2-i0}} \bigg[ \arctan\left( \delta \over\sqrt{m^2-\delta^2-i0}\right) - {\pi\over 2} 
\bigg]  + \order(\epsilon)
\bigg\} \,,
\nl
I_3(\delta,m) &= \int (dL) {1\over v\cdot L - \delta + i0} {1\over L^2-m^2+i0} 
\nl
&=[c_\epsilon] m^{-2\epsilon} \bigg\{ -{2\delta\over \epsilon} 
+ 4\sqrt{m^2-\delta^2-i0} \bigg[ 
\arctan\left( \delta \over\sqrt{m^2-\delta^2-i0}\right) - {\pi\over 2} 
\bigg]
- 4 \delta  + \order(\epsilon)
\bigg\} \,,
\nl
I_4(\delta_1,\delta_2,m) &= \int (dL) {1\over v\cdot L - \delta_1 + i0} {1\over v\cdot L - \delta_2 + i0}
{1\over L^2 - m^2 + i0}  \,.
\end{align}
For $I_4(\delta_1, \delta_2, m)$, let us specialize to $\delta_2=0$ or $\delta_1=\delta_2$,
\begin{align}
I_4(\delta,0,m) &= {1\over \delta} \big[ I_3(\delta,m) - I_3(0,m) \big]
\nl
&=[c_\epsilon] m^{-2\epsilon} \bigg\{ -{2\over\epsilon} + {4\sqrt{m^2-\delta^2-i0}\over \delta} 
\bigg[ \arctan\left( \delta \over\sqrt{m^2-\delta^2-i0}\right) - {\pi\over 2} 
\bigg]
\nl
&\quad 
-4 
+ {2\pi m\over \delta} 
+ \order(\epsilon)
 \bigg\} \,,
\nl
I_4(\delta,\delta,m) &= {\partial \over \partial \delta} I_3(\delta,m) 
\nl
&= 
[c_\epsilon] m^{-2\epsilon} \bigg\{ -{2\over \epsilon} 
- {4\delta \over\sqrt{m^2-\delta^2-i0}} 
\bigg[ \arctan\left( \delta \over\sqrt{m^2-\delta^2-i0}\right) - {\pi\over 2} 
\bigg]  + \order(\epsilon)
\bigg\} \,.
\end{align}
The two-point functions for the electroweak SM bosons appearing in
(\ref{eq:deltav}) are obtained by summing the fermionic and bosonic
contributions given below. Following Denner \cite{Denner:1991kt}, we
have
\begin{align}\label{eq:SMself}
\Sigma^{AA\prime}(0) &= -{\alpha\over 4\pi} \bigg\{ 3 B_0(0,m_W,m_W) 
+ 4 m_W^2 B_0^\prime(0,m_W,m_W)
\nl & \quad 
-\frac43 \sum_{f,i} \big[ N_c^f Q_f^2 B_0(0,m_{f,i},m_{f,i}) \big]
 \bigg\}  \,, 
\nl
{\Sigma^{AZ}(0) \over m_Z^2} &= -{\alpha\over 4\pi} \bigg\{ -{2 c_W \over s_W} B_0(0,m_W,m_W) 
\bigg\}  \,,
\nl
{\Sigma^{ZZ}(m_Z^2)_{\rm fermion}\over m_Z^2} &= -{\alpha\over 4\pi} 
\bigg\{ \frac23\bigg[-B_0(m_Z,0,0)+\frac13\bigg] \sum_{f,i} N_c^f [(g_f^+)^2 + (g_f^-)^2] 
\nl
&\quad + \frac23 N^t_c \bigg[ [ (g_t^+)^2 + (g_t^-)^2 ] \bigg[ -\left(1+{2m_t^2\over m_Z^2}\right) B_0(m_Z,m_t,m_t) 
+ B_0(m_Z,0,0) 
\nl
&\quad 
+ {2m_t^2\over m_Z^2} B_0(0,m_t,m_t) \bigg] 
+ {3\over 4 s_W^2 c_W^2} {m_t^2 \over m_Z^2} B_0(m_Z,m_t,m_t) 
\bigg] 
\bigg\} \,,
\nl
{\Sigma^{ZZ}(m_Z^2)_{\rm boson}\over m_Z^2} &= -{\alpha\over 4\pi}{1\over s_W^2 c_W^2}  
\bigg\{ 
\frac1{12} (4c_W^2 - 1)  ( 12 c_W^4 + 20c_W^2 + 1) B_0(m_Z,m_W,m_W) 
\nl
&\quad 
-\frac13 c_W^2 ( 12 c_W^4-4c_W^2 + 1) B_0(0,m_W,m_W) 
-\frac16 B_0(0,m_Z,m_Z)
\nl
&\quad 
-\frac{1}{12} \left( {m_h^4\over m_Z^4} - 4 {m_h^2\over m_Z^2} + 12 \right) B_0(m_Z,m_Z,m_h)
 -\frac16 {m_h^2\over m_Z^2} B_0(0,m_h,m_h)  
\nl
&\quad
+ {1\over 12} \left(1-{m_h^2\over m_Z^2} \right)^2 B_0(0,m_Z,m_h) 
-\frac{1}{9} ( 1- 2 c_W^2 ) 
\bigg\} \,,
\nl
{\Sigma^{WW}(m_W^2)_{\rm fermion}\over m_W^2} &= -{\alpha\over 4\pi}{1\over 2s_W^2} 
\bigg\{ \frac23 \bigg[\frac13 -B_0(m_W,0,0) \bigg] \sum_{f,i} {N_c^f\over 2} 
\nl
&\quad 
+ \frac23 N_c^t \bigg[ \frac12 \left( {m_t^4 \over m_W^4} + {m_t^2\over m_W^2} - 2 \right) B_0(m_W,m_t,0) 
+ B_0(m_W,0,0) 
\nl
&\quad
+ {m_t^2\over m_W^2} B_0(0,m_t,m_t) 
- {m_t^4\over 2 m_W^4} B_0(0,m_t,0) 
\bigg]
\bigg\} \,,
\nl
{\Sigma^{WW}(m_W^2)_{\rm boson}\over m_W^2} &= -{\alpha\over 4\pi}
\bigg\{ 
4 B_0(m_W,m_W,\lambda) -\frac43 B_0(0,m_W,m_W) + \frac23 B_0(0,m_W,\lambda) + \frac29 
\nl
&\quad 
+ {1\over 12 s_W^2} \bigg[ {1\over c_W^4} (4c_W^2-1)(12 c_W^4 + 20c_W^2 + 1) B_0(m_W,m_W,m_Z) 
\nl
&\quad
- 2(8 c_W^2+ 1) B_0(0,m_W,m_W) - {2\over c_W^2}(8c_W^2+1) B_0(0,m_Z,m_Z) 
\nl
&\quad
+ {s_W^4\over c_W^4}(8c_W^2+1) B_0(0,m_W,m_Z)  - \frac23 (1-4 c_W^2) 
\bigg] 
\nl
&\quad
+{1\over 12 s_W^2} \bigg[ -\left( {m_h^4\over m_W^4} -4 {m_h^2\over m_W^2} + 12 \right) B_0(m_W,m_W,m_h) 
- 2 B_0(0,m_W,m_W) 
\nl
&\quad 
-2 {m_h^2\over m_W^2} B_0(0,m_h,m_h) 
+ \left( 1 - {m_h^2\over m_W^2} \right)^2 B_0(0,m_W,m_h)  
- \frac23 
\bigg]
\bigg\} \,, 
\nl
\Sigma^{HH\prime}(m_h^2)_{\rm fermion}
&= 
-{\alpha\over 4\pi} {3m_t^2 \over 2 s_W^2 m_W^2} 
\bigg[ (4 m_t^2 - m_h^2)B_0^\prime(m_h,m_t,m_t) - B_0(m_h,m_t,m_t)  \bigg] 
\,,
\nl
\Sigma^{HH\prime}(m_h^2)_{\rm boson} 
&= 
-{\alpha\over 4\pi} \bigg\{ 
-{1\over 2 s_W^2} \bigg[ 
 \left(6 m_W^2 -2m_h^2 + {m_h^4 \over 2 m_W^2} \right) B_0^\prime(m_h,m_W,m_W)
 \nl &\quad  
 -2B_0(m_h,m_W,m_W) 
\bigg]
-{1\over 4 s_W^2 c_W^2} \bigg[ 
 \left(6 m_Z^2 - 2m_h^2 + {m_h^4\over 2 m_Z^2} \right) B_0^\prime(m_h,m_Z,m_Z) 
\nl
&\quad
-2 B_0(m_h,m_Z,m_Z)
\bigg]
- {9 m_h^4 \over 8 s_W^2 m_W^2} B_0^\prime(m_h,m_h,m_h) 
\bigg\} \,,
\end{align}
where the sums over indices $f$ and $i$ are for SM fermion flavors and
generations, respectively. Above, $N_c^f$ and $Q_f$ respectively
denote the number of colors and the electric charge of fermion $f$. We have
also used
\begin{align}
\alpha &= { g_2^2 s_W^2 \over 4 \pi} \, ,\quad g_f^+ = {1 \over 8 s_W^2 c_W^2 } \big[c_V^{(f)2} + c_A^{(f)2} \big] \, , \quad g_f^- = {1 \over 8 s_W^2 c_W^2 } \big[c_V^{(f)2} - c_A^{(f)2} \big] \, ,
\end{align}
where
\be
c_V^{(\ell)} = -1 + 4 s_W^2\, , \quad c_A^{(\ell)} = 1\, , \quad c_V^{(\nu)} = 1\, , \quad c_A^{(\nu)} = -1\, , 
\ee
with $\ell$ and $\nu$ denoting charged lepton and neutrino, respectively. The
coefficients $c_V^{(f)}$ and $c_A^{(f)}$ for quarks can be found in
(\ref{eq:quarkscvca}). The basic integral appearing above is
\begin{align}
{i \over (4\pi)^2} B_0(M,m_0,m_1) &= \int (dL) { 1 \over L^2 - m_0^2 + i0} {1 \over (L+p)^2 - m_1^2 + i0 }  
\nl
&=  [c_\epsilon] 
\bigg[ 
{1\over \epsilon} + 2 - \log(m_0 m_1) + {m_0^2 - m_1^2 \over M^2}\log{m_1\over m_0} - {m_0 m_1 \over M^2}\left({1\over r}-r\right)\log r 
+ \order(\epsilon) 
\bigg] \,,
\end{align} 
where $p^2=M^2$ and 
\be
r = X + \sqrt{X^2 - 1} \,, \quad 
{1\over r} = X - \sqrt{X^2 -1} \,, \quad 
X = { m_0^2 + m_1^2 - M^2 -i0 \over 2 m_0 m_1 } \,. 
\ee
We find the following limits,
\begin{align} 
B_0(0,m,m) &= (4\pi)^\epsilon \Gamma(1+\epsilon) \bigg[ {1\over \epsilon} - 2 \log{m} + \order(\epsilon) \bigg] \,,
\nl
B_0(0,m,0) &= (4\pi)^\epsilon \Gamma(1+\epsilon) \bigg[ {1\over \epsilon} - 2\log{m} + 1 + \order(\epsilon) \bigg] \,,
\nl
B_0(0,m_0,m_1) &= (4\pi)^\epsilon \Gamma(1+\epsilon) \bigg[ {1\over \epsilon} - {m_0^2 \over m_0^2 -m_1^2} \log m_0^2 + {m_1^2\over m_0^2 -m_1^2} \log m_1^2 + 1+ \order(\epsilon) \bigg] \,,
\nl
B_0(M,m,0) &= (4\pi)^\epsilon \Gamma(1+\epsilon) \bigg[ {1\over \epsilon} + 2 - {m^2\over M^2} \log m^2 + {m^2-M^2 \over M^2} \log( m^2- M^2 - i0) + \order(\epsilon) \bigg] \,,
\nl
B_0(M,0,0) &= (4\pi)^\epsilon \Gamma(1+\epsilon) \bigg[ {1\over \epsilon} + 2 - \log(-M^2 - i0) + \order(\epsilon) \bigg] \,, 
\nl
\lim_{\lambda\to 0} B_0(m,m,\lambda) &=(4\pi)^\epsilon \Gamma(1+\epsilon) \bigg[ {1\over \epsilon} + 2 - \log{m^2} + \order(\epsilon) \bigg] \,. 
\end{align} 
In the present application, only the real parts of the integrals are relevant. 
For the derivative of the integral we have,
\begin{align}
B_0^\prime(M,m,m) &\equiv {\partial \over \partial p^2} B_0(M,m,m) 
\nl
&= (4\pi)^{\epsilon} \Gamma(1+\epsilon) 
\bigg[ {m^2 \over M^4} \left({1\over r} - r\right) \log r - {1\over M^2}\left( 1 + {r^2+1\over r^2-1} \log r \right) 
+ \order(\epsilon)
\bigg] \,,
\end{align}
which has the following limits,
\begin{align}
B_0^\prime(0,m,m) &= (4\pi)^\epsilon \Gamma(1+\epsilon)\bigg[ {1\over 6 m^2} + \order(\epsilon) \bigg] \,,
\nl
B_0^\prime(M,0,0) &= (4\pi)^\epsilon \Gamma(1+\epsilon)\bigg[ - {1\over M^2} + \order(\epsilon) \bigg] \,. 
\end{align} 

\section{Box integrals \label{sec:box}}

The integrals required for the two-boson exchange amplitudes in
Sec.~\ref{sec:2BEquark} may be written in terms of the integral
operators ${\cal I}_{\rm even}$ and ${\cal I}_{\rm odd}$ defined in
(\ref{eq:bosonloop}) as
\begin{align} 
J(m_V, M, \delta) 
&= {\cal I}_{\rm even}(\delta,m_V) {1\over L^2 - M^2+i0} \, ,
\nl
J^\mu(p,m_V,M,\delta) 
&= {\cal I}_{\rm even}(\delta, m_V) {1 \over L^2 + 2 L\cdot p - M^2 + i0} L^\mu
\nl
&= v\cdot p v^\mu J_1(m_V,M,\delta) + p^\mu J_2(m_V,M,\delta)  + \order(p^3)\,,
\nl
J_-(p,m_V,M,\delta) 
&= - {\cal I}_{\rm odd}(\delta, m_V) {1\over L^2 + 2L\cdot p - M^2 + i0} 
= v\cdot p J_-(m_V,M,\delta) + \order(p^3) 
\,,
\nl
J_-^\mu(m_V,M,\delta) 
&= - {\cal I}_{\rm odd}(\delta, m_V) {1\over L^2 - M^2 + i0} L^\mu 
= v^\mu J_{1-}(m_V,M,\delta) 
\,.
\end{align}
Note that $J^\mu(p,m_V,M,\delta)$ and $J_-(p,m_V,M,\delta)$ vanish
when $p^\mu$ vanishes since the integrands are then odd in $L^\mu$. By
standard manipulations, we may express the integrals $J_1$, $J_2$,
$J$, and $J_-$, as
\begin{align} \label{eq:Jfeyn}
J_1(m_V,M,\delta) &= -8[c_\epsilon] (1+\epsilon) {\partial \over \partial m_V^2} 
\int_0^\infty d\rho \int_0^1 dx \rho^2 (1-x) 
\nl
&\qquad
\big[ x m_V^2 + (1-x) M^2
+ \rho^2 
+ 2\rho\delta -i 0 \big]^{-2-\epsilon} 
\,,
\nl
J_2(m_V,M,\delta) &= 4 [c_\epsilon] {\partial \over \partial m_V^2} 
\int_0^\infty d\rho \int_0^1 dx  (1-x) 
\big[ x m_V^2 + (1-x) M^2 + \rho^2 + 2\rho\delta -i 0  \big]^{-1-\epsilon} 
\,,
\nl
J(m_V,M,\delta) &= -4 [c_\epsilon] {\partial \over \partial m_V^2} 
\int_0^\infty d\rho \int_0^1 dx 
\big[ x m_V^2 + (1-x) M^2 + \rho^2 + 2\rho\delta -i 0  \big]^{-1-\epsilon} 
\,,
\nl
J_-(m_V,M,\delta) &= 4 [c_\epsilon] 
{\partial \over \partial \delta} 
{\partial \over \partial m_V^2} 
\int_0^\infty d\rho \int_0^1 dx (1-x)
\big[ x m_V^2 + (1-x) M^2 + \rho^2 + 2\rho\delta -i 0  \big]^{-1-\epsilon} 
\,.
\end{align}
Let us introduce the integral
\be 
\hat{J}(m_V,M,\delta) = 
[c_\epsilon] \int_0^\infty d\rho \int_0^1 dx  (1-x) 
\big[ x m_V^2 + (1-x) M^2 + \rho^2 + 2\rho\delta -i 0  \big]^{-1-\epsilon}\, ,
\ee
and write the above integrals in terms of $\hat{J}(m_V,M,\delta)$ as
\begin{align}\label{eq:Jexp}
J_2(m_V,M,\delta)& = 4 {\partial \over \partial m_V^2} \hat{J}(m_V,M,\delta) \,,
\nl
J_-(m_V,M,\delta) &= 4 {\partial\over\partial \delta} {\partial \over \partial m_V^2} \hat{J}(m_V,M,\delta) \,, 
\nl
J(m_V,M,\delta) &= -4 {\partial\over \partial m_V^2} 
\big[ \hat{J}(m_V,M,\delta) + \hat{J}(M,m_V,\delta) \big] \, ,
\nl
J_1(m_V,M,\delta) &= 4 {\partial \over \partial m_V^2} 
\bigg[  - \hat{J}(m_V,M,\delta) + {\partial \over \partial A} \hat{J}(m_V,M,\delta/A) \big|_{A=1} \bigg]
\,.
\end{align} 
For $J_{1-}$, we may use the identity (\ref{eq:relationPM}) to write
\begin{align} \label{eq:J1minus}
J_{1-}(m_V,M,\delta) &= - 2\int(dL) {1\over (L^2-m_V^2 +i0)^2}{1\over L^2-M^2+i0} - \delta J(m_V,M,\delta) 
\nl
&=  {2 [c_\epsilon] m_V^{-2-2\epsilon} \over \epsilon (1- \epsilon)}  \left( 1- {M^2\over m_V^2}\right)^{-2}
\bigg[
\epsilon + {M^2\over m_V^2} \left( 1- \epsilon -  {m_V^{2 \epsilon} \over M^{2 \epsilon} } \right) \bigg] 
- \delta \, J(m_V,M,\delta)
\,.
\end{align} 
Having determined the above integrals in terms of $\hat{J}(m_V,M,\delta)$, it remains
to compute this function. Let us write
\begin{align} \label{eq:Jhatexpress}
\hat{J}(m_V,M,\delta) &= 
-{[c_\epsilon]\over\epsilon}{\partial\over \partial M^2} 
\int_0^\infty d\rho \int_0^1 dx 
\big[ x m_V^2 + (1-x) M^2 + \rho^2 + 2\rho \delta -i 0  \big]^{-\epsilon} 
\nl
&= 
-{[c_\epsilon]\over\epsilon}{\partial\over \partial M^2} 
\int_0^\infty d\rho 
{1\over m_V^2 -M^2} {1\over 1-\epsilon} 
\nl
&\qquad
\bigg\{ [m_V^2 + \rho^2 + 2\rho\delta-i 0 ]^{1-\epsilon} 
- [ M^2 + \rho^2 + 2\rho\delta -i 0 ]^{1-\epsilon} 
\bigg\} \
\nl
&=
-{[c_\epsilon]\over\epsilon}{\partial\over \partial M^2} 
{1\over m_V^2 -M^2} {1\over 1-\epsilon} 
\bigg\{ 
m_V^{3-2\epsilon} f_1(\delta/m_V,1-\epsilon) 
- M^{3-2\epsilon} f_1(\delta/M,1-\epsilon) 
\bigg\}  \,,
\end{align}
where
\begin{align}\label{eq:f1}
f_1(\delta,a) &= \int_0^\infty d\rho (1 + \rho^2 + 2\rho\delta -i0)^a  \
\nl
&= (1-\delta^2-i0)^{a + \frac12 } {\sqrt{\pi}\over 2}{\Gamma(-a-\frac12)\over\Gamma(-a)} 
- \delta^{2a+1} \int_0^1 dx \left[ {\delta^{-2}}-1+ x^2-i0 \right]^{a} \,. 
\end{align}
Although for the present application we require only $\delta>0$, the expression is 
for general sign of $\delta$.  
We presently need $f_1(\delta,a)$ for $a=1-\epsilon$, and hence consider
\begin{align}
&{\sqrt{\pi}\over 2}{\Gamma(-\frac32+\epsilon)\over\Gamma(-1+\epsilon)} 
= -{2\pi\over 3}\epsilon + {2\pi\over 9}(6\log 2 - 5) \epsilon^2 + \order(\epsilon^3) \,,
\nl
&\int_0^1 dx\, \left[  \delta^{-2}-1+ x^2-i0 \right]^{1-\epsilon}
= B^2 + \frac13 
+ \epsilon \bigg\{ \frac29 + \frac43 B^2   -\frac43 B^3 \arccot B
-\left(B^2+\frac13\right) \log(B^2+1) \bigg\} 
\nl
&\quad  
+ \epsilon^2 \bigg\{ 
\frac{4}{27} + \frac{20}{9} B^2 +  
\frac49 B^3 (6\log{2B} - 5 ) \arccot B
+ \frac43 B^3 i \bigg[ {\rm Li}_2\left(1+i B \over 1-iB\right)  -\arccot^2 B
+ {\pi^2\over 12} 
\bigg]
\nl
&\quad
+ \frac12 \left( B^2+\frac13 \right) \log^2(B^2+1) 
-\left(\frac43 B^2 + \frac29 \right) \log(B^2+1) 
\bigg\} + \order(\epsilon^3) \,,
\end{align}
where $B^2= 1/\delta^2 -1 -i0$. 
For $B^2 > 0$, the bracket involving dilogarithm may be written 
\be
i\bigg[ {\rm Li}_2\left(1+i B \over 1-iB\right)  -\arccot^2 B
+ {\pi^2\over 12} \bigg]
= -{\rm Im}\,{\rm Li}_2\left(1+i B \over 1-iB\right) 
= -{\rm Cl}_2\left[ \arccos\left(1-B^2\over 1+B^2\right) \right] \,,
\ee
where ${\rm Cl}_2$ is the Clausen function of order two. 
The general expression is required for continuing to arbitrary mass parameters. 
Having determined $f_1(\delta,1-\epsilon)$, we may proceed to compute 
$\hat{J}(m_V,M,\delta)$ using (\ref{eq:Jhatexpress}), 
and then $J_2(m_V,M,\delta)$, $J(m_V,M,\delta)$, $J_-(m_V,M,\delta)$ and
$J_1(m_V,M,\delta)$ using (\ref{eq:Jexp}), and $J_{1-}(m_V,M,\delta)$ using (\ref{eq:J1minus}).

For $M=0$, the expressions in (\ref{eq:Jfeyn}), the expressions for $J_2(m_V,M,\delta)$, $J_1(m_V,M,\delta)$, and $J_-(m_V,M,\delta)$ in (\ref{eq:Jexp}), and the expression for $J_{1-}(m_V,M,\delta)$ in (\ref{eq:J1minus}), remain valid. The integral $J(m_V,0,\delta)$ is now given by 
\begin{align}
J(m_V, 0,\delta) &= 
-4 [c_\epsilon] {\partial\over \partial m_V^2}
\bigg\{ -{1\over \epsilon} m_V^{-1-2\epsilon} \bigg[ f_1(\delta/m_V,-\epsilon) - f_0(\delta/m_V,-\epsilon) \bigg]\bigg\} \,, 
\end{align}
and the integral $\hat{J}(m_V,0,\delta)$ by 
\begin{align}
\hat{J}(m_V,0,\delta) &= { [c_\epsilon] m_V^{-2} \over \epsilon}  \int_0^\infty d\rho \bigg\{ 
 (\rho^2+2\rho\delta -i 0 )^{-\epsilon} 
 \nl
 &\qquad
- {m_V^{-2} \over 1-\epsilon}
\bigg[ (m_V^2+\rho^2+2\rho\delta-i 0 )^{1-\epsilon} - (\rho^2+2\rho\delta-i 0 )^{1-\epsilon} \bigg] 
\bigg\} 
\nl
&= { [c_\epsilon] m_V^{-1-2\epsilon} \over \epsilon} \bigg\{  f_0(\delta/m_V, -\epsilon) 
- {1\over (1-\epsilon)} \big[ f_1(\delta/m_V,1-\epsilon) - f_0(\delta/m_V,1-\epsilon) \big] 
\bigg\} \,,
\end{align}
where $f_1(\delta,a)$ is given by (\ref{eq:f1}) and 
\be
f_0(\delta,a) = \int_0^\infty d\rho (\rho^2+2\rho\delta -i 0 )^a = { \delta^{1+2a} \Gamma(1+a) \Gamma\left( -a - \frac12 \right) \over 2\sqrt{\pi}} \,.
\ee
We also need $f_1(\delta/m_V,a)$  for $a=-\epsilon$, which we may write as
\be
f_1(\delta/m_V,-\epsilon) = {1\over 1-\epsilon} m_V^{-1+2\epsilon} {\partial\over \partial m_V^2} 
\bigg[ m_V^{3-2\epsilon} f_1(\delta/m_V,1-\epsilon) \bigg] \,.
\ee

At vanishing residual mass, $\delta =0 $, only the integrals $J(m_V,M,0)$, $J_1(m_V,M,0)$ and $J_2(m_V,M,0)$ are required, and from (\ref{eq:Jfeyn}) they can be easily represented in closed form, 
\begin{align}\label{eq:JforMzero}
J(m_V,M,0) &=  
[c_\epsilon] { 2 \sqrt{\pi}  \over (1 -2\epsilon)}{\Gamma(\frac12+\epsilon)\over\Gamma(1+\epsilon)}
{m_V^{1-2\epsilon} \over (M^2-m_V^2)^2} 
\bigg[ 1+2\epsilon - 2\left( { M \over m_V } \right)^{1-2\epsilon}  + (1-2\epsilon)\left( {  M\over m_V} \right)^2 \bigg] \,,
\nl
J_2(m_V,M,0) &= - J_1(m_V,M,0) = [c_\epsilon] 
{4\sqrt{\pi} \over (3-2\epsilon)(1-2\epsilon) }{\Gamma(\frac12+\epsilon)\over\Gamma(1+\epsilon)}
{ m_V^{3-2\epsilon} \over (M^2-m_V^2)^3} 
\nl
&\qquad \bigg[ 
 1+2\epsilon - (3-2\epsilon) \left(M\over m_V\right)^{1-2\epsilon} 
+ (3-2\epsilon) \left(M\over m_V\right)^2 
- (1+2\epsilon) \left(M\over m_V\right)^{3-2\epsilon} 
\bigg] \, .
\end{align}
The result $J_2(m_V,M,0) = - J_1(m_V,M,0)$ follows from the observation that when $\delta = 0$ the identity in (\ref{eq:relationPM}) implies $v_\mu J^\mu (p,m_V,M,0) =0$. The case $\delta=M=0$ is simply obtained by substitution in (\ref{eq:JforMzero}).

\section{Heavy particle integrals with electroweak polarization tensor insertion  \label{sec:appC}}

The two-boson exchange amplitudes for gluon matching
require the integrals $H(n)$, $F(n)$, ${H}^{\mu \nu}(n)$, and $H^\mu(n)$ defined in
(\ref{eq:tildebasis}). Let us parameterize the last two as
\begin{align} H^{\mu \nu}(n) &= H_1(n) v^\mu v^\nu +
H_2(n) g^{\mu \nu} \, ,  \quad H^\mu(n) ={H}_3(n) v^\mu\, .
\end{align}
Upon contracting the above expressions with $v_\mu$ and $g_{\mu \nu}$, we may solve for the relations
\begin{align}
H_1(n) &= {1 \over 3-2\epsilon} \big[ (4-2\epsilon) v_\mu v_\nu H^{\mu \nu}(n) - H^{\mu}_{\ \mu}(n) \big] \, ,
\nl
H_2(n) &= {1 \over 3-2\epsilon} \big[H^{\mu}_{\ \mu}(n) - v_\mu v_\nu H^{\mu \nu}(n) \big] \, , 
\nl 
H_3(n) &= v_\mu H^\mu(n) \, .
\end{align}
Using the identities in (\ref{eq:replaceL}) and (\ref{eq:relationPM}), we further obtain
\begin{align}
v_\mu H^\mu(n) &= \delta H(n) + 2 F(n) \, ,
\nl
v_\mu v_\nu H^{\mu \nu} (n) &= \delta^2 H(n) + 2 \delta F(n) \, ,
\nl
H^{\mu}_{\ \mu}(n) &= \bigg[ {\mymu^2 \over x} + {\mymd^2 \over (1-x) } \bigg] H(n)  -  {H(n-1)  \over x(1-x) } \, ,
\end{align}
and hence the boson loops are completely specified by $H(n)$
and $F(n)$. In evaluating these functions it may be
advantageous to relate to more basic integrals by means of
derivatives. Let us write,
\begin{align}
H(n) &= 2 {\partial \over \partial m_V^2} \int (dL) 
{1\over v \cdot L - \delta  + i0} {1\over L^2 - m_V^2  + i0} 
\Delta^{-n-\epsilon}  \, ,
\nl
F(n) &= {\partial \over \partial m_V^2}\int (dL) 
{1\over L^2 - m_V^2 + i0} 
\Delta^{-n-\epsilon}  \, ,
\end{align}
with $\Delta$ as defined in (\ref{eq:zDelta}).
The singularity structure and evaluation of the above integrals can be
classified into three cases, corresponding to zero, one, or two heavy
fermions contributing to the electroweak polarization tensor. For pure states we obtain analytic expressions for all integrals, while for mixed states we encounter several integrals that require numerical evaluation of one Feynman parameter integral.

\subsection{Case of zero heavy fermions} 

Upon setting $\mymu=\mymd=0$ in $\Delta$ and performing the integration in $d=4-2\epsilon$ dimensions, 
we obtain
\begin{align}
F(n) &=  [c_\epsilon] {\Gamma(2-n-2\epsilon)  \Gamma(n+2 \epsilon) \over \Gamma(2- \epsilon)\Gamma(1+\epsilon)} 
 [ x(1-x) ]^{-n-\epsilon}   m_V^{-2n-4\epsilon}
  \,,
\nl
H(n) &= [c_\epsilon] {4 \Gamma(n+2\epsilon)  \over \Gamma(n+\epsilon) \Gamma(1+\epsilon)} [x(1-x)]^{-n-\epsilon} 
{\partial \over \partial m_V^2}  I(n) \,,
\end{align}
where 
\begin{align}
I(n) &= \int_0^1dy\, (1-y)^{n-1+\epsilon} \int_0^\infty d\rho (\rho^2+2\rho \delta + ym_V^2 -i 0 )^{-n-2\epsilon} \,.
\end{align}
We may reduce to the case of $I(1)$ by noticing that 
\begin{align}
I(n+1) &= -{m_V^{-2}\over n+2\epsilon} \int_0^1 dy (1-y)^{n+\epsilon}  
 {d\over dy} 
\int_0^\infty d\rho ( \rho^2 + 2\rho\delta + y m_V^2 -i 0 )^{-n-2\epsilon}  
\nl
&= {m_V^{-2} \over n+2\epsilon} \bigg[ 
 \int_0^\infty d\rho (\rho^2+2\rho\delta-i 0 )^{-n-2\epsilon} 
+ (n+\epsilon) I(n) 
\bigg]
\nl
&= {m_V^{-2}  \over n+2\epsilon} \bigg[ 
 \delta^{1 -2n-4\epsilon} {\Gamma(1-n-2\epsilon) \Gamma\left(n-\frac12 + 2\epsilon\right) 
\over 2\sqrt{\pi} } 
+ (n+\epsilon) I(n) 
\bigg] \,.
\end{align} 
Finally, for $I(1)$ we require 
\begin{align}
I(1) &= \delta^{-1-4\epsilon} \int_0^1 dy (1 + \epsilon\log(1-y) + \dots) 
 \int_1^\infty d\rho ( \rho^2 + \alpha^2 )^{-1}
\left( 1  -2\epsilon \log(\rho^2 + \alpha^2 ) + \dots \right) \,, 
\end{align}
where $\alpha = \left(ym_V^2/\delta^2 -1 - i0 \right)^{\frac12}$. 
The relevant integrals are
\begin{align}
\int_1^\infty d\rho {1\over \rho^2+\alpha^2} &= {1\over \alpha}\arctan{\alpha} \,,
\nl
\int_1^\infty d\rho {\log(\rho^2+\alpha^2) \over \rho^2+\alpha^2} 
&= {1\over\alpha} \bigg[ 
2\log(2\alpha) \arctan\alpha 
- {1\over 2i} \bigg( {\rm Li}_2\left(1-i\alpha\over 1+i\alpha\right)
- {\rm Li}_2\left(1+i\alpha\over 1-i\alpha\right) 
\bigg)
\bigg] \, .
\end{align}
We perform the remaining integral over Feynman parameter $y$ numerically.

\subsection{Case of one heavy fermion} 

Let us set $\mymu=M$ (not to be confused with heavy WIMP mass $M$ used elsewhere in the paper)
and $\mymd=0$ in $\Delta$, and consider
separately the finite integrals for $a$- and $c$-type contributions, and the IR divergent integrals for $b$-type contributions.

\subsubsection{Finite integrals for $a$- and $c$-type contributions}

For the finite $a$- and $c$-type contributions we may take
$d=4$. Let us evaluate the required integrals $F(2)$ and $H(1)$, and
obtain the remaining integrals by differentiating with respect to $M$. We find 
\begin{align}
F(2) &= {i \over (4\pi)^2} {\partial \over \partial m_V^2} \Bigg\{  \Bigg[ x(1-x)  m_V   \left(1- {M^2\over xm_V^2}  \right) \Bigg]^{-2}
\Bigg[ -  \log{M^2\over x m_V^2} + {M^2\over x m_V^2} - 1 \Bigg] \Bigg\} \,,
\nl
H(1) &= {i\over (4\pi)^2}  {\partial \over \partial m_V^2 } \Bigg\{ 8   \Bigg[ x(1-x) m_V^2 \left( 1 - {M^2\over x m_V^2} \right) \Bigg]^{-1} \Bigg[ 
\sqrt{m_V^2-\delta^2} \arctan\left( \sqrt{ {m_V^2\over \delta^2}-1} \right) 
\nl
&\quad 
- \sqrt{{M^2\over x} - \delta^2} \arctan\left( \sqrt{{M^2\over x \delta^2} - 1 } \right) 
- {\delta\over 2} \log{x m_V^2 \over M^2} 
\Bigg] \Bigg\} \,.
\end{align} 
The integrals have been obtained by breaking an integration region into pieces, e.g.,
\begin{align}\label{eq:vareps}
& \int_\delta^\infty d\rho \left[ \log(\rho^2+m_V^2-\delta^2) - \log\left(\rho^2+{M^2\over x} -\delta^2\right) \right]
\nl
&= \lim_{\varepsilon\to 0} 
\int_\delta^\infty d\rho \left[ \log(\rho^2+m_V^2-\delta^2 -i\varepsilon) 
- \log\left(\rho^2+{M^2\over x} -\delta^2 -i\varepsilon\right) \right]
\nl
&= \delta \lim_{\varepsilon\to 0} 
 \int_1^\infty d\rho  \left[ \log\left( \rho^2+ {m_V^2\over\delta^2} -1 - i\varepsilon \right)
- \log\left( \rho^2+ {M^2\over x\delta^2} -1 - i\varepsilon \right) \right]
\nl
&= \delta \lim_{\varepsilon\to 0} \bigg\{ \int_0^\infty d\rho 
 \left[ \log\left( \rho^2+ {m_V^2\over\delta^2} -1 - i\varepsilon \right)
- \log\left( \rho^2+ {M^2\over x\delta^2} -1 - i\varepsilon \right) \right] 
\nl
 &\qquad 
 -  \int_0^1 d\rho \left[ \log\left( \rho^2+ {m_V^2\over\delta^2} -1 - i\varepsilon \right)
 - \log\left( \rho^2+ {M^2\over x\delta^2} -1 - i\varepsilon \right) \right]
 \bigg\}  \, .
\end{align}
Since the original integral is independent of $\varepsilon$, either
choice of ${\rm sgn}(\varepsilon)$ is correct provided it is used consistently in both terms.
The continuation away from $\delta\to 0$ is thus obtained above by
taking, e.g., $\delta \to \delta + i\varepsilon$ everywhere.   For the
evaluation of integrals over $x$ involving $H(1)$, let us write 
\be
H(1) \equiv  2 { \partial \over \partial m_V^2} K(1) \equiv 2 { \partial \over \partial m_V^2}  \Bigg\{ {M^2 \over x m_V^2 - M^2} k(1)  \Bigg\} \,.
\ee
We then have 
\be
x^n K(1)  = \left( M^2\over m_V^2 \right)^n K(1) + { \left( M^2\over m_V^2 \right)^n - x^n \over {M^2\over m_V^2} - x}
{M^2\over m_V^2} k(1) \,,
\ee
so that all powers $x^n K(1)$ can be reduced to the case $n=0$, in
addition to the remaining  straightforward integral involving a
polynomial in $x$ times $k(1)$, which in practice is evaluated numerically.  
The remaining integrals involving $F(2)$ are
similarly straightforward to evaluate. 

\subsubsection{Infrared divergent integrals for $b$-type contributions}

Let us now turn to the integrals for $b$-type contributions, where we
work in $d=4-2\epsilon$ spacetime dimensions to account for singular
behavior at the endpoints of the $x$ integration. 
We find,
\begin{align}
F(1) 
&= [c_\epsilon] [x(1-x)]^{-1-\epsilon} {\Gamma(1+2\epsilon) \over [\Gamma(1+\epsilon)]^2}  \Bigg\{ 
m_V^{-2-4\epsilon} \bigg[ 
\left({r^2} - 1\right)^{-2} \bigg( {r^2}\log{r^2} - {r^2} + 1 \bigg) 
\nl
&\quad
+ {\epsilon} \left(r^2 - 1\right)^{-2} \bigg( 2{r^2} \log{r^2} - {r^2}\log^2{r^2}
- r^2 + 1 + {r^2}{\rm Li}_2\left(1-{r^2}\right) 
\bigg)
\bigg]
\nl
&\quad 
+ m_V^{-2}\Bigg[ \left({r^2\over x} - 1\right)^{-2}\left( {r^2\over x}\log{r^2\over x} - {r^2\over x} + 1 
\right) 
- \left({r^2} - 1\right)^{-2}\bigg(  {r^2}\log{r^2} - {r^2} + 1 
\bigg)
\Bigg] \Bigg\} \,,
\end{align}
where $r\equiv M/m_V$. 
The first term in curly braces is obtained by taking $x=1$ inside the $\int dy$ integral, and 
the second term is the remainder having no singularity in the final $\int dx$ integral at $x=1$. 

Similarly we find,
\begin{align}
H(1)&=  [c_\epsilon] [x(1-x)]^{-1-\epsilon} {4 \Gamma(1+2\epsilon) \over [\Gamma(1+\epsilon)]^2} 
{\partial \over \partial m_V^2} \Bigg\{ 
\delta^{-1-4\epsilon} \Bigg[ Y_0(1) + \epsilon\bigg( Y_1 + Y_2 \bigg)
\Bigg] 
+\delta^{-1} \Bigg[ Y_0(x) - Y_0(1) \Bigg]
\Bigg\} \,,
\end{align}
where
\be
Y_0(x) = {2 \over r_V^2 - {r_M^2\over x} } \bigg\{ \sqrt{r_V^2-1}\arctan\left(\sqrt{{r_V^2}-1}\right)
- \sqrt{{r_M^2 \over x} -1} \arctan\left(\sqrt{{r_M^2 \over x}-1}\right)
- \frac12 \log{x m_V^2\over M^2} 
\bigg\} \, ,
\ee
with $r_V\equiv m_V/\delta$ and $r_M\equiv M/\delta$. 
As in the discussion after (\ref{eq:vareps}), 
continuation away from $\delta=0$ is given by taking $\delta\to \delta + i\varepsilon$
with arbitrary choice of ${\rm sgn}(\varepsilon)$.  
The remaining terms $Y_1$ and $Y_2$ are given by  
\begin{align}
Y_1 &= \int_0^1 dy \int_0^\infty d\beta \big( r_M^2-r_V^2 \big)^{-1} {d\over dy}
 \log^2\Big[ \beta^2 + 2\beta + y r_V^2 + (1-y)r_M^2 \Big]
\nl
&=  \big( r_M^2-r_V^2 \big)^{-1} \bigg\{ 
-4\pi \sqrt{r_V^2-1} \Big[ 1 - \log\Big( 2\sqrt{r_V^2-1}\Big) \Big] 
+ 4\pi \sqrt{r_M^2-1} \Big[ 1 - \log\Big(2\sqrt{r_M^2-1}\Big) \Big] 
\nl
&\quad
- y_1\Big(\sqrt{r_V^2-1}\Big) + y_1\Big(\sqrt{r_M^2-1}\Big) 
\bigg\} \,, 
\nl
Y_2 &= \int_0^1 dy\, \log(1-y) \Big( y r_V^2 + (1-y)r_M^2 - 1\Big)^{-1} \arctan\left(\sqrt{y r_V^2 + (1-y)r_M^2 - 1}\right) \, ,
\end{align}
where
\be
y_1(A) \equiv \int_0^1 dx \log^2(x^2+A^2) \,.
\ee
For $Y_2$, we evaluate the remaining integral over Feynman parameter $y$ numerically. 

\subsection{Case of two heavy fermions} 

Let us set $\mymu=\mymd=M$ 
(not to be confused with heavy WIMP mass $M$ used elsewhere in the paper) 
in $\Delta$, and work in $d=4$ dimensions. Naming $x(1-x) \equiv z$, we find,
\begin{align}
F(1) &= {i\over (4\pi)^2} \Bigg[ z m_V^2   \left(1 - {M^2\over z m_V^2}\right)^{2} \Bigg]^{-1}  
\Bigg[ {M^2\over z m_V^2} \log{M^2\over z m_V^2} - {M^2\over z m_V^2} + 1 \Bigg] 
\,,
\nl
H(1) &= {i\over (4\pi)^2}  {\partial \over \partial m_V^2 } \Bigg\{ 8   \Bigg[ z m_V^2 \left( 1 - {M^2\over zm_V^2} \right) \Bigg]^{-1} \Bigg[ 
\sqrt{m_V^2-\delta^2} \arctan\left( \sqrt{ {m_V^2\over \delta^2}-1} \right) 
\nl
&\quad 
- \sqrt{{M^2\over z} - \delta^2} \arctan\left( \sqrt{{M^2\over z \delta^2} - 1 } \right) 
- {\delta\over 2} \log{z m_V^2 \over M^2} 
\Bigg] \Bigg\} \,.
\end{align}
The remaining integrals can be obtained by differentiating the above results with respect to $M$.
In practice, we evaluate the remaining integral over Feynman parameter $x$ (or $z$) numerically. 

\section{Numerical inputs}\label{app:inputs}

\begin{table}[h]
\begin{center}
\begin{tabular}{c|c|c}
Parameter & Value & Reference\\
\hline
$|V_{td}|, |V_{ts}|$ & $\sim 0$ & -\\
$|V_{tb}|$ & $\sim 1$ & -\\
$m_e$ & $0.511 \,{\rm MeV}$ &\cite{Beringer:1900zz} \\
$m_\mu$ & $106 \,{\rm MeV}$ &\cite{Beringer:1900zz} \\
$m_\tau$ & $1.78 \,{\rm GeV}$ &\cite{Beringer:1900zz} \\
$m_h$ & $126\,{\rm GeV}$ & \cite{Aad:2012tfa, Chatrchyan:2012ufa} \\
$m_W$ & $80.4\,{\rm GeV}$ & \cite{Beringer:1900zz}\\
$m_Z$ & $91.188\,{\rm GeV}$ &\cite{Beringer:1900zz}\\
\end{tabular} 
\qquad
\begin{tabular}{c|c|c}
Parameter & Value & Reference\\
\hline
$m_t$ & $172\,{\rm GeV}$ & \cite{Martin:2009iq}\\
$m_b$ & $4.75\,{\rm GeV}$ & \cite{Martin:2009iq}\\
$m_c$ & $1.4\,{\rm GeV}$ & \cite{Martin:2009iq}\\
$m_s$ & $93.5\,{\rm MeV}$ & \cite{Beringer:1900zz}\\
$m_d$ & $4.70\,{\rm MeV}$ & \cite{Beringer:1900zz}\\
$m_u$ & $2.15\,{\rm MeV}$ & \cite{Beringer:1900zz}\\
$c_W$ & $m_W/m_Z$ & -\\
$\alpha_s(m_Z)$ & 0.118 & \cite{Beringer:1900zz}\\
\end{tabular} 
\end{center}
\caption{\label{tab:inputs}
Inputs to the numerical analysis.
} 
\end{table}

We use the inputs of Table~\ref{tab:inputs} in the numerical analysis 
of coefficients appearing in Fig.~\ref{fig:cplots}.   Light fermion masses
enter the analysis indirectly via the onshell renormalization scheme.  
The matching in (\ref{eq:deltav}) requires a limit of the photon two-point function which 
receives contributions from momentum regions of
light ($u$, $d$ and $s$) quark loops that are outside the domain of validity of QCD perturbation theory.  
A complete nonperturbative treatment of this function is
not numerically relevant to the present analysis; for definiteness, we model these contributions 
using $\overline{\rm MS}$ light quark masses (cf. Table~\ref{tab:inputs}) in the one-loop evaluation of 
the two-point function.   Varying these mass inputs 
by an order of magnitude in either direction does not appreciably 
change the numerical matching coefficients of Fig.~\ref{fig:cplots}. 

\end{appendix}

\end{fmffile} 

\end{document}